# Data- and Physics-driven Deep Learning Based Reconstruction for Fast MRI: Fundamentals and Methodologies

Jiahao Huang, Yinzhe Wu, *Student Member, IEEE*, Fanwen Wang, Yingying Fang, Yang Nan, Cagan Alkan, Daniel Abraham, Congyu Liao, Lei Xu, Zhifan Gao, Weiwen Wu, Lei Zhu, Zhaolin Chen, Peter Lally, Neal Bangerter, Kawin Setsompop, Yike Guo, *Fellow, IEEE*, Daniel Rueckert, *Fellow, IEEE*, Ge Wang, *Fellow, IEEE*, and Guang Yang, *Senior Member, IEEE*

***Abstract*—Magnetic Resonance Imaging (MRI) is a pivotal clinical diagnostic tool, yet its extended scanning times often compromise patient comfort and image quality, especially in volumetric, temporal and quantitative scans. This review elucidates recent advances in MRI acceleration via data and physics-driven models, leveraging techniques from algorithm unrolling models, enhancement-based methods, and plug-and-play models to the emerging full spectrum of generative model-based methods. We also explore the synergistic integration of data models with physics-based insights, encompassing the advancements in multi-coil hardware accelerations like parallel imaging and simultaneous multi-slice imaging, and the optimization of sampling patterns. We then focus on domain-specific challenges and opportunities, including image redundancy exploitation, image integrity, evaluation metrics, data heterogeneity, and model generalization. This work also discusses potential solutions and future research directions, with an emphasis on the role of data harmonization and federated learning for further improving the general applicability and performance of these methods in MRI reconstruction.**

## I. Introduction

MAGNETIC Resonance Imaging (MRI) is a non-invasive, ionizing radiation-free, and highly versatile imaging modality that has emerged as an essential tool for assessing a range of diseases over the past few decades. Unlike computed tomography (CT) scans, which capture images via projection, MRI sequentially acquires images through sampling in frequency space, commonly referred to as k-space [1]. The MR image can be easily reconstructed from k-space acquisition by applying the inverse Fourier transform, provided that the k-space data is sampled at least at the Nyquist rate. However, the sequential nature of MRI acquisition often results in prolonged durations, posing a major limitation to MRI and hindering its broader clinical application. This inherently lengthy process can cause significant discomfort for patients, compromise the spatiotemporal resolution, and increase susceptibility to motion artifacts.

Often in practice, a sub-Nyquist sampling rate is used to speed up imaging, where the aforementioned naive reconstruction approach by inverse Fourier transform can result in aliasing artifacts. Overcoming these artifacts necessitates additional domain knowledge and information to reconstruct the image from the undersampled k-space. A prevalent hardware-based technique proposed for addressing these artifacts is parallel imaging [2], which takes advantage of the redundancy in spatial information from measurements across coils in different spatial locations. Nevertheless, its achievable acceleration rates are often constrained [2].

From a signal processing perspective, compressed sensing (CS) was subsequently proposed as one such strategy that leverages the inherent compressibility of MRI to reconstruct the image from k-space data subsampled with a pseudo-random sparse sampling pattern [3]–[5]. The key to a high-quality yet accelerated MRI scan through compressed sensing is to

Manuscript received January 29, 2024. This study was supported in part by the Imperial College London President's PhD Scholarship, the ERC IMI (101005122), the H2020 (952172), the MRC (MC/PC/21013), the Royal Society (IEC\NSFC\211235), the NVIDIA Academic Hardware Grant Program, the SABER project supported by Boehringer Ingelheim Ltd, NIHR Imperial Biomedical Research Centre (RDA01), Wellcome Leap Dynamic Resilience, and the UKRI Future Leaders Fellowship (MR/V023799/1). (Jiahao Huang and Yinzhe Wu are co-first authors. Correspondence authors: Jiahao Huang, Yinzhe Wu and Guang Yang).

Jiahao Huang, Yinzhe Wu, Fanwen Wang, Yingying Fang, Yang Nan, Peter Lally and Guang Yang are with the Department of Bioengineering, Faculty of Engineering, Imperial College London, Exhibition Road, London, SW7 2AZ, UK. (email: j.huang21@imperial.ac.uk;yinzhe.wu18@imperial.ac.uk; fanwen.wang@imperial.ac.uk;y.fang@imperial.ac.uk;y.nan20@imperial.ac.uk; p.lally@imperial.ac.uk; g.yang@imperial.ac.uk)

Guang Yang is also with the Imperial-X and School of Biomedical Engineering & Imaging Sciences, King's College London.

Cagan Alkan and Daniel Abraham are with Department of Electrical Engineering, Stanford University, CA 94305, United States of America (email: calkan@stanford.edu, abrahamd@stanford.edu).

Lei Xu is with the Department of Radiology, Beijing Anzhen Hospital, Capital Medical University, 2nd Anzhen Road, Chaoyang District, Beijing, 100029, China. (email: leixu2001@hotmail.com).

Zhifan Gao and Weiwen Wu are with the School of Biomedical Engineering, Sun Yat-sen University, Guangzhou, Guangdong, 510006, China. (email: gaozhifan@gmail.com; wuweiw7@mail.sysu.edu.cn).

Lei Zhu and Yike Guo are with the Hong Kong University of Science and Technology, Hongkong, China (email: leizhu@ust.hk; yikeguo@ust.hk).

Zhaolin Chen is with the Department of Data Science and AI, Faculty of Information Technology, Monash University, Wellington Rd, Clayton, Victoria 3800, Australia (email: zhaolin.chen@monash.edu).

Neal Bangerter is with Electrical and Computer Engineering Department, Boise State University, ID 83725, United States of America (email: nealbangerter@boisestate.edu).

Yinzhe Wu, Congyu Liao and Kawin Setsompop are with the Department of Radiology, Stanford University, CA 94305, United States of America (email: cyliao@stanford.edu, kawins@stanford.edu).

Daniel Rueckert is with the Department of Computing, Imperial College London, South Kensington Campus London, SW7 2AZ, UK. and Technical University of Munich, Munich, 80333, Germany. (email: d.rueckert@imperial.ac.uk).

Ge Wang is with the Biomedical Imaging Center, Rensselaer Polytechnic Institute (RPI), New York, USA. (email: wangg6@rpi.edu).

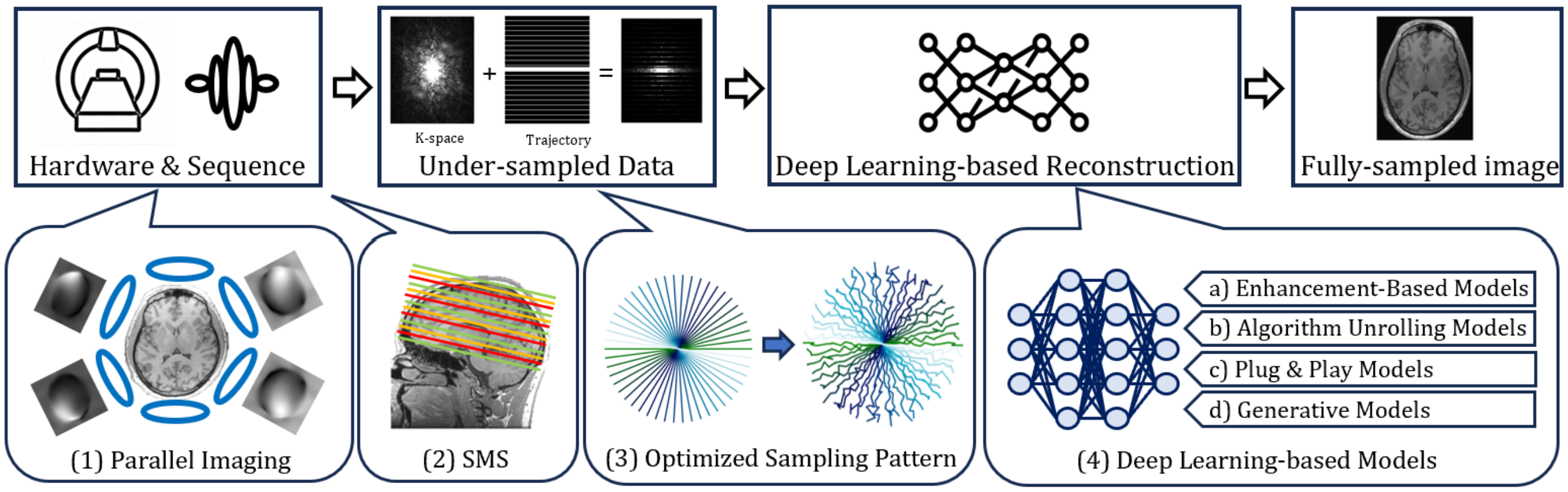

**Figure 1.** An overview of the overall pipeline of MRI acquisition and reconstruction work pipeline. Deep learning methods can improve the quality of the final reconstructed image by benefitting from advanced hardware and sequence setting and deep learning based reconstruction. Data- and physics-driven deep learning methods can be adapted to multi-coil **(1)** standard parallel imaging and **(2)** Simultaneous Multi-Slice (SMS) imaging, **(3)** further enhance the subsampled k-space data by optimizing the sampling pattern, and **(4)** enhance the acquired undersampled images by a range of deep learning based models (including (a) Enhancement-Based Models, (b) Algorithm Unrolling Models, (c) Plug & Play (PnP) models, and (d) Generative Models).

incorporate prior knowledge, often in the form of sparsity [3], low-rank [6], and/or structured low-rank [7], to recover the unknown signal from limited measurements.

Alternatively, low-rank models are also highly effective in MRI reconstruction because various aspects of MRI data are often highly correlated. This correlation can be observed in the image domain where spatio-temporal (or purely spatial) images [8], or their patches [9], frequently reside in a low-dimensional linear subspace. Constructing a matrix from such spatial images or patches typically results in a low-rank matrix. The nuclear norm is commonly employed to promote low-rankness in this matrix, which can be efficiently implemented using proximal methods to recover images from limited measurements within a compressed sensing framework. Low-rank models also extend to k-space. Techniques such as SAKE [10], LORAKS [7], [11] and ESPIRIT [12] exploit low-rank properties by assembling k-space patches into a matrix. In ESPIRIT [12], the matrix's null space is used to extract information about coil sensitivity maps. SAKE [10] and LORAKS [7], [11] enforce low-rank structure in a carefully designed matrix to impose reasonable assumptions on the image, such as slowly varying phase and limited spatial support, which further constrains ill-posed reconstructions.

A number of iterative optimization algorithms have been put forth for reconstructing such undersampled MRI data. Notable methods among them include proximal gradient descent and the alternating-direction method of multipliers (ADMM) [4], [13]. However, the iterative nature of these optimization algorithms and the high computational complexity of these algorithms significantly extends the reconstruction time [14]. Moreover, these algorithms rely on hand-crafted sparsity priors (e.g., wavelet transform and total variation (TV)), which, due to their inherent assumptions [3], [15], can result in residual artifacts appearing in reconstructed images at high acceleration rates [5]. This could be in the form of small high-frequency oscillatory artifacts for wavelet-based reconstruction methods [3] and streaking artifacts for TV-based reconstruction methods [15]. Furthermore, these hand-crafted sparsity priors and their assumptions often fail to represent the complex attributes of MR images. For example, the theory of TV assumes that the scanned object is piecewise constant, restricting the number of edges and intensity jumps that can present in the image [15], which does not follow the anatomy of humans.

To address these limitations, apart from conventional machine learning methods [16] (e.g., blind CS [17], transform learning [18], dictionary learning [19], analysis operator learning [20]), deep learning methods for MRI reconstruction have been proposed to learn the distributional properties of MR images from large datasets, avoiding manual assumptions and thereby facilitating more accurate reconstruction outcomes with further reduced artifacts [14]. Capable of modeling complex nonlinear relationships, deep learning models – spanning from algorithm unrolling models, enhancement-based methods and plug-and-play models to the rapidly emerging generative models – have demonstrated impressive outcomes, enhancing both the quality and accuracy of reconstructed MRI images over conventional methods. For example, in cardiac cine MRI, MoDL can produce reconstructed images with higher PSNR compared to GRAPPA when accelerated by 8-fold or higher [21]. As a data-driven approach leveraging large amounts of training data, these deep learning networks can be trained to efficiently reconstruct high-quality images from heavily subsampled data, reducing the need for lengthy scans and enhancing both patient comfort and imaging results [22].

This work aims to provide an overview of recent advances in MRI acceleration via data and physics-driven models **(Figure 1)**, including:

**(a)** The evolution of deep learning networks in MRI reconstruction, including their landscape across **1)** algorithm unrolling models, **2)** Plug-and-Play (PnP) models, **3)** enhancement-based methods, and a full spectrum of emerging **4)** generative model-based methods (**4.i.** Variational Autoencoders (VAE), **4.ii.** Generative Adversarial Networks (GAN), **4.iii.** Deep Energy-based Models (DEM), **4.iv.** Diffusion Models, **4.v.** Normalizing Flow-based (NF) models) for the reconstruction of undersampled images.

**(b)** The synergistic integration of data models with physics-based insights. This encompasses the advancements in multi-coil hardware accelerations such as parallel imaging and simultaneous multi-slice (SMS) imaging, and the optimization of sampling patterns to enhance imaging results.

**(c)** The exploration of domain-specific challenges and opportunities. This includes **1)** mitigating data scarcity**, 2) exploring decision-making for acquisition and reconstruction by reinforcement learning, 3)** exploiting imaging redundancy, **4)** selecting evaluation metrics, **5)** ensuring image integrity, **6)** addressing data heterogeneity, **7)** improving model generalization and domain shift problems and **8)** alleviating memory constraints.

This article offers a comprehensive review of recent advancements in MRI technology, detailing their capabilities and constraints, with a particular focus on their applications in clinical environments. It aims to assist researchers and practitioners by providing a nuanced understanding of the integration of sophisticated data and physics-based learning models, thereby enhancing the efficacy of MRI procedures.

While numerous reviews have contributed to the domain of physics- and data-driven MRI reconstruction [4], [23]–[28], there are distinct parts of the landscape that remain underexplored. Fessler et al. [4], Ahmad et al. [23] and Monga et al. [24] provided detailed reviews and comprehensive analysis, of which subject was however limited to a subclass of reconstruction models, i.e., classic optimization methods, PnP models, algorithm unrolling models and parallel image-based models. Knoll et al. [25] and Christodoulou et al. [26] focused on a rather specific kind of MR data, i.e., parallel image and dynamic MRI. Although Liang et al. [27] presents a thorough analysis of deep learning methods for MRI reconstruction available up to its publication, its taxonomy only considered unrolling and non-unrolling methods and failed to encompass the recent advancements that have since evolved. In contrast to Hammernik et al. [28], which provided a comprehensive and insightful analysis of physics-driven computational MRI, sharing a similar research object with **Section II** in our work, we instead provided and extended a more detailed and comprehensive discussion about the up-to-date landscape of modern fast-moving generative model-based methods. Moreover, our study further investigates the synergistic integration of data models with physics-based insights, encompassing the advancements in multi-coil hardware accelerations like parallel imaging and SMS imaging, as well as the sampling patterns optimization.

To guide the reader through this review, we provide an outline of the subsequent sections:

- **Section II** delves into the evolution of deep learning networks in MRI reconstruction, covering algorithm unrolling models, Plug-and-Play (PnP) models, and enhancement-based methods. It also explores a comprehensive spectrum of emerging generative model-based methods for reconstructing undersampled images, including Variational Autoencoders (VAEs), Generative Adversarial Networks (GANs), Deep Energy-based Models (DEMs), Diffusion Models, and Normalizing Flow-based (NF) models.
- **Section III** discusses the synergistic integration of data-driven models with physics-based insights, highlighting advancements in multi-coil hardware accelerations such as parallel imaging and simultaneous multi-slice (SMS) imaging, as well as the optimization of sampling patterns to enhance imaging results.
- **Section IV** explores domain-specific challenges and opportunities, including mitigating data scarcity, exploring decision-making for acquisition and reconstruction by reinforcement learning, exploiting imaging redundancy, selecting appropriate evaluation metrics, ensuring image integrity, dealing with data heterogeneity, improving model generalization and domain shift problems, and alleviating memory constraints.

## II. SYNERGISTIC DATA- AND PHYSICS-DRIVEN MODELS

The physics-based forward model for MRI acquisition, following discretization, can usually be expressed as (1).

$$\mathbf{y} = \mathbf{A}\mathbf{x} + \mathbf{n}, \quad (1)$$

where $\mathbf{x} \in \mathbb{C}^n$ is the image of interest, $\mathbf{y} \in \mathbb{C}^m$ is its corresponding frequency space (k-space) measurement, $\mathbf{n} \in \mathbb{C}^m$ is discretized measurement noise, and $\mathbf{A}: \mathbb{C}^n \rightarrow \mathbb{C}^m$ is the forward MRI encoding operator, which, if in its simplest form in a single-coil case, could be a subsampled discrete Fourier transform $\boldsymbol{\mathcal{F}}_{\boldsymbol{\Omega}}: \mathbb{C}^n \rightarrow \mathbb{C}^m$ sampling the k-space locations as specified by $\boldsymbol{\Omega}$.

Upon acquisition of the undersampled k-space, the next step is to reconstruct the acquired image space, aiming to eliminate the aliasing artifacts resulting from sparse subsampling and to restore the original image quality.

Recently, deep learning based approaches have emerged as powerful tools for MRI reconstruction. These encompass a variety of deep learning models, each with its own distinct features: **1) Algorithm Unrolling** models merge model-based algorithms with deep learning, improving interpretability but increasing the computational load. **2) PnP** models gain flexibility by using denoisers as implicit regularizers, with their performance critically depending on the chosen denoising algorithm. **3) Enhancement-based** methods swiftly convert sub-sampled to fully sampled images, relying on large paired datasets that are typically limited in medicine. The increasingly popular **4) Generative Model**-based methods learn the prior of MRI to solve reconstruction challenges, necessitating complex model design for realistic outputs (specifically through **4.i.** VAEs, **4.ii.** GANs, **4.iii.** EBMs, **4.iv**. Diffusion Models, **4.v.** NF-based models).

The interplay between data and physics plays a crucial role in the design and formulation of deep learning based approaches. These models incorporate the physics information at different stages. Specifically, the physics information is either integrated only at the test time or during both training and testing phases. Unconditional generative models and PnP models disentangle physics-based forward models from network training and learn useful data-driven priors or denoisers, respectively. They subsequently incorporate MRI physics through optimization at the inference stage. Conversely, algorithm unrolling, image enhancement methods and conditional generative models embed the physics information directly at the training phase by invoking the forward model as a part of the model architecture or at the input of the model. In general, models that separate physics from training can effectively reconstruct data collected under various acquisition settings such as different acceleration factors. On the contrary, the models that incorporate physics during the training process excel at the specific acquisition scenario for which they were trained, but require retraining or fine-tuning for different acquisition scenarios.

In this section, we first introduce the classical optimization methods and then provide an extensive overview of deep learning methods in these four different categories.

This section serves as an overview of the technical aspects of various methodological categories. For readers seeking more detailed technical settings (parameter settings, code availability, runtime, etc.), please see **Supplementary Table S1** for extended comparisons, although be reminded that conclusions in any specific article may not hold if the case scenario (e.g., modality) changes. For further benchmarking performance comparison, please refer to a benchmarking article (e.g., [29]).

### *A. Preliminaries: Classical Optimization Approach*

The classical approach to solving the inverse problem of MRI reconstruction through an optimization approach targets a regularized least-square as formulated below (2).

$$\hat{\mathbf{x}} = \arg\min_{\mathbf{x}} \frac{1}{2}\|\mathbf{A}\mathbf{x}-\mathbf{y}\|_2^2 + \lambda\mathcal{R}(\mathbf{x}), \quad (2)$$

where $\mathcal{R}(\cdot)$ denotes a class of regularizers balanced by $\lambda$. In the conventional method for compressed sensing, such $\mathcal{R}(\cdot)$ regularizers could be explicitly defined as, for example, $l_1$ norm of the sparsifying linear transform domain coefficients [4]. For conciseness, the sparse transformation $\mathbf{D}$ (e.g., wavelet transformation or total variation transformation) is subsumed under the regularizer $\mathcal{R}(\cdot)$, although it is often explicitly included in other literature. Likewise, apart from regularizing the model using sparsity, other studies exploit low-rank structure [6], structured low-rank constraint [7], [30], or a combination of both sparsity and low-rank models [8], [31], [32]. A classical approach to minimize (2) through an iterative algorithm is based on gradient descent (3):

$$\mathbf{x}_k = \mathbf{x}_{k-1} - \eta\mathbf{A}^{\mathrm{H}}(\mathbf{A}\mathbf{x}_{k-1}-\mathbf{y}) - \eta\nabla_{\mathbf{x}}\mathcal{R}(\mathbf{x}_{k-1}), \quad (3)$$

where $\mathbf{x}_k$ is the image of interest at the $k$-th iteration, and $\eta$ is the step size. $\mathbf{A}^{\mathrm{H}}$ denotes the Hermitian transpose of the encoding operator $\mathbf{A}$. $\nabla_{\mathbf{x}}$ denotes the gradient operator with respect to $\mathbf{x}$, assuming the differentiability of $\mathcal{R}(\mathbf{x})$ with respect to all $\mathbf{x}$. Otherwise, if $\mathcal{R}(\mathbf{x})$ becomes non-differentiable, a valid sub-gradient operator could replace $\nabla_{\mathbf{x}}$.

However, to facilitate convergence, proximal gradient descent is often considered as a more suitable candidate to generalize the abovementioned differentiability assumption for non-differentiable terms in (3), as formulated in (4a-b).

$$\begin{cases} \mathbf{z}_k = \mathbf{x}_{k-1} - \eta\mathbf{A}^{\mathrm{H}}(\mathbf{A}\mathbf{x}_{k-1}-\mathbf{y}), & (4a) \\ \mathbf{x}_k = \arg\min_{\mathbf{x}} \frac{1}{2}\|\mathbf{z}_k-\mathbf{x}\|_2^2 + \lambda\mathcal{R}(\mathbf{x}) \triangleq \mathrm{prox}_{\lambda\mathcal{R}}(\mathbf{z}_k), & (4b) \end{cases}$$

where $\mathbf{x}_k$ and $\mathbf{z}_k$ are respectively the image of interest and intermediate image at the $k$-th iteration. (4a) poses the restriction via data consistency (DC), and (4b) corresponds to the proximal operator $\mathrm{prox}_{\lambda\mathcal{R}}$ on regularizer $\lambda\mathcal{R}(\cdot)$. $\eta$ denotes the step size. In the case where $l_1$ regularizer with sparse transformation $\mathbf{D}$, e.g., wavelet transformation, employed ($\mathcal{R}(\cdot) = \|\mathbf{D}\cdot\|_1$), the proximal gradient descent specifically evolves into an iterative shrinkage-thresholding algorithm (ISTA) [33]–[35] (5a-b):

$$\begin{cases} \mathbf{z}_k = \mathbf{x}_{k-1} - \eta\mathbf{A}^{\mathrm{H}}(\mathbf{A}\mathbf{x}_{k-1}-\mathbf{y}), & (5a) \\ \mathbf{x}_k = \mathbf{D}^{\mathrm{T}}\,\mathrm{soft}(\mathbf{D}\mathbf{z}_k, \lambda), & (5b) \end{cases}$$

where $\mathrm{soft}(\cdot,\lambda) = \mathrm{sign}(\cdot)\max(|\cdot| - \lambda, 0)$ denotes the soft thresholding operator, and $\mathbf{D}^{\mathrm{T}}$ is the transpose transformation of $\mathbf{D}$, assuming that $\mathbf{D}$ is unitary.

Another class of conventional popular approaches comes from variable splitting, such as alternating direction method of multipliers (ADMM) [4], [13]. ADMM offers a strategy to decompose complex problems into simpler sub-problems, while retaining desirable convergence properties (6a-c).

$$\begin{cases} \mathbf{x}_k = \arg\min_{\mathbf{x}} \frac{\rho}{2}\|\mathbf{x}+\mathbf{u}_{k-1}-\mathbf{z}_{k-1}\|_2^2 + \frac{1}{2}\|\mathbf{A}\mathbf{x}-\mathbf{y}\|_2^2, & (6a) \\ \mathbf{z}_k = \arg\min_{\mathbf{z}} \frac{\rho}{2}\|\mathbf{x}_k+\mathbf{u}_{k-1}-\mathbf{z}\|_2^2 + \lambda\mathcal{R}(\mathbf{z}), & (6b) \\ \mathbf{u}_k = \mathbf{u}_{k-1} + \mathbf{x}_k - \mathbf{z}_k, & (6c) \end{cases}$$

where $\mathbf{x}_k$ is the image of interest, $\mathbf{z}_k$ and $\mathbf{u}_k$ denote intermediate images at the $k$-th iteration. $\rho$ is a penalty weight. (6a), (6b) and (6c) correspond to DC, proximal operator and dual-update subproblems respectively. By alternating between updating different variable sets, ADMM efficiently tackles problems where traditional methods might falter, illustrating the diverse optimization strategies in MRI reconstruction.

Apart from these, several other optimization methods have proven effective in MRI reconstruction. **Alternating minimization** [36] (a form of **block coordinate descent** [37]) involves breaking down the optimization problem into smaller subproblems that are solved in an alternating fashion. For MRI reconstruction, one may alternate between separately updating the magnitude and the phase of the complex-valued image [36]. **Majorize-Minimize (MM) framework** [38] constructs a surrogate function that majorizes the objective function and minimizes this surrogate iteratively [36].

### *B. Enhancement-based Method*

Recent deep learning advancements have introduced a range of powerful neural architectures such as Convolutional Neural Networks (CNN) [39], [40], Graph Neural Networks (GNN) [41], and Transformers [42], [43], all significant in natural image and MRI enhancement. Enhancement models (**Figure 2**) leverage these to map undersampled $\mathbf{y}$ to paired fully sampled $\mathbf{x}$, through $\hat{\mathbf{x}} = \mathrm{f}(\mathbf{A}^{\mathrm{H}}\mathbf{y};\,\hat{\boldsymbol{\theta}})$ or $\mathrm{f}(\mathbf{y};\,\hat{\boldsymbol{\theta}})$, where $\mathrm{f}(\cdot\,;\,\hat{\boldsymbol{\theta}})$ denotes the network with trained parameters $\hat{\boldsymbol{\theta}}$. A widely used technique for further refined reconstructed process is residual learning [39] which enhances reconstruction by adding $\mathbf{A}^{\mathrm{H}}\mathbf{y}$ to the output $\hat{\mathbf{x}}$.

Each of these wide range of architectures possesses its own unique set of specialties and features. Specifically, CNN, known for feature extraction, widely applies architectures like ResNet (residual learning strategy) [39] and U-Net [40] in MRI reconstruction [44]. Recent ConvXNet [45] further explored the potential of CNNs by "modernizing" a standard ResNet [39] with the design of a Vision Transformer [42]. GNN processes graph-based data, handling complex relationships, and is utilized in MRI for analyzing patch-based graphs from pixel arrays and capturing the non-local self-similarity and non-Euclidean relationships guided by the graph connection [46]. Lately, Transformer [42], [43], initially designed for natural language processing, has transcended its original domain and found applications in the realm of image enhancement [47]. With their multi-head self-attention, Transformer excels in capturing long-range dependencies in images, applied in tasks like MRI reconstruction and super-resolution [48]–[51]. Specifically, Feng et al. developed an integrated framework [48] for simultaneous reconstruction and super-resolution, subsequently further improved with task-specific cross-attention modules. Huang et al. [49] introduced a sensitive map coupled shifting-window (Swin) Transformer-based network for MRI reconstruction.

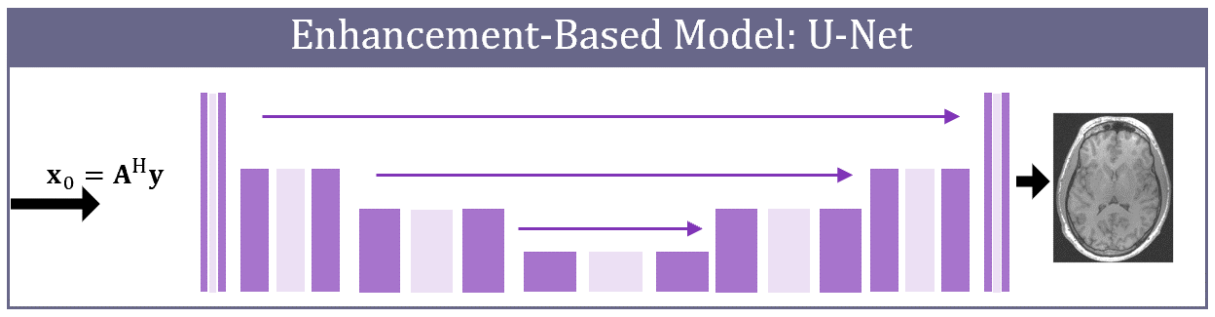

**Figure 2.** A typical Enhancement-based Model (e.g., U-Net [40]) maps a subsampled **y** to fully sampled **x** by a chain of convolutional blocks. U-Net [40] is a CNN architecture designed for image segmentation, featuring an encoder-decoder structure where skip connections link (indicated as the arrows in the figure) corresponding layers in the encoder and decoder, allowing detailed spatial information from earlier layers to be preserved and integrated into the reconstruction process.

The models mentioned above are not mutually exclusive. Often, these models are integrated with one another, thereby extending a diverse toolkit for MRI reconstruction. For instance, Huang et al. [49], [51] integrated CNN modules for latent space projection with Swin Transformer blocks for feature extraction. These advancements highlight the rich, evolving landscape of neural networks in MRI reconstruction, emphasizing their broad applicability.

### *C. Algorithm Unrolling Model*

Algorithm Unrolling Models [24], [52]–[54] (**Figure 3**) integrate iterative algorithms with deep learning, merging conventional model-based methods and deep learning's adaptability. They unfold each iterative algorithm step into a corresponding neural network layer with learnable parameters, optimized during training.

A line of algorithm unrolling models is based on generalized CS algorithms [52], [55], where the conventional CS algorithm, e.g., ISTA, ADMM, is unfolded into a network with a fixed number of stages. For example, ISTA-Net [52] unrolls ISTA iterations into discrete network stages, each handling data consistency and regularization with learnable networks and parameters, which facilitates the end-to-end optimization of parameters to streamline the learning process.

Specifically, in ISTA-Net [52], $N$ stages of deep networks represent $N$ ISTA iterations. Each stage, $k$, comprises a data consistency module, $\mathbf{z}_k = \mathbf{x}_{k-1} - \eta_k \mathbf{A}^{\mathrm{H}}(\mathbf{A}\mathbf{x}_{k-1} - \mathbf{y})$, and a regularization module, $\mathbf{x}_k = \tilde{\mathrm{h}}_k(\mathrm{soft}(\mathrm{h}_k(\mathbf{z}_k; \theta_k), \lambda_k); \tilde{\theta}_k)$, paralleling ISTA's structure (5a-b). Here, $\mathrm{h}_k(\cdot)$ and $\tilde{\mathrm{h}}_k(\cdot)$ are learnable networks at $k$-th stage, potentially consisting of convolution and ReLU layers or transformer blocks [56], for forward and backward transformations, respectively. $\tilde{\mathrm{h}}_k(\cdot; \tilde{\theta}_k)$ is designed as the left inverse of $\mathrm{h}_k(\cdot; \theta_k)$, mirroring wavelet transformation's invertibility and imposing $\tilde{\mathrm{h}}_k \circ \mathrm{h}_k = \mathbf{I}$ through the loss function. Differing from traditional ISTA, ISTA-Net introduces learnable parameters, $\eta_k$ and $\lambda_k$, for step size and thresholding control, allowing end-to-end optimization of $\mathbf{\Theta} = \{\theta_k, \tilde{\theta}_k, \eta_k, \lambda_k\}_{k=1}^{N}$ during training.

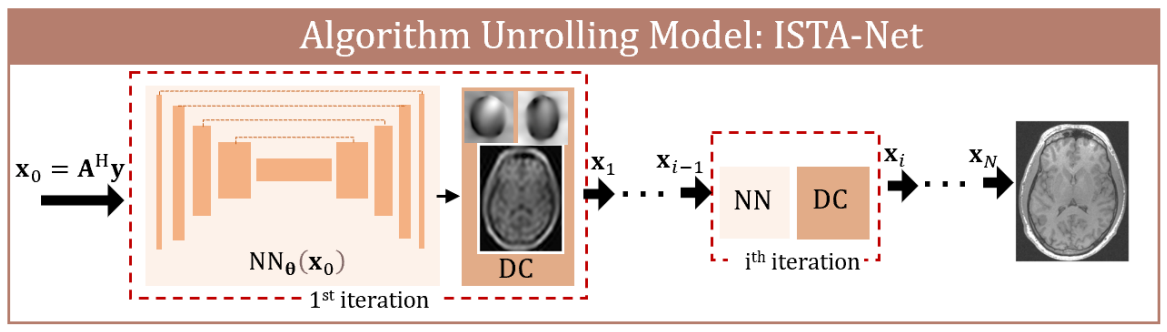

**Figure 3.** A typical Algorithm Unrolling Model (e.g., ISTA-Net [52]) unrolls a block of a neural network (NN) and a data consistency (DC) layer for a fixed number of iterations ($N$). The DC layer here enforces fidelity with the raw k-space (5a). The whole iteration chain is trained end-to-end with pairs of subsampled and fully sampled images.

In another approach, works like DCCNN [53] and MoDL [54] introduce residual-based regularization, framing the optimization as minimizing the discrepancy between the target image $\mathbf{x}$ and its denoised, de-aliased version $\mathrm{h}(\mathbf{x}; \boldsymbol{\theta})$. The optimization, rewriting the regularizer from (2) to give $\hat{\mathbf{x}} = \arg\min_{\mathbf{x}} \frac{1}{2}\|\mathbf{A}\mathbf{x} - \mathbf{y}\|_2^2 + \lambda\|\mathrm{h}(\mathbf{x}; \boldsymbol{\theta}) - \mathbf{x}\|_2^2$, leverages a neural network $\mathrm{h}(\cdot; \boldsymbol{\theta})$ with learnable parameters $\boldsymbol{\theta}$ for regularization.

DCCNN [53] addresses this optimization problem by unrolling the proximal gradient descent algorithm, substituting the proximal operator with a CNN. It structures the network, $\mathrm{f}(\cdot, \mathbf{\Theta})$, as a cascade of $N$ alternately arranged CNN modules $\mathbf{z}_k = \mathrm{h}(\mathbf{x}_{k-1}; \theta_k)$ and DC modules $\mathbf{x}_k = \mathbf{z}_k - \eta_k \mathbf{A}^{\mathrm{H}}(\mathbf{A}\mathbf{z}_k - \mathbf{y})$. Throughout training, it adapts the parameters $\mathbf{\Theta} = \{\eta_k, \theta_k\}_{k=1}^{N}$ learnable in an end-to-end fashion.

Similarly, MoDL [54] solves the optimization problem using a sequence of steps that involve a CNN module $\mathbf{z}_k = \mathrm{h}(\mathbf{x}_{k-1}; \theta)$, and a conjugate gradient (CG)-based data consistency module $\mathbf{x}_k = (\eta\mathbf{I} + \mathbf{A}^{\mathrm{H}}\mathbf{A})^{-1}(\eta\mathbf{z}_k + \mathbf{A}^{\mathrm{H}}\mathbf{y})$, particularly suited for multi-coil data where direct matrix inversion is computationally challenging. Unlike DCCNN, MoDL shares parameters across stages, setting $\mathbf{\Theta} = \{\eta_k, \theta_k\}_{k=1}^{N} = \{\eta, \theta\}$, promoting consistency and efficiency in parameter updating.

Going further, bilevel optimization [57] can be employed in MRI reconstruction [58], which involves iteratively solving an inner problem to find task-invariant regularization and an outer problem to optimize task-specific regularization, further enhancing the model's robustness and generalizability. Moreover, unrolling methods can also be seen as an approximation to tackle the bilevel problem, as they iteratively solve the inner problem given by the regularization modules and the outer problem defined by the DC modules in an alternating fashion.

Additionally, besides the aforementioned methods applied in the image space, some methods employ networks in only k-space [59] or both image and k-space [60].

Traditional MRI reconstruction methods rely on iterative optimization and manually predefined regularizers, which may not fully capture the intricacies of MR images. Algorithm Unrolling Models address this by converting each iteration into learnable modules within a deep learning framework, thereby enhancing representation accuracy.

These models offer better theoretical guarantees and interpretability in comparison to pure enhancement-based models by providing provable convergence rates under certain assumptions about the data and learning parameters. Furthermore, their structured design, which mirrors the iterations of the algorithm they are based on, allows for enhanced interpretability. This systematic and theoretically grounded approach ensures that the reconstruction process incorporates vital prior information through the regularizer and maintains fidelity to the original optimization problem, promoting a more robust understanding of the model's behavior and outcomes [27].

Different approaches to unrolling algorithms offer unique advantages and disadvantages depending on the task and the structure of the optimization problem. Broadly, unrolling can follow various paths: one can choose to unroll algorithms using distinct modules for each iteration, where the parameters are independently learned at each stage, or employ a shared-parameter approach, where the same set of parameters is reused across all stages. For instance, models like DCCNN use stage-specific parameters, allowing for greater flexibility in learning iteration-dependent transformations but increasing the network parameters. On the other hand, approaches like MoDL, which share parameters across iterations, reduce the network parameters but may sacrifice flexibility in learning stage-specific dynamics. Thus, the trade-off between flexibility and efficiency becomes a crucial factor when deciding on the unrolling strategy.

Even when optimizing the same loss function, different unrolling methods can lead to varying outcomes, making it difficult to definitively determine which unrolling is "better". For example, ISTA-Net and DCCNN both unroll iterative algorithms, but they differ in how they handle regularization and data consistency, with ISTA-Net maintaining distinct parameters for each stage and DCCNN employing CNN-based regularization. While both aim to minimize a similar loss function, the architecture and unrolling design significantly affect their convergence rates, computational complexity, and performance. A shared-parameter approach like that of MoDL might work well for problems with consistent behavior across iterations, while stage-specific parameterization might excel in tasks where each iteration plays a distinct role in optimization.

Ultimately, the best unrolling strategy depends on the problem characteristics, the computational resources available, and the balance between interpretability and performance. Shared-parameter unrolling offers better parameter efficiency, making it ideal for resource-constrained settings, but it may struggle with more complex, non-uniform problems. Conversely, unrolling methods that use independent parameters per stage provide more flexibility and better fit complex data but may require greater computational power and fine-tuning. Hence, selecting the right unrolling technique is context-dependent, and no single method can be universally considered superior across all applications.

### *D. Plug-and-Play (PnP) Model*

Plug-and-Play (PnP) approaches [61] (**Figure 4**) in MRI reconstruction utilize advanced algorithms as modular components substituted (i.e., “plugged”) into an iterative process, without extensive modification of the reconstruction pipeline [23].

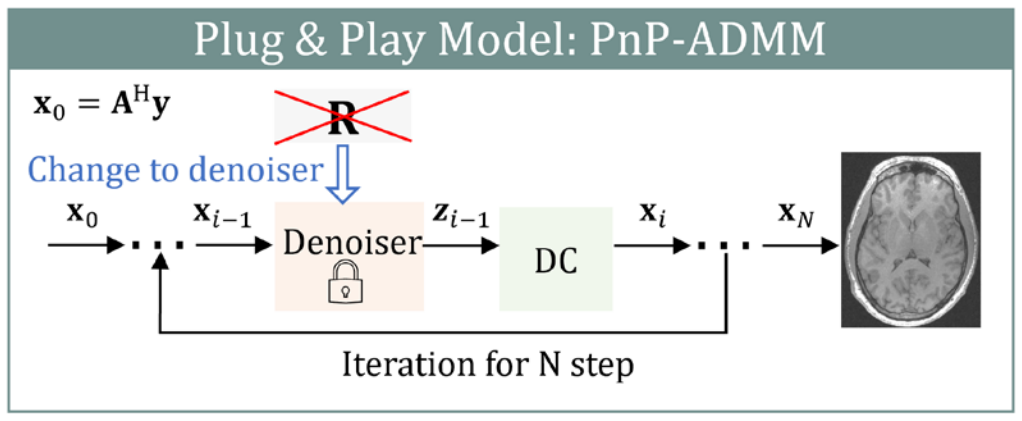

**Figure 4.** A typical Plug-and-Play (PnP) Model (e.g., PnP-ADMM) substitutes (i.e., “plugged”) conventional proximal denoiser (6b) (“**R**” in figure) in a conventional iterative algorithm (e.g., ADMM (6a-c)) with a pre-trained deep learning based denoiser (where the “lock” icon in the figure means that the denoiser is pre-trained and is not updated during iterations). This approach preserves the iterative architectures and regularizing functions of conventional iterative model-based algorithms and facilitates a modular and flexible structure composed of deep learning blocks.

For example, in Proximal-based PnP, such as PnP-ADMM [62], the traditional proximal denoiser is substituted with a powerful pretrained deep learning denoiser, $f(\cdot;\,\hat{\boldsymbol{\theta}})$. In the ADMM framework (6a-c) [63], this substitution occurs in the update step (6b), reformulating the optimization process. Here, (6a) minimizes a combination of fidelity and regularization terms, then in the **substituted (6b):** $\mathbf{z}_k = f(\mathbf{x}_k + \mathbf{u}_{k-1};\,\hat{\boldsymbol{\theta}})$, and finally (6c) updates as the difference between $\mathbf{x}_k$ and $\mathbf{z}_k$. This sequence restructures the traditional optimization steps into a coherent set of modular updates.

This approach has been empirically validated in various iterations, including PnP-FISTA [23], PnP-ADMM [64], and PnP- Half-Quadratic Splitting (HQS) [65], [66], showcasing its effectiveness and efficiency. Notable implementations include PnP-ADMM with a U-Net denoiser and GRAPPA for multi-coil MRI [64], and PnP-HQS with a CNN denoiser [66]. For a more detailed exploration, we direct readers to the extensive review by Ahmad et al. [23], which offers an in-depth analysis of PnP models for MRI reconstruction, including comparative studies of PnP-FISTA and PnP-ADMM models.

Beyond PnP, regularization by denoising (RED) [67] and regularization by artifact-removal (RARE) [68] offer nuanced alternatives. Instead of simply replacing the proximal operator with a deep denoiser, RED [67] incorporates the denoiser as a regularization term in the form $\mathcal{R}(\mathbf{x}) = \frac{1}{2}\mathbf{x}^T[\mathbf{x} - f(\mathbf{x};\,\hat{\boldsymbol{\theta}})]$, linking the penalty to the inner product between the image $\mathbf{x}$ and its denoising residual. This grants RED greater flexibility in choosing optimization methods, unlike the classic PnP model, where the modular denoiser can limit the choice of optimization techniques. RARE [68] further extends this concept, training the denoiser for broader artifact removal instead of just additive Gaussian noise, enhancing MRI reconstruction applicability. Collectively, these models leverage the "plug-and-play" concept, offering the flexibility to incorporate advanced denoisers and update denoising modules without reworking the entire reconstruction algorithm.

Notably, the algorithmic structures of PnP, RED, and algorithm unrolling models are distinct, even though they all employ denoising neural networks in their methodologies. PnP cycles between image updates and denoising, using an independent, pre-trained denoiser with variable iteration counts. RED differs by incorporating the denoiser as a regularizer within the optimization objective itself. In contrast, algorithm unrolling structurally embeds the denoising process within each network layer, with a fixed number of iterations equal to the network's depth, and updates the entire network end-to-end during training.

### *E. Generative Model-based Method*

The rising popularity of generative models presents a diverse array of options, including **1)** variational autoencoders (VAEs), **2)** generative adversarial networks (GANs), **3)** deep energy-based models (EBMs), **4)** diffusion models, and **5)** normalizing flow (NF)-based models. This section provides a summary of the application of these models in MRI reconstruction [69].

Generative models combine physics of MRI with the data in different ways. Unconditional generative models directly learn the data distribution from fully-sampled MR image datasets and MRI physics is only incorporated at test time. Conditional generative models, on the other hand, learn the distribution of fully sampled images conditioned on undersampled reconstructions during training time. The conditioning can be done in ways similar to those used in image enhancement models or algorithm unrolling. Therefore, there is an inherent overlap between the model families based on how MRI physics is conjoined with the data.

Notably, generative models are also commonly used in unsupervised learning, which is discussed in detail in Section IV-A. For readers seeking further insights, we have included an extensive technical discussion with mathematical derivations in the **Supplementary Section S-I**.

#### *1) Variational Autoencoder*

Variational Autoencoders (**Figure 5**) are probabilistic models that use variational inference to learn data distributions $p_{\boldsymbol{\theta}}(\mathbf{x})$. They encode data $\mathbf{x}$ into a latent space $\mathcal{Z}$ as a latent vector $\mathbf{z}$ and decode it back, utilizing a probabilistic encoder and a decoder with parameters $\boldsymbol{\phi}$ and $\boldsymbol{\theta}$, respectively. In inference, the decoder samples from $p_{\boldsymbol{\theta}}(\mathbf{x} \mid \mathbf{z})$ using a latent vector $\mathbf{z}$ from a prior distribution $p(\mathbf{z})$ to yield $p_{\boldsymbol{\theta}}(\mathbf{x})$. Yet in training, directly maximizing $p_{\boldsymbol{\theta}}(\mathbf{x})$ is challenging due to high-dimensional integrals involved. VAEs address this by employing the Evidence Lower Bound (ELBO), combining negative Kullback-Leibler (KL) divergence and the expected log-likelihood of the data. The model uses reparameterization for the latent variable $\mathbf{z}$ and minimizes the negative ELBO during training, refining the data distribution approximation and preserving a meaningful latent space.

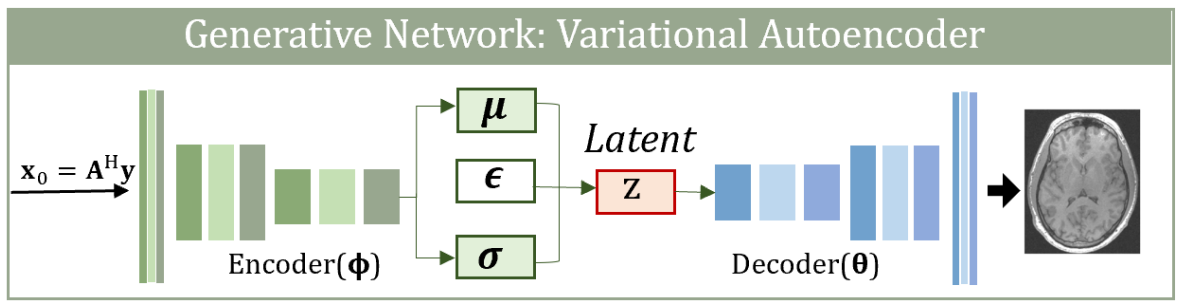

**Figure 5.** A typical Variational Autoencoder (VAE) encodes $\mathbf{x}$ into a latent space $\mathcal{Z}$ as a latent vector $\mathbf{z}$ and decode it for the reconstructed image, utilizing a probabilistic Encoder and a Decoder with parameters $\boldsymbol{\phi}$ and $\boldsymbol{\theta}$ respectively. To ease the backpropagation process for gradient descent, the reparameterization of VAE reconstructs the latent vector $\mathbf{z} \sim \mathcal{Z}$ from the mean $\boldsymbol{\mu}_{\boldsymbol{\phi}}(\mathbf{x})$ and standard deviation $\boldsymbol{\sigma}_{\boldsymbol{\phi}}(\mathbf{x})$ of the Encoder with injection of a randomness variable $\boldsymbol{\epsilon} \sim \mathcal{N}(0, \mathbf{I})$.

Specifically for MRI reconstruction, VAEs are commonly used in conjunction with Monte Carlo sampling [70] to facilitate image reconstruction [71] and uncertainty quantification [72]. The VAE model, specifically in [72], uses a loss function combining KL divergence and L2 norm, ensuring accurate reconstruction and proper latent distribution alignment. After training, the VAE's decoder, with Monte Carlo sampling, generates pixel-wise mean and variance maps, providing uncertainty quantification for reconstruction outcomes.

#### *2) Generative Adversarial Network*

Generative Adversarial Networks (**Figure 6**) [73] have emerged as a powerful tool for MRI reconstruction, addressing challenges in detail recovery and image quality. In this setup, two neural networks, a generator and a discriminator, engage in an adversarial training process. The generator reconstructs MRI images from undersampled data, striving to produce results indistinguishable from actual images. Simultaneously, the discriminator evaluates the authenticity of the generated images against the fully sampled reference. This dynamic training approach allows GANs to recover finer details and textures for MRI reconstruction, even from significantly undersampled data.

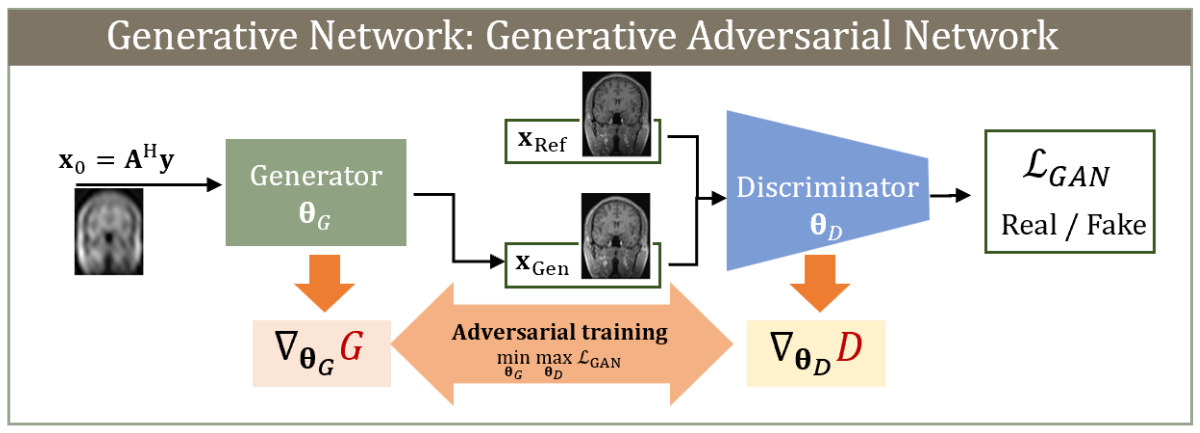

**Figure 6.** A Generative Adversarial Network (GAN) typically consists of a Generator and a Discriminator, which are jointly trained adversarially. From a subsampled aliased $\mathbf{x}_0$, the generator (typically an Enhancement based, Algorithm Unrolling, or PnP model) generates a reconstructed image $\mathbf{x}_{\text{Gen}}$. The generated $\mathbf{x}_{\text{Gen}}$ is then fed into the discriminator, which is trained to discriminate the generated $\mathbf{x}_{\text{Gen}}$ from the fully sampled reference image $\mathbf{x}_{\text{Ref}}$, to give an adversarial loss $\mathcal{L}_{GAN}$ for finer recovery of details and textures of images.

Yang et al. [74] introduced a deep de-aliasing GAN with a modified U-Net generator and a CNN classifier discriminator for MRI reconstruction. Lately, Huang et al. [75] developed a Swin Transformer-based GAN, which shows that Transformer-coupled adversarial training enhances detail perception but may risk reconstruction fidelity, especially when reconstructing highly undersampled data, consistent with similar findings in super-resolution studies of natural images [76].

#### *3) Deep Energy-based Model*

Deep Energy-Based Models (**Figure 7**) [77] are generative models that learn data distributions using a learnable energy function, which assigns lower energy to real data and higher energy to generated data [78]. The model is trained by optimizing a loss function that contrasts the energy of real and generated data. Sampling, often implemented through a Markov

Chain Monte Carlo method (often Langevin dynamics for MRI reconstruction [79]), iteratively adjusts data points based on the energy function and a stochastic component, enabling EBMs to emulate the training data's distribution.

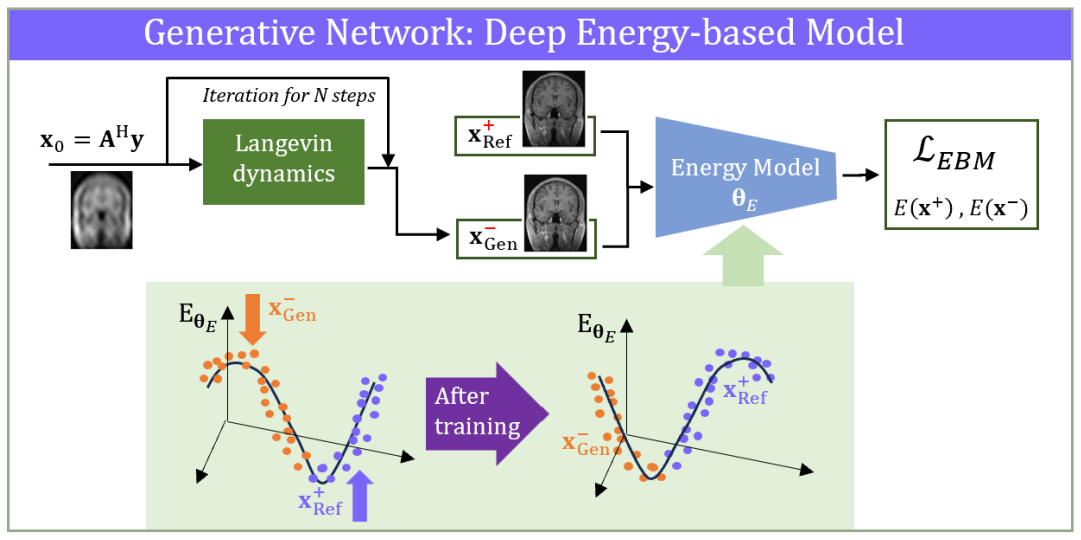

**Figure 7.** A Deep Energy-based Model (EBM) typically generates batches of reconstructed images by iterative sampling using Langevin dynamics through the energy model $E(\cdot)$. The energy model $E(\cdot)$ is then further optimized by assigning the generated images $\mathbf{x}_{\text{Gen}}^{+}$ with lower energy $E(\mathbf{x}^{-})$ and assigning the reference images $\mathbf{x}_{\text{Ref}}^{-}$ with higher energy $E(\mathbf{x}^{+})$. The two processes repeat alternately to enhance the energy model and performance of reconstruction.

In MRI reconstruction, EBMs have been utilized to enhance image quality by using deep energy-based information as an image prior, with an additional data consistency step in each iteration [79]. Additionally, methods integrating image and frequency domains through collaborative learning have been proposed for multi-coil MRI reconstruction [80].

*4) Diffusion Model*

Diffusion models (**Figure 8**) [81]–[83], including Denoising Diffusion Probabilistic Models (DDPMs) [81] and Score Matching with Langevin Dynamics (SMLD) [82], have become prominent in generative methods like MRI reconstruction [84], [85] due to their adaptability across various subsampling masks. These models operate by gradually transforming a simple initial distribution, often Gaussian, into a complex target data distribution through diffusion processes. The process is mathematically underpinned by score-based Stochastic Differential Equation (SDE) theory [83], describing how data evolves over time with specific drift and diffusion coefficients. The reverse diffusion reconstructs data from the Gaussian distribution back to the original data distribution. Neural networks approximate the score functions in this reversible SDE, allowing for numerical solutions and facilitating the transformation from simple (often Gaussian) to complex distributions, proving particularly effective in tasks like MRI reconstruction [84], [85].

Recent advances in MRI reconstruction have seen the emergence of diffusion model-based approaches. DiffuseRecon [84] is a DDPM-based model eliminating the need for training on specific acceleration factors. Chung et al. [85] introduced a score-based model using an iterative numerical SDE solver and data consistency step, suitable for multi-coil MRI with or without sensitivity maps. Cao et al. [86] developed the Complex Diffusion Probabilistic Model (CDPM) for enhanced complex-value data preservation. AdaDiff [87] offers an adaptive diffusion prior for improved reconstruction, while Cao et al. [88] focused on a high-frequency DDPM model to better retain high-frequency information in MRI data.

Despite their benefits, diffusion models typically have long inference times due to iterative processing. Chung et al.'s Come-Closer-Diffuse-Faster-MRI [89] combats this by speeding up the sampling in reverse problems, maintaining

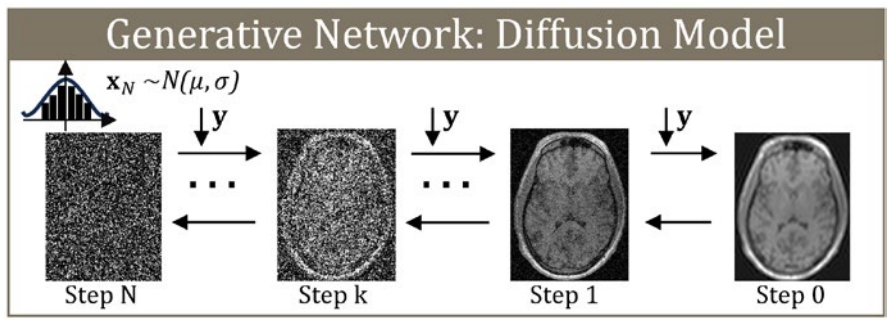

**Figure 8.** A Diffusion Model typically generates images by reversing the diffusion process, starting from a basic initial distribution, commonly Gaussian, and progressively transitioning towards a complex target data distribution through a series of diffusion steps.

reconstruction quality with fewer reverse process steps.

*5) Normalizing Flow-based Model*

Normalizing Flow-based models (**Figure 9**) [90] are generative models that learn data distributions $p(\mathbf{x})$ by transforming a simple prior distribution $p_Z(\mathbf{z})$ (e.g., Gaussian) into a complex one resembling the target data distribution [78], through an invertible function $\mathbf{f}(\cdot\,;\boldsymbol{\theta})$. This process maps a random vector $\mathbf{z}\sim p_Z(\mathbf{z})$ to a sample $\mathbf{x}$ in the target domain $\mathbf{x} = \mathbf{f}(\mathbf{z};\boldsymbol{\theta})$ and vice versa $\mathbf{z} = \mathbf{f}^{-1}(\mathbf{x};\boldsymbol{\theta})$. NFs aim to approximate the data distribution $p(\mathbf{x})$ with $p_{\boldsymbol{\theta}}(\mathbf{x})$, and training involves optimizing parameters $\boldsymbol{\theta}$ by minimizing the negative log-likelihood, which is equivalent to minimizing the KL divergence between the actual data distribution and the model's estimated distribution.

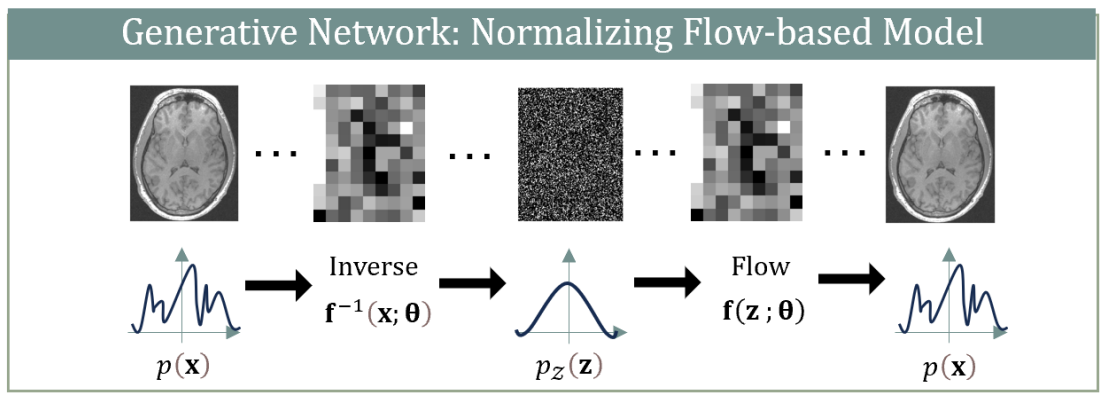

**Figure 9.** A Normalizing Flow (NF)-based model typically learn data distributions $p(\mathbf{x})$ by transforming a simple prior distribution $p_Z(\mathbf{z})$ (e.g., Gaussian) into a complex one resembling the target data distribution, through an invertible function $\mathbf{f}(\cdot\,;\boldsymbol{\theta})$. This process maps a random vector $\mathbf{z}\sim p_Z(\mathbf{z})$ to a sample $\mathbf{x}$ in the target domain $\mathbf{x} = \mathbf{f}(\mathbf{z};\boldsymbol{\theta})$ and vice versa $\mathbf{z} = \mathbf{f}^{-1}(\mathbf{x};\boldsymbol{\theta})$.

Compared to other generative models, NF-based models yield rapid inference speed, and allow exact likelihood computation thus requiring only simple maximum likelihood training [78]. The reliance on invertible transformations makes the NF-based model work with a full-dimension-based distribution, complicating the training and efficiency of deep neural networks. An effective solution involves employing a multiscale architecture by splitting and forwarding a part of the intermediate representation to the output at each scale [91], providing hierarchical features, and reducing the computational cost.

In the context of MRI reconstruction, Denker et al. [92] introduced a conditional NF-based model for medical image reconstruction, including two key components: an invertible neural network for normalizing flows and a neural network for conditioning using undersampled zero-filled images. This work further investigated the choice of invertible neural networks across a multi-scale architecture (RealNVP [91]) and an invertible UNet [93]. Wen et al. [94] also adopted multi-scale architecture from RealNVP [91], and expanded the NF-based approach for multi-coil MRI reconstruction.

## III. Incorporating Physics Information with Data-driven Models

### A. *Coupling with Multi-Coil Parallel Imaging*

Modern MRI scanners are equipped with multicoil receiver arrays that can expedite MRI scans by leveraging the distinct spatial sensitivity profiles of coils arrayed around the subject (**Figure 10**). By incorporating coil sensitivity profiles alongside a series of magnetic field gradients, spatial information can be encoded more rapidly while maintaining its integrity. This approach enables parallel imaging, allowing for faster data acquisition by skipping part of the k-space for undersampling. Although such means of k-space undersampling may introduce artifacts, the coil information can be used to remove these artifacts effectively. [25], [95].

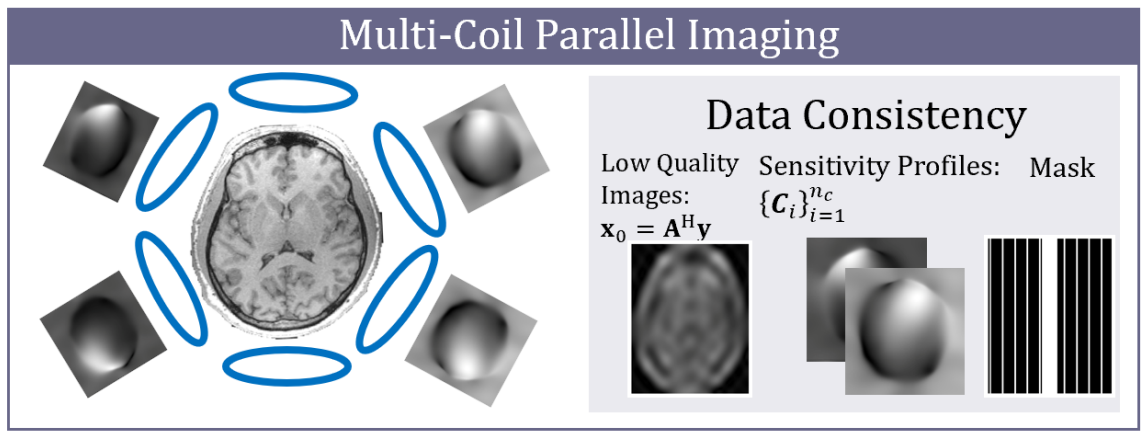

**Figure 10.** Multi-Coil Parallel Imaging expedites MRI by leveraging the distinct spatial sensitivity profiles of coils $n_c$ arrayed around the subject. By incorporating coil sensitivity profiles $\mathbf{C}_{n_c}$ alongside a series of magnetic field gradients, spatial information can be encoded more rapidly while maintaining its integrity. This approach enables parallel imaging, allowing for faster data acquisition by skipping part of the k-space. The additional sensitivity profiles are then added as part of the data consistency (DC) module to incorporate additional spatial information.

With sensitivity profiles from multiple coils, the forward encoding operator $\mathbf{A}$ in Equation (1) can be then generalized into its multi-coil form $\mathbf{A}: \mathbb{C}^n \to \mathbb{C}^{m,n_c}$, as specified in (7).

$$\mathbf{A} = \begin{bmatrix} \mathcal{F}_\Omega \mathbf{C}_1 \\ \vdots \\ \mathcal{F}_\Omega \mathbf{C}_{n_c} \end{bmatrix}, \tag{7}$$

where $n_c$ is the number of coils in the receiver array, and $\mathbf{C}_q: \mathbb{C}^n \to \mathbb{C}^n$ is the sensitivity map in the form of a diagonal matrix for the $q$-th receiver coil.

In general, there are two main types of parallel imaging methods: **(a) image domain** methods and **(b) k-space** methods. **(a)** Image domain methods, such as SENSitivity Encoding (**SENSE**) [2], reconstruct the image in the image space after inverse Fourier transform by solving an inverse problem that uses coil sensitivity maps. **(b)** K-space methods directly estimate the missing data in k-space by using linear combinations of the acquired data from different coils [96]. Techniques such as GeneRalized Autocalibrating Partial Parallel Acquisition (**GRAPPA**) [97], which reconstruct the image by training a k-space linear estimators from a fully-sampled block of data at the center of k-space, called the AutoCalibrating Signal (ACS).

However, these parallel imaging methods are limited by the quality of the calibration data and the accuracy of the model used to interpolate the missing k-space data. Taking GRAPPA as an example, one main concern about this linear approach is that it can lead to **(a) noise amplification** due to estimation or interpolation from k-space lines that are further apart, which can degrade image quality and contrast. Although multiple conventional iterative methods have been proposed, such as SPIRiT [98], which enforces self-consistency across k-space among different receiver coils for noise reduction in k-space and accepts more versatile sampling patterns, the use of deep neural networks to enhance k-space interpolation techniques utilizing non-linear approaches in a data-driven way has recently piqued interest. One typical method could be **(a.i) RAKI** (Regularized Adaptive K-space Interpolation) [99], which generalizes the linear convolutional kernels of GRAPPA. RAKI, trained on subject-specific calibration data, learns a non-linear mapping from the undersampled k-space data to the fully sampled k-space data using a convolutional neural network to interpolate the undersampled k-space. Similarly, LORAKI [100] builds upon structured low-rank modeling techniques in k-space, which have shown great performance in many MRI reconstruction applications. LORAKI integrates autocalibrated recurrent neural networks for autoregressive MRI reconstruction in k-space, effectively combining low-rank modeling with deep learning to further enhance the reconstruction quality and robustness. On the other hand, **(a.ii) GrappaNet** [101] uses a different approach by incorporating the parallel imaging method into the neural network to extend the GRAPPA method. Its staged approach uses a neural network to simplify the reconstruction problem, then applies a parallel imaging method, and finally another neural network for fine-tuning.

Besides the noise amplification problem, GRAPPA relies on the estimation of convolutional kernels from the ACS. This can reduce the effective acceleration factor and introduce errors if bulk motion leads to **(b) a mismatch between ACS** lines and other regions of the undersampled k-space. More recent deep learning methods have been proposed to help improve the estimation of coil sensitivity functions from **(b.i)** limited ACS regions in SENSE-based reconstruction [102], and also **(b.ii)** incorporate GAN based models [103].

The methods included above are a representative abstract of the recently proposed deep learning enhanced parallel imaging methods. Please see a review article by Knoll et al. [25] for more extended discussions.

### B. *Enhancing Simultaneous Multi-Slice (SMS)*

For Echo Planar Imaging (EPI), standard parallel imaging reduces EPI readout and distortion in T2* and diffusion B0 scans [104], but doesn't proportionally shorten overall scan time, particularly in Diffusion Weighted Imaging (DWI), where it cannot reduce the lengthy diffusion encoding period.

In contrast, Simultaneous Multi-slice Imaging (SMS) (**Figure 11(A)**) [104] recently emerged to accelerate the acquisition by acquiring multiple slices simultaneously within the same acquisition period without shortening it, representing a significant leap in fast MRI. SMS works by simultaneously acquiring multiple slices of an object (with varied frequencies) using a single composite Radio Frequency (RF) pulse. The signals from these concurrently excited slices are then differentiated using coil sensitivity data, a process akin to parallel imaging techniques [104], [105]. Specifically, CAIPIRINHA [106] enhances SMS imaging in structural imaging by controlled aliasing and blipped CAIPIRINHA [107] further extends SMS to EPI for greater advantage [105], [108]. Notably, SMS excels in **(a)** greatly shortening DWI overall scan time, with faster overall scan time acceleration that is directly proportional to the SMS acceleration factor [104] (a marked improvement over standard parallel imaging) and **(b)** leveraging its higher temporal sampling rate in fMRI to enable Nyquist sampling and thus filtering of physiological noise, thereby refining fMRI analysis [109].

Notably, although SMS and standard parallel imaging exhibit advantages over each other in different ways, they are not mutually exclusive. SMS can be synergistically combined with parallel imaging, as the two accelerate the scan along different encoding directions: parallel imaging accelerates spatial encoding by allowing fewer phase encoding lines (in $k_y$-directions), while SMS does it by simultaneous sampling for multiple planes (in $k_z$-directions). In fact, combining both has been a common approach to further advance the acceleration rates in, for example, diffusion MRI and fMRI [110], although combining both SMS and parallel imaging can cause much higher computational complexity [104].

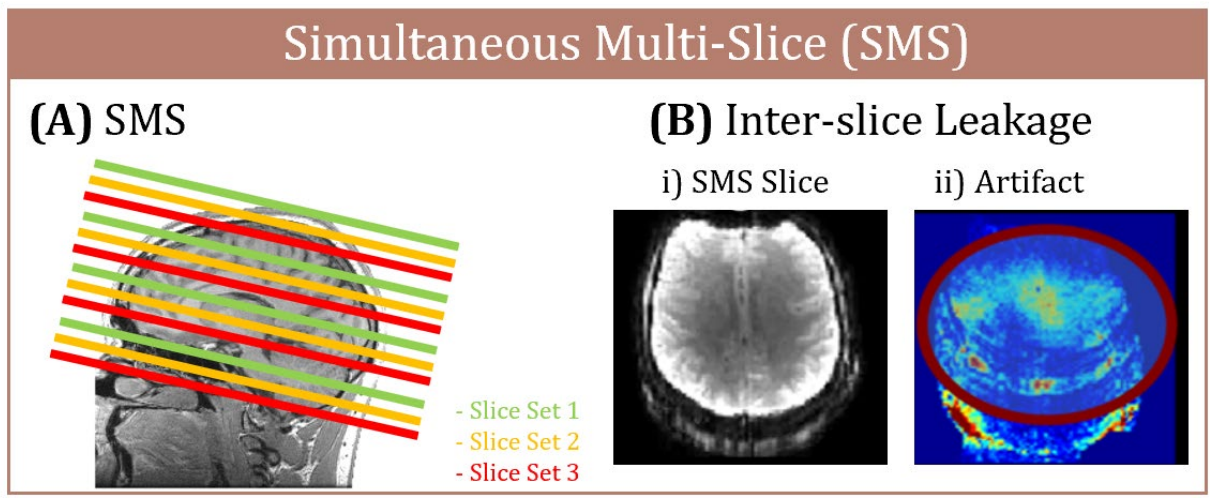

**Figure 11. (A)** Simultaneous Multi-slice Imaging (SMS) expedites MRI by acquiring multiple slices simultaneously within the same acquisition period without shortening it. SMS works by simultaneously acquiring multiple slices of an object (with varied frequencies) using a single composite Radio Frequency (RF) pulse with controlled aliasing. The signals from these concurrently excited slices are then differentiated using coil sensitivity data. **(B)** However, even with controlled aliasing in i) SMS, ii) interslice leakage could be observed, where deep learning models have been proposed to improve image quality.

However, it remains difficult to employ SMS, even if combined with controlled aliasing approaches, to image smaller organs such as the knee and the heart, where the distance between slices is too close to eliminate **interslice leakage artifacts** (**Figure 11(B)**) [111], [112], also known as residual artifacts. While conventional methods have shown effectiveness in reducing interslice leakage artifacts [113], [114], there is a growing need for data-driven models to manage closer slice gaps more effectively. Addressing this challenge, Zhang et al. [115] extended RAKI [99] to SMS knowledge of SMS acquisition model into subject-specific training of the CNNs, using readout-concatenated SMS calibration data. Similarly, Nencka et al. [116] employed "split-slice" training [113] to further enhance RAKI with its hyperparameters tuned. By tuning hyperparameters and integrating slice-specific information, they accounted for interslice interference patterns inherent in SMS acquisitions, embedding deeper MR physics principles into the neural network architecture. Besides these two MR-physics-informed neural networks that integrate specialized MR physics knowledge, pure data-driven networks have also been developed, where U-Net [40] has been a popular backbone deep learning neural network. Le et al. [117] developed a 3D U-Net [40] based end-to-end deep learning solution for fast SMS-based myocardial perfusion reconstruction. Li et al. [118] designed a U-Net [40] based CNN to process and separate complex multi-slice overlapping-echo signals realizing a reliable SMS T2 mapping. Tanzer et al. [111] also explored how different variants of U-Net [40] could respectively further enhance the image quality of SMS-accelerated cardiac DTI.

Nevertheless, acquiring a large set of fully sampled reference data for training these supervised learning based methods may be difficult, particularly for the heart and temporal scans such as fMRI. To address this, Demirel et al. [119] proposed self-supervised unrolled networks for SMS 10-fold accelerated 7T fMRI data, enabling significantly suppressed residual artifacts and improved SNR without the need to acquire fully sampled data.

### C. *Reconstruction Outcome Targeted Sampling Pattern Optimization*

Recent developments in MRI technology have introduced a variety of sampling patterns, from traditional Cartesian raster to spiral, radial, and other non-Cartesian trajectories. This variety necessitates a method to identify the optimal subsampling pattern for the best image quality. To address this, conventional adaptive methods were proposed to identify the optimal sampling pattern for the best post-reconstruction image quality [120]–[129]. Going forward with modern deep learning frameworks, the downstream reconstruction model can be further jointly optimized with the sampling pattern, collectively advancing the quality of the reconstruction outcome (**Figure 12**) [130]–[137].

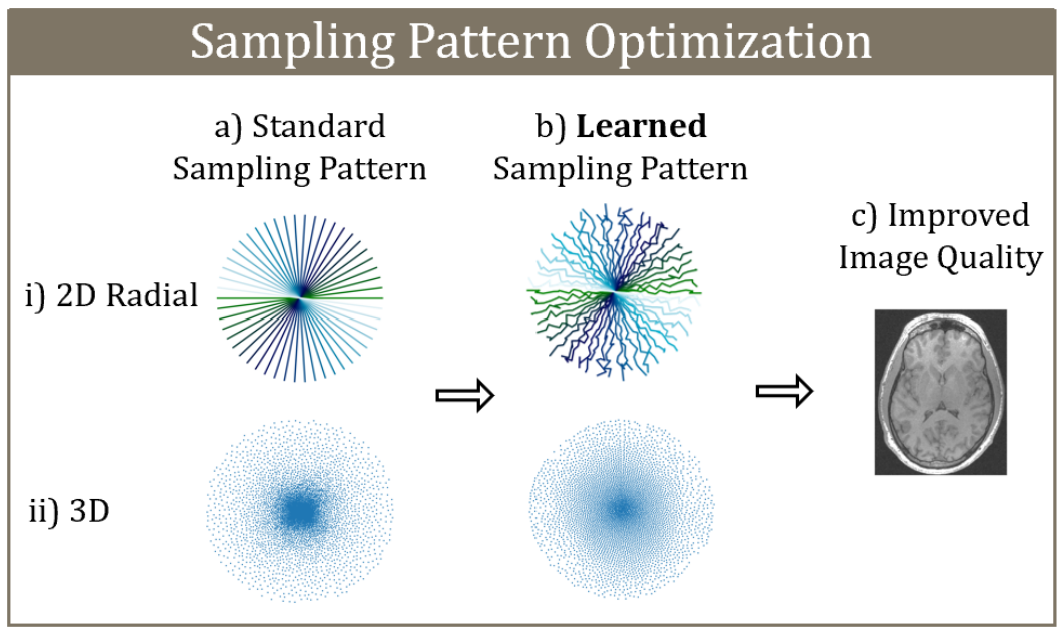

**Figure 12.** Typical optimization of k-space sampling pattern can optimize sampling trajectories **a-b)** (e.g., **i)** 2D radial [135], **ii)** 3D [148]) with reference to their corresponding **c)** image output. NB. **ii)** is a cross-sectional slice of its 3D sampling trajectories, where each dot on this cross section corresponds to a 3D fully sampled readout axis [148].

Conventional parallel imaging methods typically use controlled aliasing in their sampling design to increase the distance between aliasing voxels to effectively reduce the *g*-factor noise penalty [138]–[140]. When prior information is incorporated into the reconstruction (e.g., compressed sensing), pseudo-random sampling patterns are typically employed to provide incoherent aliasing. Various methods have been proposed, including **(a)** creating patterns from power spectra datasets [120], **(b)** using Bayesian design for sequential sample selection for sparse reconstruction in single-coil settings [122], and **(c)** employing greedy subset selection for iterative pattern optimization [123] (extendable through deep learning to parallel imaging and low-rank cases [124]), **(d)** constrainable by the Cramer-Rao bound [125]. More recent advancements include SPARKLING [126]–[129], which generates optimal 2D [126] and 3D [129] trajectories while considering MRI hardware constraints, which could be further enhanced by machine learning optimized parameterization of its sampling density field [141].

Following advances in deep learning based methods, data-driven **(A) scan-specific** sampling strategies have been proposed. Considering this optimization challenge as a path finding problem in k-space, reinforcement learning has emerged as a solution, actively suggesting the next sampling locations in real-time [142], [143]. Other active acquisition frameworks select the next sampling locations based on already acquired k-space data using single-pass sampler networks [144], sequential sampling mechanisms [145], or evaluator networks that reduce reconstruction uncertainty [130].

Alternatively, sampling patterns can be pre-determined through non-active deep learning strategies in a **(B) population-adaptive** manner, either **(a) probabilistically** [131], [132] or **(b) deterministically** [133]–[137]. **(a) Probabilistic methods** model the binary sampling mask in k-space. Focusing mainly on Cartesian sampling, they employed relaxation such as the straight-through (ST) estimator [132], [146] or Gumbell-Softmax parameterization [131], [147] to maintain differentiability for backpropagation. In contrast, **(b) deterministic methods** produce optimized defined paths for sampling in k-space. Early methods like J-MoDL [134] were limited to uniform grid patterns, but recent developments like PILOT [133], BJORK [135], SNOPY [136], 3D-FLAT [137], and AutoSamp [148] allow for Cartesian 3D [148], non-Cartesian 2D [133], [135], and non-Cartesian 3D [133], [136], [137] sampling routes through, for example, parameterization of sampling patterns [135], [136].

Importantly, generated sampling patterns must be feasible to deploy. To ensure this, numerous methods [133], [135]–[137], [149] take MRI hardware limitations into account, including the maximum gradient amplitude and slew rate constraints [133], as well as the sampling rate limits [137], when optimizing trajectory speed and acceleration. They also limit peripheral nerve stimulation (PNS) effects [136], thereby enhancing clinical applicability. Radhakrishna et al. [149] further enforced the admissibility of the generated sampling pattern by employing a projection-based method [150] instead of the commonly used penalty-based method [133], [135] to enforce the hardware constraints. This focus on practicality and clinical applicability rules out impractical sampling patterns, such as discrete dots for 2D acquisitions [151].

These aforementioned deep learning end-to-end frameworks are often capable of bridging both the sampling pattern optimization model and the image reconstruction model via co-optimization [130]–[137], [148], either through optimization of a joint loss function for both models [135] or through adversarial training of the two separate models alternately [130]. On top of that, PILOT [133] was further shown to co-optimize the segmentation model with the sampling pattern, highlighting the potential of co-optimization for a broader range of downstream tasks beyond reconstruction.

## IV. Domain Challenges And Opportunities

This section delves into the multifaceted aspects of MRI reconstruction (**Figure 13**), focusing on the use of unsupervised learning, the exploration of decision-making for acquisition and reconstruction by reinforcement learning, the strategic utilization of multi-contrast scan redundancies, the assessment of reconstructed images through diverse metrics, the intricate balance between accelerating scans and preserving image integrity, and the mitigation of data heterogeneity. Finally, it contrasts specialized versus generalizable models, highlighting issues like domain shift and data diversity, while noting solutions such as federated learning and efficient GPU memory use.

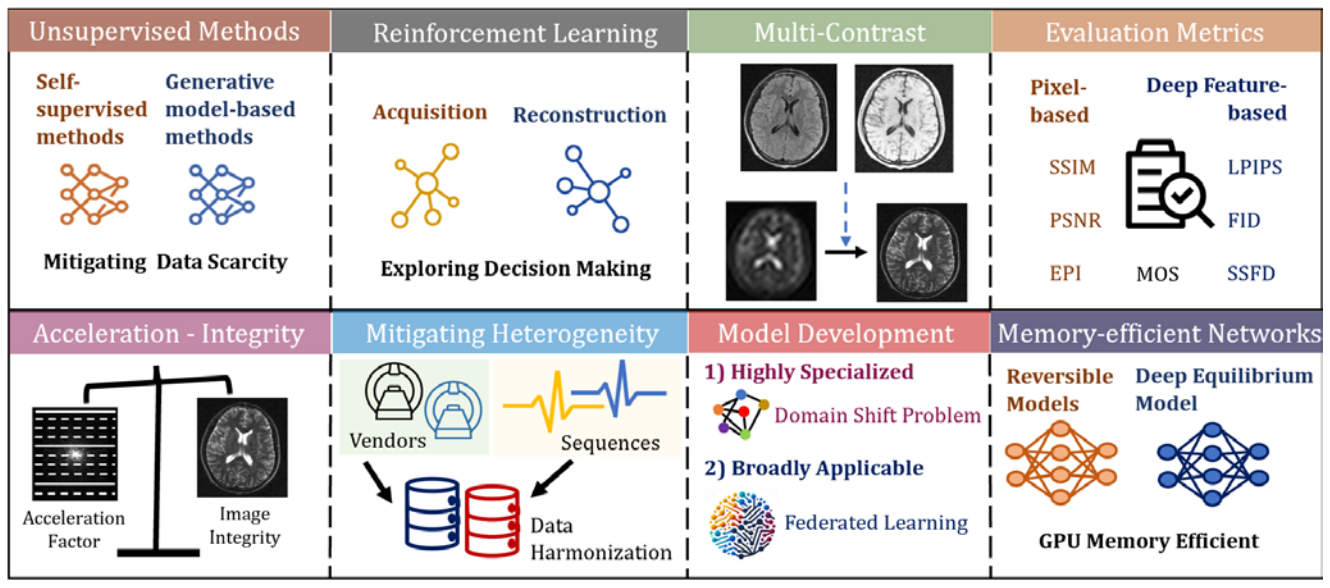

**Figure 13.** **(A)** Mitigating Data Scarcity by unsupervised learning, **(B)** Exploring decision-making for acquisition and reconstruction by reinforcement learning, **(C)** Leveraging imaging redundancy in multi-contrast scans for MR acceleration, **(D)** Evaluation metrics of different types offer different perspectives of observations. **(E)** Balancing Acceleration with Image Integrity. **(F)** Mitigating heterogeneity in multi-center/vendor/sequence images. **(G)** Model Developments: more specialized (domain shift problem) vs. more generalized (federated learning), **(H)** Alleviating GPU memory constraints by memory-efficient networks.

### *A. Mitigating Data Scarcity: Unsupervised Learning*

Supervised learning for MRI reconstruction typically relies on fully sampled MR data for model training, which is often impractical due to the slow and expensive data acquisition process. Unsupervised learning is emerging as a promising alternative to overcome data scarcity in MRI reconstruction by learning the distribution of the undersampled data [152].

Within unsupervised learning, self-supervised methods utilize the inherent structure of MRI data for training guidance, thereby improving image quality and data efficiency. One effective approach is self-supervised learning via data undersampling (SSDU) [153], which exclusively uses undersampled k-space data for MRI reconstruction. In SSDU, the undersampled measurements are split into two separate subsets for data consistency modules in the unrolling network and the loss function computation. Akçakaya et al. [99] introduced RAKI, a robust artificial neural network for scan-specific k-space interpolation, allowing nonlinear estimation of missing k-space lines, learned from a limited amount of Auto-Calibrating Signal data. Aggarwal et al. [154] introduced an Ensemble Stein's Unbiased Risk Estimate metric, which helps approximate the weighted projected mean square error, thus eliminating the need for fully sampled training datasets.

Another effective approach involves generative model-based methods. Deep Image Prior (DIP) [155] utilizes the structure of convolutional networks to inherently prefer natural image solutions, regularizing the reconstruction problem without specific training data. In the context of MRI, DIP has been adapted to reconstruct dynamic MR images by feeding random noise into a network and optimizing its parameters, until the output images match the undersampled MRI data in the measurement domain, without the need for prior training or additional [156].

Besides DIP, GANs [73] have also been utilized to advance MRI reconstruction through unsupervised learning. Lei et al. [157] developed a modified Wasserstein GAN, leveraging unpaired adversarial training with the k-space undersampled zero-filled images and fully-sampled images from two different datasets. Additionally, Oh et al. [158] introduced an unpaired cycle-consistent GAN, driven by optimal transport theory, for accelerated MRI reconstruction.

### *B. Exploring Decision-Making for Acquisition and Reconstruction: Reinforcement Learning*

Reinforcement Learning (RL) is an advanced machine learning technique in which agents learn to make decisions through interactions with a dynamic environment to maximize cumulative rewards. RL is particularly suited to applications requiring adaptive and optimized decision-making due to its ability to continuously refine strategies based on interactive feedback. It has been applied effectively in optimizing MRI acquisition and reconstruction [159].

In the realm of MRI acquisition optimization, RL-based models are trained to adaptively select sampling patterns based on the specific imaging requirements and constraints of each scan, resulting in substantial improvements in both acquisition efficiency and image quality. Pineda et al. [143] proposed to optimize the MRI acquisition pattern by formulating the problem as a partially observable Markov decision process [160] and solving it using Double Deep Q-Networks [161]. Experimental results have demonstrated that the learned sampling policies outperform the simple acquisition heuristics. Bakker et al. [142] employed a policy gradient-based approach to learn efficient experimental designs for MRI subsampling. The use of a simple greedy policy, surprisingly effective, underscored the potential of RL methods to simplify and enhance the design of subsampling strategies, often governed by heuristic methods in clinical settings. Liu et al. [162] introduced a dual-state-based RL framework that efficiently combined visible Parameter-Free states and hidden deep neural network states to optimize phase selection for accelerated MR imaging. The dual states method improves both the efficiency of phase selection and the accuracy of subsequent image reconstruction, demonstrating the potential of RL to enhance both the speed and quality of MRI procedures.

Recent advancements in RL have also shown promising results in MRI reconstruction, enhancing both the quality of reconstruction and the interpretability and efficiency of image reconstruction processes. Li et al. [163] proposed RL-based and enhancement-based MRI reconstruction methods to address the limitations of deep learning approaches, which often function as black boxes with limited interpretability. This method selected image processing operators and filter parameters by combining a discrete action policy with continuous action parameters, through a Markov decision process [160]. Further developments include the integration of RL with unrolling networks for MRI reconstruction, where model priors are incorporated into the intrinsic iterative process of RL strategies [164]. This approach not only improves the quality of image reconstruction but does so with an efficient use of computational resources.

### *C. Exploiting Redundancy in Multi-Contrast Scans*

Clinicians commonly prescribe multiple MRI contrasts, including T1-weighted (T1w), T2-weighted (T2w), and Proton Density (PD), offering varied observation perspectives. Despite different contrasts, these modalities all target the same organ, conveying similar anatomical structural information, which can be exploited to improve MRI reconstruction by leveraging their inter-contrast redundancy.

A prevalent method pairs a fully sampled modality (e.g., T1w) with an undersampled one (e.g., T2w) [165], [166], enabling the latter to learn structural details while maintaining unique contrast features. Further studies have shown that trained models can also co-reconstruct multiple undersampled modalities, even without a fully sampled reference [167], [168]. More recent research using separate models for each contrast, but allowing interaction between them, has significantly enhanced reconstruction quality [169]. However, co-optimizing various contrasts may cause contrast signal leakage across different modalities, potentially creating false features in the resulting image [168].

Apart from T1w, T2w and PD, a key advantage of MRI is its ability to generate quantitative parametric maps from a series of images with different contrasts. In these techniques, individual datasets are fit to a signal model to infer tissue properties in every voxel, providing a similar basis of redundancy for accelerating the reconstruction of quantitative maps. However, the multi-dimensionality of these data greatly increases computational demands. To alleviate this, taking cardiac T1 mapping as an example, common reconstruction methods include using deep learning networks in reduced temporal dimensions [170], integrating recurrent modules in these networks [171], and projection of data onto low-dimensional manifolds [172] to manage the computational load.

### *D. Evaluation Metrics: Perspectives of Observations*

Quantitative metrics representing post-reconstruction image quality are crucial for evaluating their fidelity and perceptual quality, offering an assessment of image outcomes from varied perspectives **(Figure 14)**. In terms of the reference image availability (*i.e., whether or not a metric requires a ground truth reference image for evaluation of the result*), these metrics can be categorized into (i) Full-Reference, (ii) No-Reference, and (iii) Distribution-based metrics [173]. In this part, we discuss these metrics from the perspective of evaluation focus, where metrics can be classified into **(A) Human-Perceived Metrics** provided by radiologists, **(B) Pixel-based Metrics** and recently emerging **(C) Deep Feature-based Metrics** [174]. It is also pertinent to consider the extensive literature on model-based observers, which assess image quality by simulating human interpretation processes in clinical environments [175]. Details of all these metrics, along with their reference image availabilities, are summarized in **Supplementary Table S2**.

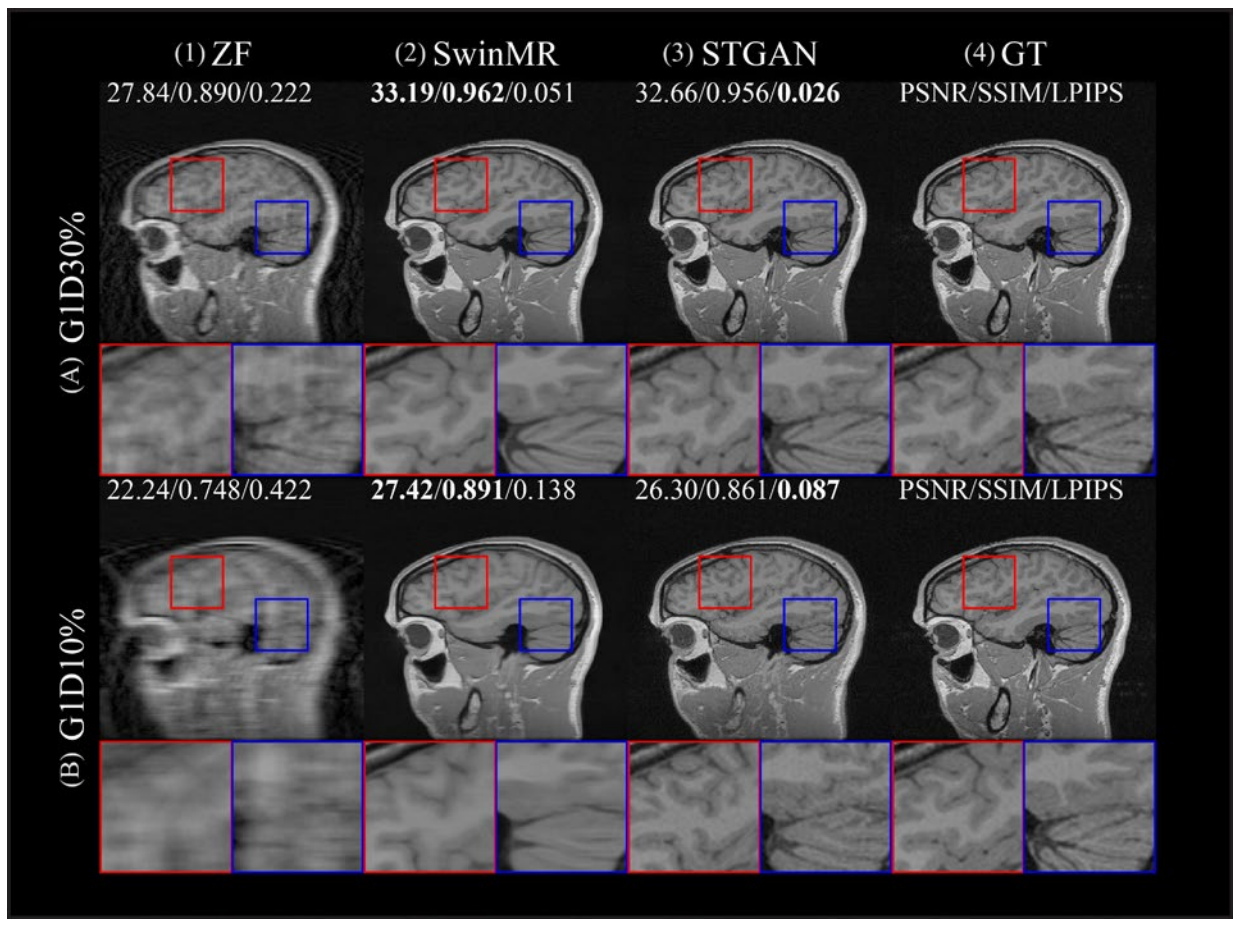

**Figure 14.** Performance of **(2)** an image enhancement model SwinMR [49] and **(3)** a generative model STGAN [75] compared to the **(4)** fully sampled ground truth (GT) and **(1)** zero-filled (ZF) subsampled MRI, reconstructing Gaussian 1D (G1D) subsampled MRI of different subsampling ratios **(A)** 30% and **(B)** 10%. PSNR, SSIM and LPIPS were listed on top of each image. **For metrics**, we can see these pixel-based metrics (PSNR and SSIM) were sensitive to the accuracy and any false positives generated, whereas the deep feature-based metric LPIPS was more sensitive to perceptual quality, akin to that in human perception (**B.2 vs. B.3**). **For models**, image enhancement and generative models behave similarly at moderate subsampling ratio (**A.** GID30%). However, when heavily subsampled (**B.** G1D10%), the generative model **(3)** STGAN tended to generate perceptually convincing yet false features (higher LPIPS, **B.3 vs. B.4**) with lower specificity (lower PSNR/SSIM), whereas the image enhancement **(2)** SwinMR failed to generate perceptually convincing texture details (lower LPIPS) but maintained relatively higher specificity and pixel-level accuracy (moderately higher PSNR/SSIM, **B.2 vs. B.3**)

**(A) Human-Perceived Metrics**, e.g., **Mean Opinion Scores (MOS)** [176], capture human visual perception by gathering subjective impressions from skilled clinical observers. These metrics, often utilized with a Likert scale with a certain range, provide a measure of the observer's perception on the reconstruction outcome, also allowing the observer to identify any notable artifacts. However, the subjective basis can negatively impact its reproducibility, both inter- and intra-observers. This subjectivity also makes MOS sensitive to the specifics of the experimental setup and the varying experience levels of the raters [176].

**(B) Pixel-based Metrics**, e.g., **Normalized Root Mean Square Error (NRMSE)** and **Peak Signal-to-Noise Ratio (PSNR)** [177], are routinely utilized for reconstruction quality assessment. These metrics offer a direct indication of an image's pixel-wise fidelity when compared to the original ground truth. Some metrics focus on hard-crafted statistical features including **Structural Similarity Index (SSIM)** [178] and **Blind/Referenceless Image Spatial Quality Evaluator (BRISQUE)** [179]. Additionally, some metrics that focus on the qualification of some specific factors like edge preservation have been proposed [180], including **Edge Strength Similarity Index Metric (ESSIM)** and **Edge Preservation Index (EPI)**.

However, pixel-based metrics' simplicity and the reliance on shallow features mean they cannot completely replicate the complex, hierarchical structures of human vision. Consequently, they may not fully reflect the human perception of an image [181]. This limitation is particularly evident in scenarios like blurring effects, which human observers can easily see but these fidelity metrics may not be sensitive to.

**(C) Deep Feature-based Metrics** are developed to emulate the hierarchical structure of human visual perception and provide an objective, reliable measure. These metrics utilize the hierarchical structure of the trained neural networks to capture various levels of image abstraction and details, from basic textures and edges to more complex patterns and object representations.

For example, **Fréchet Inception Distance (FID)** [182] calculates the distance between latent features extracted from a pre-trained Inception-Net between the reconstructed and the ground truth image, and measures the Fréchet distance between these two distributions. **Learned Perceptual Image Patch Similarity (LPIPS)** [181] operates on a similar principle but focuses on comparing images on a patch-by-patch basis within the feature space of pre-trained neural networks. This approach helps in capturing local variations and details in patches that might be lost in a more global analysis. These metrics show improved correlation with professional clinical observers' assessments. However, they are criticized for their insensitivity to minor pixel differences and 'hallucinations', hiding image integrity problems. Additionally, these metrics rely on the neural network trained on natural images such as ImageNet, which lacks domain knowledge in the anatomical structures. Consequently, 'in-domain' deep feature-based metrics, e.g., **Self-Supervised Feature Distance (SSFD)** [183], which involve neural network trained on MR dataset, have been proposed.

To bridge the subjective and objective assessment, research has also discussed the correlation between subjective radiologist perceptions and objective quantitative metrics [173], and explored the potential to predict the quality perceived by radiologists using computational assessments [184], [185].

Apart from quantitative evaluation metric, models could also be evaluated qualitatively. Local perturbation response [186] was introduced to evaluate nonlinear reconstruction models by analyzing the images generated before and after introducing controlled perturbations to the model input. A null space 'hallucination' map [187] was proposed to indicate the 'hallucination' by decomposing the reconstructed images into measurement and null components and calculating the null space error map. Uncertainty maps [188] are also commonly used for assessment and are discussed in details in **Section IV.D.**

### *E. Balancing Acceleration with Image Integrity*

The pursuit of faster MRI scans through acceleration techniques presents a complex trade-off between speed and image integrity. While heavy acceleration is desirable for reducing scan times and enhancing patient comfort, it often leads to much poorer image quality, offering a challenging starting point for the subsequent reconstruction process.

With high acceleration rates, the resulting undersampled MRI images often suffer from much lower initial quality, characterized by more pronounced aliasing and sparser k-space data. Early exploration of deep learning based MRI reconstruction usually relied on unrolled algorithms with data consistency modules and neural networks with shallow and simple network design [53]. Such models usually have an admirable performance on MRI reconstruction with relatively low acceleration rates; however, they struggle with these heavily accelerated datasets [49], [189]. Their reliance on data consistency modules and inability to adequately compensate for information loss often leads to outputs that fall below clinical imaging standards.

Recent progress in deep generative models in inverse problems (e.g., GAN [74], diffusion model [85]) and sophisticated network architectures (e.g., Transformer [49], [51]) has shown promising results in reconstructing high-quality images from accelerated MRI data. These networks demonstrate an enhanced capacity to 'fill in' missing information, presenting a leap forward in reconstruction technology. However, the use of these advanced networks involves a delicate perception-distortion balance [187] and introduces the risk of generating 'hallucinations' - artifacts or incorrect features that occur due to the prior that cannot be produced from the measurements [187] **(Figure 14)**. These 'hallucinations' are perceptually convincing, making it challenging for clinical observers to distinguish them from real anatomical features without comparing them to fully sampled ground truth, thereby risking the uncompromising need for maintaining image integrity and its diagnostic validity. Recent studies have developed hybrid models, merging the physics-driven methods with recent diffusion models [190], or transformers [191], which provide a potential solution to mitigate this problem.

Uncertainty can quantify the integrity of reconstructed images. It originates from two primary sources: aleatoric uncertainty, which results from the missing k-space data, and epistemic uncertainty, which arises from the probability distributions of the network's weights [188].

Uncertainty in image reconstruction can be estimated using Bayesian Neural Networks (BNNs) and Monte Carlo (MC) dropouts [192]. BNNs [193], provide a probabilistic interpretation of the reconstruction process by placing a prior distribution over the model's weights and using variational inference to estimate pixel-wise uncertainty. MC dropouts [166] simulate multiple reconstructions by randomly dropping units during inference, generating a distribution of possible outcomes from which uncertainty can be derived. The combination of MC dropout sampling, BNNs, and generative models offers a robust approach for uncertainty estimation [188], [194], [195]. Generative models, such as Variational Autoencoders (VAEs) [193], capture the latent space representation of input data, including inherent uncertainties. Similarly, Diffusion Models [188] progressively denoise random samples to produce a variety of potential reconstructions.

## *F. Mitigating Heterogeneity in MRI Data*

The heterogeneity of MRI data from various centers and vendors presents a significant challenge in the development of robust MRI reconstruction models. Despite extensive efforts to standardize scanning protocols from both MR physics [196] and clinical perspectives [197] for quality assurance, inevitable variations in MR hardware, scanning sequences, and patient-specific protocols often result in datasets with heterogeneous features. These variations in scanning parameters can lead to differences in image quality, contrast, and noise levels.

For deep learning models, even minor perturbations can sometimes cause significant artifacts in reconstruction outcomes **(Figure 15)** [198], due to the instability phenomenon. This issue not only compromises the integrity and diagnostic validity of the images but also impacts the performance and generalizability of sensitive reconstruction algorithms. For instance, they may obscure minor yet clinically important changes in imaging [198]. The recent trend towards developing larger models trained with greater amounts of data further exacerbates this problem [198]. Ensuring the reliability and accuracy of these models remains a critical challenge in MRI reconstruction.

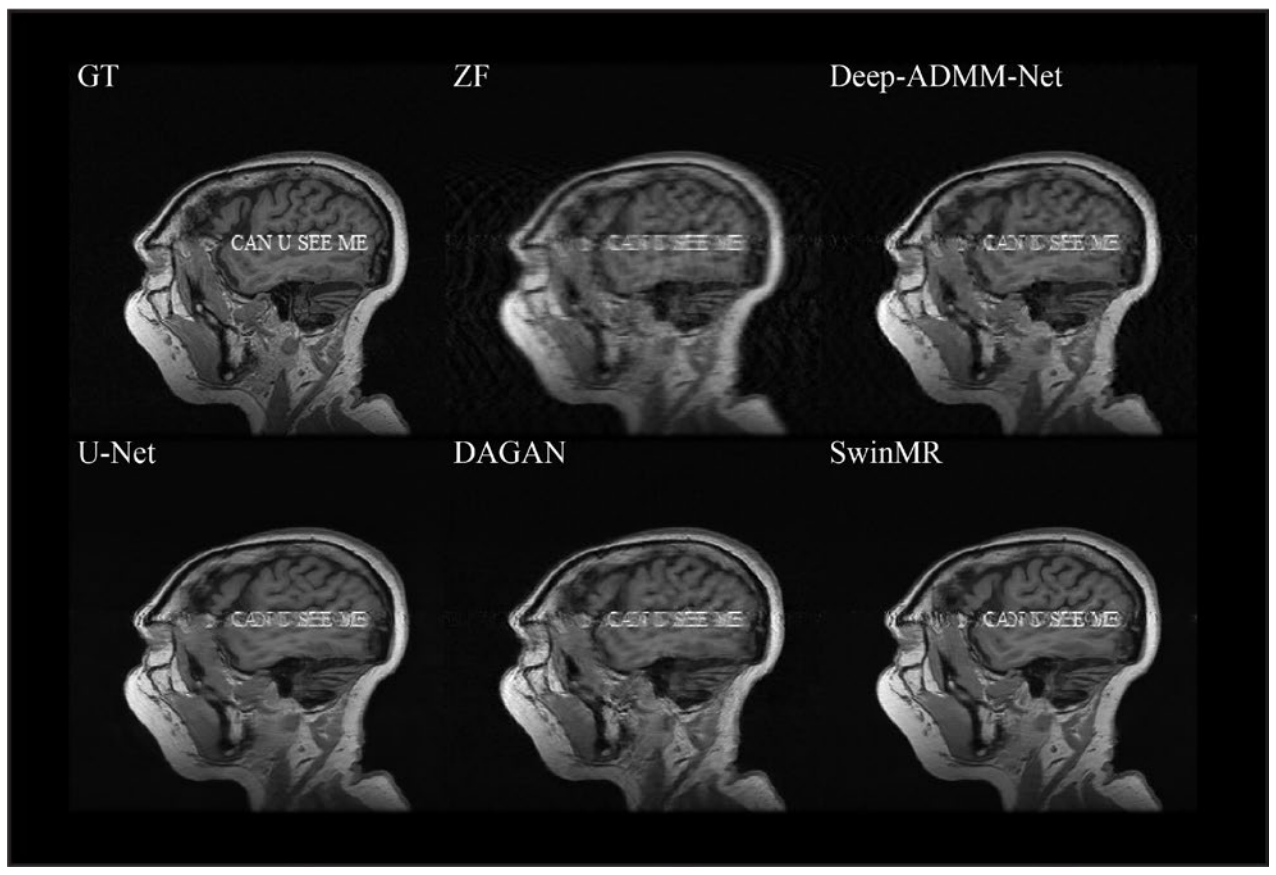

**Figure 15.** The reconstruction results of different models with structured perturbations (for this case: text, in line with [198]) in the fully sampled image. From left to right: the ground truth (GT) image, and the reconstruction results of zero-filled (ZF), Deep ADMM Net, U-Net, DAGAN, and SwinMR.

To enhance model stability across various data sources, several strategies have been proposed, focusing on either **(a) model development** or **(b) data handling**.

**(a) Model development:** One effective strategy is the use of **(a.i) data augmentation** [199] methods for model training. This could involve simulating variations in imaging conditions for model training, which can enhance the robustness of models to handle data from diverse MRI scanners. Additionally, employing **(a.ii) domain adaptation** [200] techniques enables models to learn more universal features that are less influenced by vendor-specific traits. Transfer learning can be particularly effective, where a model is first trained on a large, diverse dataset and then fine-tuned on specific datasets.

**(b) Data handling**: **Data harmonization** is a crucial approach to resolving data heterogeneity. It aims to align data distributions, either **(b.i) sample-wise** or **(b.ii) feature-wise**. **(b.i) Sample-wise harmonization** involves image synthesis or processing techniques to explicitly standardize raw image scans, making them more uniform across different sources [201], [202]. Alternatively, **(b.ii) feature-wise harmonization** typically uses statistical strategies after feature extraction to align cohort distributions [201], ensuring that features from different datasets are comparable. It could also be done implicitly by associating source domain and reference domain through knowledge distillation and transfer to the reconstruction model [202], helping the model understand and adjust to the variations in data from different sources.

These strategies represent a multifaceted approach to addressing the challenges posed by dataset heterogeneity in MRI reconstruction, ensuring the development of models that are not only technically advanced but also clinically reliable and widely applicable.

### *G. Model Development: Highly Specialized or Broadly Generalizable?*

In medical imaging, the development of **(A) highly specialized models** tailored to specific contrasts, modalities, patient groups, and scanners offers remarkable precision and accuracy.

**(a) Domain shift problem:** However, these models face limitations in versatility and may struggle with performance outside their intended domain, a phenomenon known as domain shift [203]. Although transfer learning has been proposed for domain adaptation to mitigate impacts of domain shifts, we can observe 'hallucinations' and artificial artifacts generated from the source domain in heavily undersampled k-space of the target domain [203], confusing clinical observers.

In contrast, the trend towards creating **(B) broadly applicable models** addresses these limitations. These models cater to a wider range of imaging scenarios and patient conditions, making them suitable for multi-vendor or multi-center studies. However, their development is complex and resource-intensive, requiring large datasets that are often challenging to obtain due to privacy concerns and logistical issues.

**(b) Federated learning** [204] has emerged as a solution, enabling privacy-preserving studies through cryptographic protocols, while enhancing performance of reconstruction models, particularly including MRI-specific multi-coil reconstruction models for complex-valued k-space data [205]. Yet, **challenges** remain in managing **(b.i) Data Diversity**, where inherent complexities and disparities in MRI data, due to varied acquisition protocols and scanner technologies, affect consistency; **(b.ii) System Architecture**, necessitating robust data integrity and encryption across multiple nodes, while balancing computational resources to prevent performance issues like the straggler effect; and **(b.iii) Traceability**, as the indirect handling of data can obscure the learning process, raising concerns about transparency and auditability. Addressing these issues is vital for the effective use of federated learning in this domain.

### *H. Alleviating Memory Constraints: Memory-efficient Networks*

The limited GPU memory often deters the optimization of larger models, where larger models could promote higher input resolution, temporal/volumetric dimensions and the use of modern model structures such as Transformer. To achieve these goals in the context of limited GPU memory, many memory-efficient networks have been developed

One effective strategy involves using reversible networks [206]. Reversible networks can recalculate the activations from subsequent layers during backpropagation, thus eliminating the need retain intermediate feature maps for most layers, to reduce memory usage. In the context of MRI, Putzky et al. [207] proposed an iterative inverse model with constant memory based on invertible networks.

Another innovative approach is the deep equilibrium model (DEQ) [208]. DEQ directly finds the fixed point of the nonlinear transformation by root-finding methods, which is equivalent to an infinite depth feedforward network. This infinite-depth modeling enhances stability in unrolled optimization processes. Its backpropagation can be directly conducted at the equilibrium point using implicit differentiation, without the need to store any intermediate activation values and thus mitigating memory constraints. Gilton et al. [209] developed a DEQ architecture for inverse problems including MRI reconstruction. Gan et al. [210] further proposed a self-supervised DEQ, mitigating memory constraints and data scarcity at the same time.

Additionally, its backpropagation can be conducted at the equilibrium point using implicit differentiation, eliminating the need to store any intermediate activation values and thus mitigating memory constraints.

## V. Conclusion

This review has spotlighted the significant strides made in recent years through data- and physics-driven models. It comprehensively categorizes and evaluates deep learning methods for image reconstruction, including enhancement-based methods, algorithm unrolling models, PnP models, and the full spectrum of generative model-based methods, and integrates critical physics information through coupling with hardware accelerations and sampling pattern optimization. The discussion extends to the challenges and opportunities in the domain, including data scarcity mitigation, redundancy exploitation, evaluation metrics, image integrity preservation, data heterogeneity mitigation, model generalization and memory-efficient models. Looking ahead, the article underscores the transformative potential of these methods in improving MRI efficiency and accuracy, and their integration into clinical practice, while suggesting promising directions for future research.

## Acknowledgements

We appreciate Hwihun Jeong for supplying a subfigure and checking Section II.E.5) Normalizing Flow-based Model and Chi Zhang for checking Section III.B. Enhancing Simultaneous Multi-Slice (SMS).

## References

[1] K. G. Hollingsworth, "Reducing acquisition time in clinical MRI by data undersampling and compressed sensing reconstruction," *Phys. Med. Biol.*, vol. 60, no. 21, pp. R297–R322, Nov. 2015, doi: 10.1088/0031-9155/60/21/R297.

[2] K. P. Pruessmann *et al.*, "SENSE: Sensitivity encoding for fast MRI," *Magn. Reson. Med.*, vol. 42, no. 5, pp. 952–962, Nov. 1999, doi: 10.1002/(SICI)1522-2594(199911)42:5<952::AID-MRM16>3.0.CO;2-S.

[3] M. Lustig *et al.*, “Sparse MRI: The application of compressed sensing for rapid MR imaging,” *Magn. Reson. Med.*, vol. 58, no. 6, pp. 1182–1195, Dec. 2007, doi: 10.1002/mrm.21391.

[4] J. A. Fessler, “Optimization Methods for Magnetic Resonance Image Reconstruction: Key Models and Optimization Algorithms,” *IEEE Signal Process. Mag.*, vol. 37, no. 1, pp. 33–40, Jan. 2020, doi: 10.1109/MSP.2019.2943645.

[5] S. Wang *et al.*, “Deep learning for fast MR imaging: a review for learning reconstruction from incomplete k-space data,” *Biomed. Signal Process. Control*, vol. 68, p. 102579, 2021.

[6] Z. Liang, “SPATIOTEMPORAL IMAGINGWITH PARTIALLY SEPARABLE FUNCTIONS,” in *2007 4th IEEE International Symposium on Biomedical Imaging: From Nano to Macro*, IEEE, 2007, pp. 988–991. doi: 10.1109/ISBI.2007.357020.

[7] J. P. Haldar, “Low-rank modeling of local k-space neighborhoods (LORAKS) for constrained MRI.,” *IEEE Trans. Med. Imaging*, vol. 33, no. 3, pp. 668–81, Mar. 2014, doi: 10.1109/TMI.2013.2293974.

[8] R. Otazo *et al.*, “Low-rank plus sparse matrix decomposition for accelerated dynamic MRI with separation of background and dynamic components.,” *Magn. Reson. Med.*, vol. 73, no. 3, pp. 1125–36, Mar. 2015, doi: 10.1002/mrm.25240.

[9] F. Ong and M. Lustig, “Beyond low rank + sparse: Multi-scale low rank matrix decomposition,” in *2016 IEEE International Conference on Acoustics, Speech and Signal Processing (ICASSP)*, IEEE, Mar. 2016, pp. 4663–4667. doi: 10.1109/ICASSP.2016.7472561.

[10] P. J. Shin *et al.*, “Calibrationless parallel imaging reconstruction based on structured low-rank matrix completion,” *Magn. Reson. Med.*, vol. 72, no. 4, pp. 959–970, Oct. 2014, doi: 10.1002/mrm.24997.

[11] J. P. Haldar and K. Setsompop, “Linear Predictability in MRI Reconstruction: Leveraging Shift-Invariant Fourier Structure for Faster and Better Imaging.,” *IEEE Signal Process. Mag.*, vol. 37, no. 1, pp. 69–82, Jan. 2020, doi: 10.1109/msp.2019.2949570.

[12] M. Uecker *et al.*, “ESPIRiT—an eigenvalue approach to autocalibrating parallel MRI: Where SENSE meets GRAPPA,” *Magn. Reson. Med.*, vol. 71, no. 3, pp. 990–1001, Mar. 2014, doi: 10.1002/mrm.24751.

[13] J. Eckstein and D. P. Bertsekas, “On the Douglas—Rachford splitting method and the proximal point algorithm for maximal monotone operators,” *Math. Program.*, vol. 55, no. 1–3, pp. 293–318, Apr. 1992, doi: 10.1007/BF01581204.

[14] J. C. Ye, “Compressed sensing MRI: a review from signal processing perspective,” *BMC Biomed. Eng.*, vol. 1, no. 1, p. 8, Dec. 2019, doi: 10.1186/s42490-019-0006-z.

[15] K. T. Block *et al.*, “Undersampled radial MRI with multiple coils. Iterative image reconstruction using a total variation constraint,” *Magn. Reson. Med.*, vol. 57, no. 6, pp. 1086–1098, Jun. 2007, doi: 10.1002/mrm.21236.

[16] G. Wang *et al.*, “Image Reconstruction is a New Frontier of Machine Learning,” *IEEE Trans. Med. Imaging*, vol. 37, no. 6, pp. 1289–1296, Jun. 2018, doi: 10.1109/TMI.2018.2833635.

[17] S. G. Lingala and M. Jacob, “Blind compressive sensing dynamic MRI.,” *IEEE Trans. Med. Imaging*, vol. 32, no. 6, pp. 1132–45, Jun. 2013, doi: 10.1109/TMI.2013.2255133.

[18] B. Wen *et al.*, “Transform Learning for Magnetic Resonance Image Reconstruction: From Model-Based Learning to Building Neural Networks,” *IEEE Signal Process. Mag.*, vol. 37, no. 1, pp. 41–53, Jan. 2020, doi: 10.1109/MSP.2019.2951469.

[19] S. Ravishankar and Y. Bresler, “MR Image Reconstruction From Highly Undersampled k-Space Data by Dictionary Learning,” *IEEE Trans. Med. Imaging*, vol. 30, no. 5, pp. 1028–1041, May 2011, doi: 10.1109/TMI.2010.2090538.

[20] I. Y. Chun and J. A. Fessler, “Convolutional Analysis Operator Learning: Acceleration and Convergence,” *IEEE Trans. Image Process.*, vol. 29, pp. 2108–2122, 2020, doi: 10.1109/TIP.2019.2937734.

[21] C. Wang *et al.*, “CMRxRecon: An open cardiac MRI dataset for the competition of accelerated image reconstruction,” p. arXiv:2309.10836, Sep. 2023.

[22] S. S. Chandra *et al.*, “Deep learning in magnetic resonance image reconstruction,” *J. Med. Imaging Radiat. Oncol.*, vol. 65, no. 5, pp. 564–577, 2021.

[23] R. Ahmad *et al.*, “Plug-and-Play Methods for Magnetic Resonance Imaging: Using Denoisers for Image Recovery,” *IEEE Signal Process. Mag.*, vol. 37, no. 1, pp. 105–116, Jan. 2020, doi: 10.1109/MSP.2019.2949470.

[24] V. Monga *et al.*, “Algorithm Unrolling: Interpretable, Efficient Deep Learning for Signal and Image Processing,” *IEEE Signal Process. Mag.*, vol. 38, no. 2, pp. 18–44, Mar. 2021, doi: 10.1109/MSP.2020.3016905.

[25] F. Knoll *et al.*, “Deep-Learning Methods for Parallel Magnetic Resonance Imaging Reconstruction: A Survey of the Current Approaches, Trends, and Issues,” *IEEE Signal Process. Mag.*, vol. 37, no. 1, pp. 128–140, Jan. 2020, doi: 10.1109/MSP.2019.2950640.

[26] A. G. Christodoulou and S. G. Lingala, “Accelerated Dynamic Magnetic Resonance Imaging Using Learned Representations: A New Frontier in Biomedical Imaging,” *IEEE Signal Process. Mag.*, vol. 37, no. 1, pp. 83–93, Jan. 2020, doi: 10.1109/MSP.2019.2942180.

[27] D. Liang *et al.*, “Deep Magnetic Resonance Image Reconstruction: Inverse Problems Meet Neural Networks,” *IEEE Signal Process. Mag.*, vol. 37, no. 1, pp. 141–151, Jan. 2020, doi: 10.1109/MSP.2019.2950557.

[28] K. Hammernik *et al.*, “Physics-Driven Deep Learning for Computational Magnetic Resonance Imaging: Combining physics and machine learning for improved medical imaging,” *IEEE Signal Process. Mag.*, vol. 40, no. 1, pp. 98–114, Jan. 2023, doi: 10.1109/MSP.2022.3215288.

[29] J. Lyu *et al.*, “The state-of-the-art in Cardiac MRI Reconstruction: Results of the CMRxRecon Challenge in MICCAI 2023,” p. arXiv:2404.01082, Apr. 2024.

[30] T. H. Kim *et al.*, “LORAKS makes better SENSE: Phase-constrained partial fourier SENSE reconstruction without phase calibration,” *Magn. Reson. Med.*, vol. 77, no. 3, pp. 1021–1035, Mar. 2017, doi: 10.1002/mrm.26182.

[31] S. G. Lingala *et al.*, “Accelerated Dynamic MRI Exploiting Sparsity and Low-Rank Structure: k-t SLR,” *IEEE Trans. Med. Imaging*, vol. 30, no. 5, pp. 1042–1054, May 2011, doi: 10.1109/TMI.2010.2100850.

[32] B. Zhao *et al.*, “Image reconstruction from highly undersampled ( k , t )-space data with joint partial separability and sparsity constraints,” *IEEE Trans. Med. Imaging*, vol. 31, no. 9, pp. 1809–1820, Sep. 2012, doi: 10.1109/TMI.2012.2203921.

[33] A. Beck and M. Teboulle, “A Fast Iterative Shrinkage-Thresholding Algorithm for Linear Inverse Problems,” *SIAM J. Imaging Sci.*, vol. 2, no. 1, pp. 183–202, Jan. 2009, doi: 10.1137/080716542.

[34] Jian Zhang *et al.*, “Structural Group Sparse Representation for Image Compressive Sensing Recovery,” in *2013 Data Compression Conference*, IEEE, Mar. 2013, pp. 331–340. doi: 10.1109/DCC.2013.41.

[35] I. Daubechies *et al.*, “An iterative thresholding algorithm for linear inverse problems with a sparsity constraint,” *Commun. Pure Appl. Math.*, vol. 57, no. 11, pp. 1413–1457, Nov. 2004, doi: 10.1002/cpa.20042.

[36] Y. Liu and J. P. Haldar, “PALMNUT: An Enhanced Proximal Alternating Linearized Minimization Algorithm With Application to Separate Regularization of Magnitude and Phase,” *IEEE Trans. Comput. Imaging*, vol. 7, pp. 530–518, 2021, doi: 10.1109/TCI.2021.3077806.

[37] S. J. Wright, “Coordinate descent algorithms,” *Math. Program.*, vol. 151, no. 1, pp. 3–34, Jun. 2015, doi: 10.1007/s10107-015-0892-3.

[38] D. R. Hunter and K. Lange, “A tutorial on MM algorithms,” *Am. Stat.*, vol. 58, no. 1, pp. 30–37, 2004.

[39] K. He *et al.*, “Deep Residual Learning for Image Recognition,” in *2016 IEEE Conference on Computer Vision and Pattern Recognition (CVPR)*, IEEE, Jun. 2016, pp. 770–778. doi: 10.1109/CVPR.2016.90.

[40] O. Ronneberger *et al.*, “U-Net: Convolutional Networks for Biomedical Image Segmentation,” 2015, pp. 234–241. doi: 10.1007/978-3-319-24574-4_28.

[41] K. Han *et al.*, “Vision GNN: An Image is Worth Graph of Nodes,” in *Advances in Neural Information Processing Systems*, A. H. Oh, A. Agarwal, D. Belgrave, and K. Cho, Eds., 2022.

[42] A. Dosovitskiy *et al.*, “An Image is Worth 16x16 Words: Transformers for Image Recognition at Scale,” in *International Conference on Learning Representations*, 2021.

[43] Z. Liu *et al.*, “Swin Transformer: Hierarchical Vision Transformer using Shifted Windows,” in *2021 IEEE/CVF International*

*Conference on Computer Vision (ICCV)*, IEEE, Oct. 2021, pp. 9992–10002. doi: 10.1109/ICCV48922.2021.00986.
[44] C. M. Hyun *et al.*, "Deep learning for undersampled MRI reconstruction," *Phys. Med. Biol.*, vol. 63, no. 13, p. 135007, Jun. 2018, doi: 10.1088/1361-6560/aac71a.
[45] Z. Liu *et al.*, "A ConvNet for the 2020s," in *2022 IEEE/CVF Conference on Computer Vision and Pattern Recognition (CVPR)*, IEEE, Jun. 2022, pp. 11966–11976. doi: 10.1109/CVPR52688.2022.01167.
[46] J. Huang *et al.*, "ViGU: Vision GNN U-Net for fast MRI," in *2023 IEEE 20th International Symposium on Biomedical Imaging (ISBI)*, IEEE, Apr. 2023, pp. 1–5. doi: 10.1109/ISBI53787.2023.10230600.
[47] J. Liang *et al.*, "SwinIR: Image Restoration Using Swin Transformer," in *Proceedings of the IEEE/CVF International Conference on Computer Vision (ICCV) Workshops*, 2021, pp. 1833–1844.
[48] C.-M. Feng *et al.*, "Task Transformer Network for Joint MRI Reconstruction and Super-Resolution," 2021, pp. 307–317. doi: 10.1007/978-3-030-87231-1_30.
[49] J. Huang *et al.*, "Swin transformer for fast MRI," *Neurocomputing*, vol. 493, pp. 281–304, Jul. 2022, doi: 10.1016/j.neucom.2022.04.051.
[50] Y. Korkmaz *et al.*, "Unsupervised MRI Reconstruction via Zero-Shot Learned Adversarial Transformers," *IEEE Trans. Med. Imaging*, p. 1, May 2022, doi: 10.1109/TMI.2022.3147426.
[51] J. Huang *et al.*, "Swin Deformable Attention U-Net Transformer (SDAUT) for Explainable Fast MRI," in *Medical Image Computing and Computer Assisted Intervention – MICCAI 2022*, 2022. doi: 10.1007/978-3-031-16446-0_51.
[52] J. Zhang and B. Ghanem, "ISTA-Net: Interpretable optimization-inspired deep network for image compressive sensing," in *Proceedings of the IEEE conference on computer vision and pattern recognition*, 2018, pp. 1828–1837.
[53] J. Schlemper *et al.*, "A Deep Cascade of Convolutional Neural Networks for Dynamic MR Image Reconstruction," *IEEE Trans. Med. Imaging*, vol. 37, no. 2, pp. 491–503, Feb. 2018, doi: 10.1109/TMI.2017.2760978.
[54] H. K. Aggarwal *et al.*, "MoDL: Model-Based Deep Learning Architecture for Inverse Problems," *IEEE Trans. Med. Imaging*, vol. 38, no. 2, pp. 394–405, Feb. 2019, doi: 10.1109/TMI.2018.2865356.
[55] Y. Yang *et al.*, "Deep ADMM-Net for Compressive Sensing MRI," in *Proceedings of the 30th International Conference on Neural Information Processing Systems*, in NIPS'16. 2016, pp. 10–18.
[56] M. Shen *et al.*, "TransCS: A Transformer-Based Hybrid Architecture for Image Compressed Sensing," *IEEE Trans. Image Process.*, vol. 31, pp. 6991–7005, 2022, doi: 10.1109/TIP.2022.3217365.
[57] Y. Zhang *et al.*, "An Introduction to Bilevel Optimization: Foundations and applications in signal processing and machine learning," *IEEE Signal Process. Mag.*, vol. 41, no. 1, pp. 38–59, Apr. 2024, doi: 10.1109/MSP.2024.3358284.
[58] W. Bian *et al.*, "An Optimization-Based Meta-Learning Model for MRI Reconstruction with Diverse Dataset," *J. Imaging*, vol. 7, no. 11, p. 231, Oct. 2021, doi: 10.3390/jimaging7110231.
[59] Q. Ding and X. Zhang, "MRI Reconstruction by Completing Under-sampled K-space Data with Learnable Fourier Interpolation," 2022, pp. 676–685. doi: 10.1007/978-3-031-16446-0_64.
[60] B. Zhou and S. K. Zhou, "DuDoRNet: Learning a Dual-Domain Recurrent Network for Fast MRI Reconstruction With Deep T1 Prior," in *2020 IEEE/CVF Conference on Computer Vision and Pattern Recognition (CVPR)*, IEEE, Jun. 2020, pp. 4272–4281. doi: 10.1109/CVPR42600.2020.00433.
[61] S. V. Venkatakrishnan *et al.*, "Plug-and-Play priors for model based reconstruction," in *2013 IEEE Global Conference on Signal and Information Processing*, IEEE, Dec. 2013, pp. 945–948. doi: 10.1109/GlobalSIP.2013.6737048.
[62] S. H. Chan *et al.*, "Plug-and-Play ADMM for Image Restoration: Fixed-Point Convergence and Applications," *IEEE Trans. Comput. Imaging*, vol. 3, no. 1, pp. 84–98, Mar. 2017, doi: 10.1109/TCI.2016.2629286.
[63] S. Boyd, "Distributed Optimization and Statistical Learning via the Alternating Direction Method of Multipliers," *Found. Trends® Mach. Learn.*, vol. 3, no. 1, pp. 1–122, 2010, doi: 10.1561/2200000016.
[64] A. Pour Yazdanpanah *et al.*, "Deep Plug-and-Play Prior for Parallel MRI Reconstruction," in *2019 IEEE/CVF International Conference on Computer Vision Workshop (ICCVW)*, IEEE, Oct. 2019, pp. 3952–3958. doi: 10.1109/ICCVW.2019.00489.
[65] A. Rasti-Meymandi *et al.*, "Plug and play augmented HQS: Convergence analysis and its application in MRI reconstruction," *Neurocomputing*, vol. 518, pp. 1–14, Jan. 2023, doi: 10.1016/j.neucom.2022.10.061.
[66] H. H. Karaoglu and E. M. Eksioglu, "A Plug-and-Play Deep Denoiser Prior Model for Accelerated MRI Reconstruction," in *2022 45th International Conference on Telecommunications and Signal Processing (TSP)*, IEEE, Jul. 2022, pp. 260–263. doi: 10.1109/TSP55681.2022.9851376.
[67] Y. Romano *et al.*, "The Little Engine That Could: Regularization by Denoising (RED)," *SIAM J. Imaging Sci.*, vol. 10, no. 4, pp. 1804–1844, Jan. 2017, doi: 10.1137/16M1102884.
[68] J. Liu *et al.*, "RARE: Image Reconstruction Using Deep Priors Learned Without Groundtruth," *IEEE J. Sel. Top. Signal Process.*, vol. 14, no. 6, pp. 1088–1099, Oct. 2020, doi: 10.1109/JSTSP.2020.2998402.
[69] Z. Zhao *et al.*, "Generative Models for Inverse Imaging Problems: From mathematical foundations to physics-driven applications," *IEEE Signal Process. Mag.*, vol. 40, no. 1, pp. 148–163, Jan. 2023, doi: 10.1109/MSP.2022.3215282.
[70] C. P. Robert and G. Casella, *Monte Carlo Statistical Methods*. in Springer Texts in Statistics. New York, NY: Springer New York, 2004. doi: 10.1007/978-1-4757-4145-2.
[71] C. Zhang *et al.*, "Conditional Variational Autoencoder for Learned Image Reconstruction," *Computation*, vol. 9, no. 11, p. 114, Oct. 2021, doi: 10.3390/computation9110114.
[72] V. Edupuganti *et al.*, "Uncertainty Quantification in Deep MRI Reconstruction," *IEEE Trans. Med. Imaging*, vol. 40, no. 1, pp. 239–250, Jan. 2021, doi: 10.1109/TMI.2020.3025065.
[73] I. J. Goodfellow *et al.*, "Generative adversarial nets," in *Advances in Neural Information Processing Systems*, Jan. 2014, pp. 2672–2680. doi: 10.3156/jsoft.29.5_177_2.
[74] G. Yang *et al.*, "DAGAN: Deep De-Aliasing Generative Adversarial Networks for Fast Compressed Sensing MRI Reconstruction," *IEEE Trans. Med. Imaging*, vol. 37, no. 6, pp. 1310–1321, Jun. 2018, doi: 10.1109/TMI.2017.2785879.
[75] J. Huang *et al.*, "Fast MRI Reconstruction: How Powerful Transformers Are?," in *2022 44th Annual International Conference of the IEEE Engineering in Medicine & Biology Society (EMBC)*, IEEE, Jul. 2022, pp. 2066–2070. doi: 10.1109/EMBC48229.2022.9871475.
[76] S. Anwar *et al.*, "A Deep Journey into Super-resolution," *ACM Comput. Surv.*, vol. 53, no. 3, pp. 1–34, May 2021, doi: 10.1145/3390462.
[77] Y. Lecun *et al.*, "A tutorial on energy-based learning," 2006.
[78] S. Bond-Taylor *et al.*, "Deep Generative Modelling: A Comparative Review of VAEs, GANs, Normalizing Flows, Energy-Based and Autoregressive Models," *IEEE Trans. Pattern Anal. Mach. Intell.*, vol. 44, no. 11, pp. 7327–7347, Nov. 2022, doi: 10.1109/TPAMI.2021.3116668.
[79] Y. Guan *et al.*, "Magnetic resonance imaging reconstruction using a deep energy-based model," *NMR Biomed.*, vol. 36, no. 3, Mar. 2023, doi: 10.1002/nbm.4848.
[80] Z. Tu *et al.*, "K-space and image domain collaborative energy-based model for parallel MRI reconstruction," *Magn. Reson. Imaging*, vol. 99, pp. 110–122, Jun. 2023, doi: 10.1016/j.mri.2023.02.004.
[81] J. Ho *et al.*, "Denoising Diffusion Probabilistic Models," in *Advances in Neural Information Processing Systems*, Curran Associates, Inc., 2020.
[82] Y. Song and S. Ermon, "Generative Modeling by Estimating Gradients of the Data Distribution," in *Advances in Neural Information Processing Systems*, Curran Associates, Inc., 2019.
[83] Y. Song *et al.*, "Score-Based Generative Modeling through Stochastic Differential Equations," in *International Conference on Learning Representations*, 2021.
[84] C. Peng *et al.*, "Towards Performant and Reliable Undersampled MR Reconstruction via Diffusion Model Sampling," in *Medical Image Computing and Computer Assisted Intervention -- MICCAI 2022*, Cham: Springer Nature Switzerland, 2022, pp. 623–633.
[85] H. Chung and J. C. Ye, "Score-based diffusion models for accelerated MRI," *Med. Image Anal.*, vol. 80, p. 102479, 2022.

[86] Y. Cao *et al.*, "Accelerating multi-echo MRI in k-space with complex-valued diffusion probabilistic model," in *2022 16th IEEE International Conference on Signal Processing (ICSP)*, 2022, pp. 479–484.

[87] A. Güngör *et al.*, "Adaptive diffusion priors for accelerated MRI reconstruction," *Med. Image Anal.*, vol. 88, p. 102872, Aug. 2023, doi: 10.1016/j.media.2023.102872.

[88] C. Cao *et al.*, "High-Frequency Space Diffusion Model for Accelerated MRI," *IEEE Trans. Med. Imaging*, vol. 43, no. 5, pp. 1853–1865, May 2024, doi: 10.1109/TMI.2024.3351702.

[89] H. Chung *et al.*, "Come-Closer-Diffuse-Faster: Accelerating Conditional Diffusion Models for Inverse Problems Through Stochastic Contraction," in *Proceedings of the IEEE/CVF Conference on Computer Vision and Pattern Recognition (CVPR)*, 2022, pp. 12413–12422.

[90] D. J. Rezende and S. Mohamed, "Variational Inference with Normalizing Flows," in *Proceedings of the 32nd International Conference on International Conference on Machine Learning - Volume 37*, in ICML'15. JMLR.org, 2015, pp. 1530–1538.

[91] L. Dinh *et al.*, "Density estimation using Real NVP," in *International Conference on Learning Representations*, 2017.

[92] A. Denker *et al.*, "Conditional Invertible Neural Networks for Medical Imaging," *J. Imaging*, vol. 7, no. 11, p. 243, Nov. 2021, doi: 10.3390/jimaging7110243.

[93] C. Etmann *et al.*, "iUNets: Learnable Invertible Up- and Downsampling for Large-Scale Inverse Problems," in *2020 IEEE 30th International Workshop on Machine Learning for Signal Processing (MLSP)*, IEEE, Sep. 2020, pp. 1–6. doi: 10.1109/MLSP49062.2020.9231874.

[94] J. Wen *et al.*, "A Conditional Normalizing Flow for Accelerated Multi-Coil MR Imaging.," *Proc. Mach. Learn. Res.*, vol. 202, pp. 36926–36939, Jul. 2023.

[95] J. Hamilton *et al.*, "Recent advances in parallel imaging for MRI," *Prog. Nucl. Magn. Reson. Spectrosc.*, vol. 101, pp. 71–95, Aug. 2017, doi: 10.1016/j.pnmrs.2017.04.002.

[96] R. A. Lobos *et al.*, "New Theory and Faster Computations for Subspace-Based Sensitivity Map Estimation in Multichannel MRI," *IEEE Trans. Med. Imaging*, vol. 43, no. 1, pp. 286–296, Jan. 2024, doi: 10.1109/TMI.2023.3297851.

[97] M. A. Griswold *et al.*, "Generalized autocalibrating partially parallel acquisitions (GRAPPA)," *Magn. Reson. Med.*, vol. 47, no. 6, pp. 1202–1210, Jun. 2002, doi: 10.1002/mrm.10171.

[98] M. Lustig and J. M. Pauly, "SPIRiT: Iterative self-consistent parallel imaging reconstruction from arbitrary k -space," *Magn. Reson. Med.*, vol. 64, no. 2, pp. 457–471, Aug. 2010, doi: 10.1002/mrm.22428.

[99] M. Akçakaya *et al.*, "Scan-specific robust artificial-neural-networks for k-space interpolation (RAKI) reconstruction: Database-free deep learning for fast imaging," *Magn. Reson. Med.*, vol. 81, no. 1, pp. 439–453, Jan. 2019, doi: 10.1002/mrm.27420.

[100] T. H. Kim *et al.*, "LORAKI: Autocalibrated Recurrent Neural Networks for Autoregressive MRI Reconstruction in k-Space," p. arXiv:1904.09390, Apr. 2019.

[101] A. Sriram *et al.*, "GrappaNet: Combining parallel imaging with deep learning for multi-coil MRI reconstruction," in *Proceedings of the IEEE/CVF Conference on Computer Vision and Pattern Recognition*, 2020, pp. 14315–14322.

[102] X. Peng *et al.*, "DeepSENSE: Learning coil sensitivity functions for SENSE reconstruction using deep learning," *Magn. Reson. Med.*, vol. 87, no. 4, pp. 1894–1902, 2022.

[103] J. Lv *et al.*, "PIC-GAN: A Parallel Imaging Coupled Generative Adversarial Network for Accelerated Multi-Channel MRI Reconstruction," *Diagnostics*, vol. 11, no. 1, p. 61, Jan. 2021, doi: 10.3390/diagnostics11010061.

[104] M. Barth *et al.*, "Simultaneous multislice (SMS) imaging techniques," *Magn. Reson. Med.*, vol. 75, no. 1, pp. 63–81, Jan. 2016, doi: 10.1002/mrm.25897.

[105] D. A. Feinberg and K. Setsompop, "Ultra-fast MRI of the human brain with simultaneous multi-slice imaging," *J. Magn. Reson.*, vol. 229, pp. 90–100, Apr. 2013, doi: 10.1016/j.jmr.2013.02.002.

[106] F. A. Breuer *et al.*, "Controlled aliasing in parallel imaging results in higher acceleration (CAIPIRINHA) for multi-slice imaging," *Magn. Reson. Med.*, vol. 53, no. 3, pp. 684–691, Mar. 2005, doi: 10.1002/mrm.20401.

[107] K. Setsompop *et al.*, "Blipped-controlled aliasing in parallel imaging for simultaneous multislice echo planar imaging with reduced g -factor penalty," *Magn. Reson. Med.*, vol. 67, no. 5, pp. 1210–1224, May 2012, doi: 10.1002/mrm.23097.

[108] K. Setsompop *et al.*, "Improving diffusion MRI using simultaneous multi-slice echo planar imaging," *Neuroimage*, vol. 63, no. 1, pp. 569–580, Oct. 2012, doi: 10.1016/j.neuroimage.2012.06.033.

[109] J. Xu *et al.*, "Evaluation of slice accelerations using multiband echo planar imaging at 3T," *Neuroimage*, vol. 83, pp. 991–1001, Dec. 2013, doi: 10.1016/j.neuroimage.2013.07.055.

[110] S. Moeller *et al.*, "Multiband multislice GE-EPI at 7 tesla, with 16-fold acceleration using partial parallel imaging with application to high spatial and temporal whole-brain fMRI," *Magn. Reson. Med.*, vol. 63, no. 5, pp. 1144–1153, May 2010, doi: 10.1002/mrm.22361.

[111] M. Tänzer *et al.*, "Review of Data Types and Model Dimensionality for Cardiac DTI SMS-Related Artefact Removal," 2022, pp. 123–132. doi: 10.1007/978-3-031-23443-9_12.

[112] M. Kim *et al.*, "Deep Learning-Enhanced Parallel Imaging and Simultaneous Multislice Acceleration Reconstruction in Knee MRI," *Invest. Radiol.*, vol. 57, no. 12, pp. 826–833, Dec. 2022, doi: 10.1097/RLI.0000000000000900.

[113] S. F. Cauley *et al.*, "Interslice leakage artifact reduction technique for simultaneous multislice acquisitions," *Magn. Reson. Med.*, vol. 72, no. 1, pp. 93–102, Jul. 2014, doi: 10.1002/mrm.24898.

[114] O. B. Demirel *et al.*, "Improved simultaneous multislice cardiac MRI using readout concatenated k-space SPIRiT (ROCK-SPIRiT)," *Magn. Reson. Med.*, vol. 85, no. 6, pp. 3036–3048, Jun. 2021, doi: 10.1002/mrm.28680.

[115] C. Zhang *et al.*, "Accelerated Simultaneous Multi-Slice MRI using Subject-Specific Convolutional Neural Networks," in *2018 52nd Asilomar Conference on Signals, Systems, and Computers*, IEEE, Oct. 2018, pp. 1636–1640. doi: 10.1109/ACSSC.2018.8645313.

[116] A. S. Nencka *et al.*, "Split-slice training and hyperparameter tuning of RAKI networks for simultaneous multi-slice reconstruction," *Magn. Reson. Med.*, vol. 85, no. 6, pp. 3272–3280, Jun. 2021, doi: 10.1002/mrm.28634.

[117] J. Le *et al.*, "Deep learning for radial SMS myocardial perfusion reconstruction using the 3D residual booster U-net," *Magn. Reson. Imaging*, vol. 83, pp. 178–188, 2021.

[118] S. Li *et al.*, "A simultaneous multi-slice T2 mapping framework based on overlapping-echo detachment planar imaging and deep learning reconstruction," *Magn. Reson. Med.*, 2022.

[119] O. B. Demirel *et al.*, "Improved Simultaneous Multi-Slice Functional MRI Using Self-supervised Deep Learning," in *2021 55th Asilomar Conference on Signals, Systems, and Computers*, IEEE, Oct. 2021, pp. 890–894. doi: 10.1109/IEEECONF53345.2021.9723264.

[120] F. Knoll *et al.*, "Adapted random sampling patterns for accelerated MRI," *Magn. Reson. Mater. Physics, Biol. Med.*, vol. 24, no. 1, pp. 43–50, Feb. 2011, doi: 10.1007/s10334-010-0234-7.

[121] S. Ravishankar and Y. Bresler, "Adaptive sampling design for compressed sensing MRI," in *2011 Annual International Conference of the IEEE Engineering in Medicine and Biology Society*, IEEE, Aug. 2011, pp. 3751–3755. doi: 10.1109/IEMBS.2011.6090639.

[122] M. Seeger *et al.*, "Optimization of k -space trajectories for compressed sensing by Bayesian experimental design," *Magn. Reson. Med.*, vol. 63, no. 1, pp. 116–126, Jan. 2010, doi: 10.1002/mrm.22180.

[123] T. Sanchez *et al.*, "Scalable Learning-Based Sampling Optimization for Compressive Dynamic MRI," in *ICASSP 2020 - 2020 IEEE International Conference on Acoustics, Speech and Signal Processing (ICASSP)*, IEEE, May 2020, pp. 8584–8588. doi: 10.1109/ICASSP40776.2020.9053345.

[124] M. V. W. Zibetti *et al.*, "Fast data-driven learning of parallel MRI sampling patterns for large scale problems," *Sci. Rep.*, vol. 11, no. 1, p. 19312, Sep. 2021, doi: 10.1038/s41598-021-97995-w.

[125] J. P. Haldar and D. Kim, "OEDIPUS: An Experiment Design Framework for Sparsity-Constrained MRI," *IEEE Trans. Med. Imaging*, vol. 38, no. 7, pp. 1545–1558, Jul. 2019, doi: 10.1109/TMI.2019.2896180.

[126] C. Lazarus *et al.*, "SPARKLING: variable-density k-space filling curves for accelerated T 2 * -weighted MRI," *Magn. Reson. Med.*, vol. 81, no. 6, pp. 3643–3661, Jun. 2019, doi: 10.1002/mrm.27678.

[127] C. Boyer *et al.*, "On the Generation of Sampling Schemes for Magnetic Resonance Imaging," *SIAM J. Imaging Sci.*, vol. 9, no. 4, pp. 2039–2072, Jan. 2016, doi: 10.1137/16M1059205.

[128] N. Chauffert *et al.*, "A Projection Method on Measures Sets," *Constr. Approx.*, vol. 45, no. 1, pp. 83–111, Feb. 2017, doi: 10.1007/s00365-016-9346-2.

[129] C. Lazarus *et al.*, "3D variable-density SPARKLING trajectories for high-resolution T2*-weighted magnetic resonance imaging," *NMR Biomed.*, vol. 33, no. 9, Sep. 2020, doi: 10.1002/nbm.4349.

[130] Z. Zhang *et al.*, "Reducing Uncertainty in Undersampled MRI Reconstruction With Active Acquisition," in *2019 IEEE/CVF Conference on Computer Vision and Pattern Recognition (CVPR)*, IEEE, Jun. 2019, pp. 2049–2053. doi: 10.1109/CVPR.2019.00215.

[131] I. A. M. Huijben *et al.*, "Learning Sampling and Model-Based Signal Recovery for Compressed Sensing MRI," in *ICASSP 2020 - 2020 IEEE International Conference on Acoustics, Speech and Signal Processing (ICASSP)*, IEEE, May 2020, pp. 8906–8910. doi: 10.1109/ICASSP40776.2020.9053331.

[132] J. Zhang *et al.*, "Extending LOUPE for K-Space Under-Sampling Pattern Optimization in Multi-coil MRI," 2020, pp. 91–101. doi: 10.1007/978-3-030-61598-7_9.

[133] T. Weiss *et al.*, "PILOT: Physics-Informed Learned Optimized Trajectories for Accelerated MRI," *Mach. Learn. Biomed. Imaging*, vol. 1, no. April 2021, pp. 1–23, Apr. 2021, doi: 10.59275/j.melba.2021-1a1f.

[134] H. K. Aggarwal and M. Jacob, "J-MoDL: Joint Model-Based Deep Learning for Optimized Sampling and Reconstruction," *IEEE J. Sel. Top. Signal Process.*, vol. 14, no. 6, pp. 1151–1162, Oct. 2020, doi: 10.1109/JSTSP.2020.3004094.

[135] G. Wang *et al.*, "B-Spline Parameterized Joint Optimization of Reconstruction and K-Space Trajectories (BJORK) for Accelerated 2D MRI," *IEEE Trans. Med. Imaging*, vol. 41, no. 9, pp. 2318–2330, Sep. 2022, doi: 10.1109/TMI.2022.3161875.

[136] G. Wang *et al.*, "Stochastic optimization of three-dimensional non-Cartesian sampling trajectory," *Magn. Reson. Med.*, vol. 90, no. 2, pp. 417–431, Aug. 2023, doi: 10.1002/mrm.29645.

[137] J. Alush-Aben *et al.*, "3D FLAT: Feasible Learned Acquisition Trajectories for Accelerated MRI," 2020, pp. 3–16. doi: 10.1007/978-3-030-61598-7_1.

[138] F. A. Breuer *et al.*, "Controlled aliasing in parallel imaging results in higher acceleration (CAIPIRINHA) for multi-slice imaging.," *Magn. Reson. Med.*, vol. 53, no. 3, pp. 684–91, Mar. 2005, doi: 10.1002/mrm.20401.

[139] F. A. Breuer *et al.*, "Controlled aliasing in volumetric parallel imaging (2D CAIPIRINHA).," *Magn. Reson. Med.*, vol. 55, no. 3, pp. 549–56, Mar. 2006, doi: 10.1002/mrm.20787.

[140] M. Engel *et al.*, "T-Hex: Tilted hexagonal grids for rapid 3D imaging," *Magn. Reson. Med.*, vol. 85, no. 5, pp. 2507–2523, May 2021, doi: 10.1002/mrm.28600.

[141] G. R. Chaithya *et al.*, "Optimizing Full 3D SPARKLING Trajectories for High-Resolution Magnetic Resonance Imaging," *IEEE Trans. Med. Imaging*, vol. 41, no. 8, pp. 2105–2117, Aug. 2022, doi: 10.1109/TMI.2022.3157269.

[142] T. Bakker *et al.*, "Experimental design for MRI by greedy policy search," in *Advances in Neural Information Processing Systems*, H. Larochelle, M. Ranzato, R. Hadsell, M. F. Balcan, and H. Lin, Eds., Curran Associates, Inc., 2020, pp. 18954–18966.

[143] L. Pineda *et al.*, "Active MR k-space Sampling with Reinforcement Learning," 2020, pp. 23–33. doi: 10.1007/978-3-030-59713-9_3.

[144] Z. Huang and S. Ravishankar, "Single-Pass Object-Adaptive Data Undersampling and Reconstruction for MRI," *IEEE Trans. Comput. Imaging*, vol. 8, pp. 333–345, 2022, doi: 10.1109/TCI.2022.3167454.

[145] T. Yin *et al.*, "End-to-End Sequential Sampling and Reconstruction for MRI," *Proc. Mach. Learn. Res.*, vol. 158, pp. 261–279, May 2021.

[146] Y. Bengio *et al.*, "Estimating or Propagating Gradients Through Stochastic Neurons for Conditional Computation," p. arXiv:1308.3432, Aug. 2013.

[147] E. Jang *et al.*, "Categorical Reparameterization with Gumbel-Softmax," in *International Conference on Learning Representations*, 2017.

[148] C. Alkan *et al.*, "AutoSamp: Autoencoding k-space Sampling via Variational Information Maximization for 3D MRI," *IEEE Trans. Med. Imaging*, pp. 1–1, Jun. 2024, doi: 10.1109/TMI.2024.3443292.

[149] C. G. Radhakrishna and P. Ciuciu, "Jointly Learning Non-Cartesian k-Space Trajectories and Reconstruction Networks for 2D and 3D MR Imaging through Projection," *Bioengineering*, vol. 10, no. 2, p. 158, Jan. 2023, doi: 10.3390/bioengineering10020158.

[150] N. Chauffert *et al.*, "A Projection Algorithm for Gradient Waveforms Design in Magnetic Resonance Imaging," *IEEE Trans. Med. Imaging*, vol. 35, no. 9, pp. 2026–2039, Sep. 2016, doi: 10.1109/TMI.2016.2544251.

[151] F. Sherry *et al.*, "Learning the Sampling Pattern for MRI," *IEEE Trans. Med. Imaging*, vol. 39, no. 12, pp. 4310–4321, Dec. 2020, doi: 10.1109/TMI.2020.3017353.

[152] S. Wang *et al.*, "Knowledge-driven deep learning for fast MR imaging: Undersampled MR image reconstruction from supervised to un-supervised learning.," *Magn. Reson. Med.*, Apr. 2024, doi: 10.1002/mrm.30105.

[153] B. Yaman *et al.*, "Self-supervised learning of physics-guided reconstruction neural networks without fully sampled reference data," *Magn. Reson. Med.*, vol. 84, no. 6, pp. 3172–3191, Dec. 2020, doi: 10.1002/mrm.28378.

[154] H. K. Aggarwal *et al.*, "ENSURE: A General Approach for Unsupervised Training of Deep Image Reconstruction Algorithms," *IEEE Trans. Med. Imaging*, vol. 42, no. 4, pp. 1133–1144, Apr. 2023, doi: 10.1109/TMI.2022.3224359.

[155] D. Ulyanov *et al.*, "Deep Image Prior," *Int. J. Comput. Vis.*, vol. 128, no. 7, pp. 1867–1888, Jul. 2020, doi: 10.1007/s11263-020-01303-4.

[156] J. Yoo *et al.*, "Time-Dependent Deep Image Prior for Dynamic MRI," *IEEE Trans. Med. Imaging*, vol. 40, no. 12, pp. 3337–3348, Dec. 2021, doi: 10.1109/TMI.2021.3084288.

[157] K. Lei *et al.*, "Wasserstein GANs for MR Imaging: From Paired to Unpaired Training," *IEEE Trans. Med. Imaging*, vol. 40, no. 1, pp. 105–115, Jan. 2021, doi: 10.1109/TMI.2020.3022968.

[158] G. Oh *et al.*, "Unpaired Deep Learning for Accelerated MRI Using Optimal Transport Driven CycleGAN," *IEEE Trans. Comput. Imaging*, vol. 6, pp. 1285–1296, 2020, doi: 10.1109/TCI.2020.3018562.

[159] S. K. Zhou *et al.*, "Deep reinforcement learning in medical imaging: A literature review," *Med. Image Anal.*, vol. 73, p. 102193, Oct. 2021, doi: 10.1016/j.media.2021.102193.

[160] M. L. Puterman, "Chapter 8 Markov decision processes," in *Stochastic Models*, Handbooks in Operations Research and Management Science, 1990, pp. 331–434. doi: 10.1016/S0927-0507(05)80172-0.

[161] H. Van Hasselt *et al.*, "Deep Reinforcement Learning with Double Q-Learning," *Proc. AAAI Conf. Artif. Intell.*, vol. 30, no. 1, Mar. 2016, doi: 10.1609/aaai.v30i1.10295.

[162] Y. Liu *et al.*, "Dual states based reinforcement learning for fast MR scan and image reconstruction," *Neurocomputing*, vol. 568, p. 127067, Feb. 2024, doi: 10.1016/j.neucom.2023.127067.

[163] W. Li *et al.*, "MRI Reconstruction with Interpretable Pixel-Wise Operations Using Reinforcement Learning," *Proc. AAAI Conf. Artif. Intell.*, vol. 34, no. 01, pp. 792–799, Apr. 2020, doi: 10.1609/aaai.v34i01.5423.

[164] C. Wang *et al.*, "Deep Reinforcement Learning Based Unrolling Network for MRI Reconstruction," in *2023 IEEE 20th International Symposium on Biomedical Imaging (ISBI)*, IEEE, Apr. 2023, pp. 1–5. doi: 10.1109/ISBI53787.2023.10230611.

[165] L. Xiang *et al.*, "Deep-Learning-Based Multi-Modal Fusion for Fast MR Reconstruction," *IEEE Trans. Biomed. Eng.*, vol. 66, no. 7, pp. 2105–2114, Jul. 2019, doi: 10.1109/TBME.2018.2883958.

[166] S. U. H. Dar *et al.*, "Prior-Guided Image Reconstruction for Accelerated Multi-Contrast MRI via Generative Adversarial Networks," *IEEE J. Sel. Top. Signal Process.*, vol. 14, no. 6, pp. 1072–1087, Oct. 2020, doi: 10.1109/JSTSP.2020.3001737.

[167] L. Sun *et al.*, "A Deep Information Sharing Network for Multi-Contrast Compressed Sensing MRI Reconstruction," *IEEE Trans. Image Process.*, vol. 28, no. 12, pp. 6141–6153, Dec. 2019, doi: 10.1109/TIP.2019.2925288.

[168] D. Polak *et al.*, "Joint multi-contrast variational network reconstruction (jVN) with application to rapid 2D and 3D imaging," *Magn. Reson. Med.*, vol. 84, no. 3, pp. 1456–1469, Sep. 2020, doi: 10.1002/mrm.28219.

[169] W. Do *et al.*, "Reconstruction of multicontrast MR images through deep learning," *Med. Phys.*, vol. 47, no. 3, pp. 983–997, Mar. 2020, doi: 10.1002/mp.14006.

[170] Y. Chen *et al.*, “Deep Learning Within a Priori Temporal Feature Spaces for Large-Scale Dynamic MR Image Reconstruction: Application to 5-D Cardiac MR Multitasking,” 2019, pp. 495–504. doi: 10.1007/978-3-030-32245-8_55.

[171] H. Jeelani *et al.*, “A Myocardial T1-Mapping Framework with Recurrent and U-Net Convolutional Neural Networks,” in *2020 IEEE 17th International Symposium on Biomedical Imaging (ISBI)*, IEEE, Apr. 2020, pp. 1941–1944. doi: 10.1109/ISBI45749.2020.9098459.

[172] J. I. Hamilton *et al.*, “Deep learning reconstruction for cardiac magnetic resonance fingerprinting T 1 and T 2 mapping,” *Magn. Reson. Med.*, vol. 85, no. 4, pp. 2127–2135, Apr. 2021, doi: 10.1002/mrm.28568.

[173] S. Kastryulin *et al.*, “Image Quality Assessment for Magnetic Resonance Imaging,” *IEEE Access*, vol. 11, pp. 14154–14168, 2023, doi: 10.1109/ACCESS.2023.3243466.

[174] P. M. Adamson *et al.*, “Using Deep Feature Distances for Evaluating MR Image Reconstruction Quality,” in *NeurIPS 2023 Workshop on Deep Learning and Inverse Problems*, 2023.

[175] H. H. Barrett *et al.*, “Objective assessment of image quality III ROC metrics, ideal observers, and likelihood-generating functions,” *J. Opt. Soc. Am. A*, vol. 15, no. 6, p. 1520, Jun. 1998, doi: 10.1364/JOSAA.15.001520.

[176] R. C. Streijl *et al.*, “Mean opinion score (MOS) revisited: methods and applications, limitations and alternatives,” *Multimed. Syst.*, vol. 22, no. 2, pp. 213–227, Mar. 2016, doi: 10.1007/s00530-014-0446-1.

[177] S. Abdullah *et al.*, “MRI Reconstruction From Sparse K-Space Data Using Low Dimensional Manifold Model,” *IEEE Access*, vol. 7, pp. 88072–88081, 2019, doi: 10.1109/ACCESS.2019.2925051.

[178] Z. Wang *et al.*, “Image Quality Assessment: From Error Visibility to Structural Similarity,” *IEEE Trans. Image Process.*, vol. 13, no. 4, pp. 600–612, Apr. 2004, doi: 10.1109/TIP.2003.819861.

[179] A. Mittal *et al.*, “No-Reference Image Quality Assessment in the Spatial Domain,” *IEEE Trans. Image Process.*, vol. 21, no. 12, pp. 4695–4708, Dec. 2012, doi: 10.1109/TIP.2012.2214050.

[180] J. Joseph and R. Periyasamy, “An image driven bilateral filter with adaptive range and spatial parameters for denoising Magnetic Resonance Images,” *Comput. Electr. Eng.*, vol. 69, pp. 782–795, Jul. 2018, doi: 10.1016/j.compeleceng.2018.02.033.

[181] R. Zhang *et al.*, “The Unreasonable Effectiveness of Deep Features as a Perceptual Metric,” in *2018 IEEE/CVF Conference on Computer Vision and Pattern Recognition*, IEEE, Jun. 2018, pp. 586–595. doi: 10.1109/CVPR.2018.00068.

[182] M. Heusel *et al.*, “GANs Trained by a Two Time-Scale Update Rule Converge to a Local Nash Equilibrium,” in *Advances in Neural Information Processing Systems*, Curran Associates, Inc., 2017.

[183] P. M. Adamson *et al.*, “SSFD: Self-Supervised Feature Distance as an MR Image Reconstruction Quality Metric,” in *NeurIPS 2021 Workshop on Deep Learning and Inverse Problems*, 2021.

[184] S. Chabert *et al.*, “Image Quality Assessment to Emulate Experts’ Perception in Lumbar MRI Using Machine Learning,” *Appl. Sci.*, vol. 11, no. 14, p. 6616, Jul. 2021, doi: 10.3390/app11146616.

[185] L. McCullum *et al.*, “The Use of Quantitative Metrics and Machine Learning to Predict Radiologist Interpretations of MRI Image Quality and Artifacts,” p. arXiv:2311.05412, Nov. 2023.

[186] C. Chan and J. P. Haldar, “Local perturbation responses and checkerboard tests: Characterization tools for nonlinear MRI methods,” *Magn. Reson. Med.*, vol. 86, no. 4, pp. 1873–1887, Oct. 2021, doi: 10.1002/mrm.28828.

[187] S. Bhadra *et al.*, “On Hallucinations in Tomographic Image Reconstruction,” *IEEE Trans. Med. Imaging*, vol. 40, no. 11, pp. 3249–3260, Nov. 2021, doi: 10.1109/TMI.2021.3077857.

[188] G. Luo *et al.*, “Bayesian MRI reconstruction with joint uncertainty estimation using diffusion models,” *Magn. Reson. Med.*, vol. 90, no. 1, pp. 295–311, Jul. 2023, doi: 10.1002/mrm.29624.

[189] J. Huang *et al.*, “Deep learning-based diffusion tensor cardiac magnetic resonance reconstruction: a comparison study,” *Sci. Rep.*, vol. 14, no. 1, p. 5658, Mar. 2024, doi: 10.1038/s41598-024-55880-2.

[190] Y. Korkmaz *et al.*, “Self-supervised MRI Reconstruction with Unrolled Diffusion Models,” 2023, pp. 491–501. doi: 10.1007/978-3-031-43999-5_47.

[191] Z. Fabian *et al.*, “HUMUS-Net: Hybrid unrolled multi-scale network architecture for accelerated MRI reconstruction,” in *Advances in Neural Information Processing Systems*, 2022, pp. 25306–25319.

[192] A. Kendall and Y. Gal, “What uncertainties do we need in Bayesian deep learning for computer vision?,” in *Proceedings of the 31st International Conference on Neural Information Processing Systems*, in NIPS’17. Red Hook, NY, USA: Curran Associates Inc., 2017, pp. 5580–5590.

[193] D. Narnhofer *et al.*, “Bayesian Uncertainty Estimation of Learned Variational MRI Reconstruction,” *IEEE Trans. Med. Imaging*, vol. 41, no. 2, pp. 279–291, Feb. 2022, doi: 10.1109/TMI.2021.3112040.

[194] H. Chung and J. C. Ye, “Score-based diffusion models for accelerated MRI,” *Med. Image Anal.*, vol. 80, p. 102479, Aug. 2022, doi: 10.1016/j.media.2022.102479.

[195] A. Jalal *et al.*, “Robust Compressed Sensing MRI with Deep Generative Priors,” in *Advances in Neural Information Processing Systems*, 2021, pp. 14938–14954.

[196] Q. Zhang *et al.*, “Quality assurance of quantitative cardiac T1-mapping in multicenter clinical trials – A T1 phantom program from the hypertrophic cardiomyopathy registry (HCMR) study,” *Int. J. Cardiol.*, vol. 330, pp. 251–258, May 2021, doi: 10.1016/j.ijcard.2021.01.026.

[197] D. R. Messroghli *et al.*, “Clinical recommendations for cardiovascular magnetic resonance mapping of T1, T2, T2* and extracellular volume: A consensus statement by the Society for Cardiovascular Magnetic Resonance (SCMR) endorsed by the European Association for Cardiovascular Imagi,” *J. Cardiovasc. Magn. Reson.*, vol. 19, no. 1, p. 75, Dec. 2017, doi: 10.1186/s12968-017-0389-8.

[198] V. Antun *et al.*, “On instabilities of deep learning in image reconstruction and the potential costs of AI,” *Proc. Natl. Acad. Sci.*, vol. 117, no. 48, pp. 30088–30095, 2020.

[199] E. Goceri, “Medical image data augmentation: techniques, comparisons and interpretations,” *Artif. Intell. Rev.*, vol. 56, no. 11, pp. 12561–12605, Nov. 2023, doi: 10.1007/s10462-023-10453-z.

[200] H. Guan and M. Liu, “Domain Adaptation for Medical Image Analysis: A Survey,” *IEEE Trans. Biomed. Eng.*, vol. 69, no. 3, pp. 1173–1185, Mar. 2022, doi: 10.1109/TBME.2021.3117407.

[201] Y. Nan *et al.*, “Data Harmonisation for Information Fusion in Digital Healthcare: A State-of-the-Art Systematic Review, Meta-Analysis and Future Research Directions,” *Inf. Fusion*, 2022.

[202] G. Wen *et al.*, “Machine Learning for Brain MRI Data Harmonisation: A Systematic Review,” *Bioengineering*, vol. 10, no. 4, p. 397, Mar. 2023, doi: 10.3390/bioengineering10040397.

[203] K. Hammernik *et al.*, “Systematic evaluation of iterative deep neural networks for fast parallel MRI reconstruction with sensitivity-weighted coil combination,” *Magn. Reson. Med.*, vol. 86, no. 4, pp. 1859–1872, Oct. 2021, doi: 10.1002/mrm.28827.

[204] P. Guo *et al.*, “Multi-institutional collaborations for improving deep learning-based magnetic resonance image reconstruction using federated learning,” in *Proceedings of the IEEE/CVF Conference on Computer Vision and Pattern Recognition*, 2021, pp. 2423–2432.

[205] G. Elmas *et al.*, “Federated Learning of Generative Image Priors for MRI Reconstruction,” *IEEE Trans. Med. Imaging*, vol. 42, no. 7, pp. 1996–2009, Jul. 2023, doi: 10.1109/TMI.2022.3220757.

[206] A. N. Gomez *et al.*, “The Reversible Residual Network: Backpropagation Without Storing Activations,” *Adv. Neural Inf. Process. Syst.*, vol. 30, Jul. 2017.

[207] P. Putzky and M. Welling, “Invert to Learn to Invert,” in *Advances in Neural Information Processing Systems*, 2019.

[208] S. Bai *et al.*, “Deep Equilibrium Models,” in *Advances in Neural Information Processing Systems*, H. Wallach, H. Larochelle, A. Beygelzimer, F. d\textquotesingle Alché-Buc, E. Fox, and R. Garnett, Eds., 2019.

[209] D. Gilton *et al.*, “Deep Equilibrium Architectures for Inverse Problems in Imaging,” *IEEE Trans. Comput. Imaging*, vol. 7, pp. 1123–1133, 2021, doi: 10.1109/TCI.2021.3118944.

[210] W. Gan *et al.*, “Self-Supervised Deep Equilibrium Models With Theoretical Guarantees and Applications to MRI Reconstruction,” *IEEE Trans. Comput. Imaging*, vol. 9, pp. 796–807, 2023, doi: 10.1109/TCI.2023.3304475.

**Supplementary Materials**

# Supplementary Table

**TABLE S1.** COMPARISON OF REPRESENTATIVE METHODS OF DIFFERENT TYPES

| Paper Name | Year | Author Name (First Author, Given Name) | Model Name | Model Type | Backbone | Train Strategy | Metrics | | Physics Model | | Dataset* | Performance | Computational Cost | Runtime | Code Availability | Note |
|---|---|---|---|---|---|---|---|---|---|---|---|---|---|---|---|---|
| | | | | | | | Quality | Other | DC (Y/N) | Other | | | | | | |
| **Deep learning for undersampled MRI reconstruction** | 2018 | Hyun et al. | N.A. | Enhancement | SC, U-Net, CNNs | Supervised | MSE, SSIM | N.A. | YES | NA | Private | MSE: 0.0004; SSIM: 0.9039 (Cartesian AF×4) | Not Provided | Not Provided | Not Provided | |
| **ViGU: Vision GNN U-Net for Fast MRI** | 2023 | Huang et al. | ViGU | Enhancement | SC, U-Net, GNNs | Supervised | PSNR, SSIM, LPIPS | MACs | NO | NA | Calgary-Campinas Public Dataset* | PSNR: 32.95; SSIM: 0.955; FID: 22.73 (Gaussian 1D 30% / AF×3.33)<br><br>PSNR: 27.72; SSIM: 0.872; FID: 58.61 (Radial 10% / AF×10) | MACs(G): 73.02 | Not Provided | https://github.com/ayanglab/ViGU | MACs: Multiply–accumulate operations |
| **Task Transformer Network for Joint MRI Reconstruction and Super-Resolution** | 2021 | Feng et al. | T2Net | Enhancement | SC, Transformers | Supervised | NMSE, PSNR, SSIM | N.A. | NO | NA | IXI Dataset* | NMSE: 0.027; PSNR: 29.397; SSIM 0.872; HFEN 0.112 (Cartesian AF×2)<br><br>NMSE: 0.032; PSNR: 28.659; SSIM 0.850; HFEN 0.173 (Cartesian AF×4) | Not Provided | Not Provided | https://github.com/chunmeifeng/T2Net | |
| **Swin transformer for fast MRI** | 2022 | Huang et al. | SwinMR | Enhancement | SC/MC, SwinIR, Transformers | Supervised | PSNR, SSIM, FID | #PARAMs, MACs, Inference Time | NO | NA | Calgary-Campinas Public Dataset* | PSNR: 33.06; SSIM: 0.956; FID: 21.03 (Gaussian 1D 30% / AF×3.33) | #PARAMs(M): 11.40; MACs(G): 800.73 | 0.041s (NVIDIA RTX3090) | https://github.com/ayanglab/SwinMR | #PARAMs: Number of parameters |
| **Unsupervised MRI Reconstruction via Zero-Shot Learned Adversarial Transformers** | 2022 | Korkmaz et al. | SLATER | Enhancement | SC/MC, Transformers | Unsupervised | PSNR, SSIM | Inference Time | YES | NA | IXI Dataset* FastMRI Dataset* | PSNR: 38.8; SSIM: 0.979 (Gaussian 2D 4, IXI T1) PSNR: 33.2; SSIM: 0.952 (Gaussian 2D AF×8, IXI T1)<br><br>PSNR: 40.0; SSIM: 0.977 (Gaussian 2D AF×4, IXI T2)<br><br>PSNR: 34.1; SSIM: 0.948 (Gaussian 2D AF×8, IXI T2)<br><br>PSNR: 37.6; SSIM: 0.939 (Gaussian 2D AF×4, FastMRI Brain T1)<br><br>PSNR: 34.2; SSIM: 0.907 (Gaussian 2D AF×8, FastMRI Brain T1)<br><br>PSNR: 36.3; SSIM: 0.955 (Gaussian 2D AF×4, FastMRI Brain T2)<br><br>PSNR: 33.4; SSIM: 0.930 (Gaussian 2D AF×8, FastMRI Brain T2) | Not Provided | 2.63s (5×NVIDIA RTX2080Ti) | https://github.com/icon-lab/SLATER | |
| **Swin Deformable Attention U-Net Transformer (SDAUT) for Explainable Fast MRI** | 2022 | Huang et al. | SDAUT | Enhancement | SC, U-Net, Transformers | Supervised | PSNR, SSIM, FID | MACs | NO | NA | Calgary-Campinas Public Dataset* | PSNR: 33.92; SSIM: 0.963; FID: 20.45 (Gaussian 1D 30% / AF×3.33)<br>PSNR: 28.28; SSIM: 0.885; FID: 55.12 (Radial 10% / AF×10) | MACs(G): 293.02 | Not Provided | https://github.com/ayanglab/SDAUT | |
| **Deep ADMM-Net for Compressive Sensing MRI** | 2016 | Yang et al. | ADMM-Net | Algorithm Unrolling | MC | Supervised | NMSE, PSNR | Inference Time | YES | ADMM | CAF Project: Brain and Chest* | NMSE: 0.0739; PSNR: 37.17 (Radial AF×5)<br><br>NMSE: 0.0544; PSNR: 39.84 (Radial AF×3.3)<br><br>NMSE: 0.0447; PSNR: 41.56 (Radial AF×2.5)<br><br>NMSE: 0.0379; PSNR: 43.00 (Radial AF×2) | Not provided | 0.7911s (i7-4790k CPU) | https://github.com/yangyan92/Deep-ADMM-Net | |

| | | | | | | | | | | | | | | | |
|---|---|---|---|---|---|---|---|---|---|---|---|---|---|---|---|
| **A Deep Cascade of Convolutional Neural Networks for Dynamic MR Image Reconstruction** | 2018 | Schlemper et al. | DLMRI (DCCNN, D5C5) | Algorithm Unrolling | SC, CNNs | Supervised | MSE, PSNR | Inference Time, #PARAMS | YES | NA | Private | No Numerical Results Provided. (Provided in Plots.) | #PARAMS(M): 0.6~3.4 | 8.21s (NVIDIA GTX1080) | https://github.com/js3611/Deep-MRI-Reconstruction |
| **MoDL: Model-Based Deep Learning Architecture for Inverse Problems** | 2019 | Aggarwal et al. | MoDL | Algorithm Unrolling | MC, CNNs | Supervised | PSNR | Inference Time, #PARAMS | YES | SENSE | Private | PSNR: 39.08 (Variable density 2D AF x6);<br><br>PSNR: 37.35 (Variable density 2D AF×10) | #PARAMS(K): 188 | 28s (for 164 slices, unknown GPU) | https://github.com/hkaggarwal/modl |
| **An Optimization-Based Meta-Learning Model for MRI Reconstruction with Diverse Dataset,** | 2021 | Bian et al. | N.A. | Algorithm Unrolling | SC, CNNs | Supervised | NMSE, PSNR, SSIM | N.A. | YES | ADMM | BraTS 2018* | NMSE:0.0184; PSNR: 23.2672; SSIM: 0.6101 (Radial AF×10, T1)<br><br>NMSE:0.0058; PSNR: 28.2944; SSIM: 0.7640 (Radial AF×5, T1)<br><br>NMSE:0.0031; PSNR: 31.1417; SSIM: 0.8363 (Radial AF×3.3, T1)<br><br>NMSE:0.0010; PSNR: 32.8238; SSIM: 0.8659 (Radial AF×2.5, T1)<br><br>NMSE:0.0611; PSNR: 22.0434; SSIM: 0.6279 (Cartesian 1D AF×10, T2)<br><br>NMSE:0.0329; PSNR: 24.7162; SSIM: 0.6971 (Cartesian 1D AF×5, T2)<br><br>NMSE:0.0221; PSNR: 26.4537; SSIM: 0.7353 (Cartesian 1D AF×3.3, T2)<br><br>NMSE:0.0171; PSNR: 27.5367; SSIM: 0.7726 (Cartesian 1D AF×2.5, T2) | Not provided | Not provided | Not provided |
| **MRI Reconstruction by Completing Under-sampled K-space Data with Learnable Fourier Interpolation** | 2022 | Ding et al. | N.A. | Algorithm Unrolling | SC,CNNs | Supervised | PSNR, SSIM | Inference Time | YES | NA | BraTS*, ADNI* | PSNR: 39.41; SSIM: 0.98; (BraTS Radial AF×5)<br><br>PSNR: 35.62; SSIM: 0.96; (BraTS Radial AF×4)<br><br>PSNR: 33.41; SSIM: 0.95; (BraTS Radial AF×3)<br><br>PSNR: 36.36; SSIM: 0.97; (BraTS Gaussian 2D AF×5)<br><br>PSNR: 36.05; SSIM: 0.97; (BraTS Gaussian 2D AF×4)<br><br>PSNR: 34.90; SSIM: 0.96; (BraTS Gaussian 2D AF×3)<br><br>PSNR: 31.87; SSIM: 0.95; (BraTS Gaussian 1D AF×5)<br><br>PSNR: 30.68; SSIM: 0.92; (BraTS Gaussian 1D AF×4)<br><br>PSNR: 31.03; SSIM: 0.92; (BraTS Gaussian 1D AF×3) | Not provided | 0.58s (NVIDIA A100) | Not provided |
| **DuDoRNet: Learning a Dual-Domain Recurrent Network for Fast MRI Reconstruction With Deep T1 Prior** | 2020 | Zhou et al. | DuDoRNet | Algorithm Unrolling | MC, CNNs | Supervised | PSNR, SSIM, MSE | N.A. | YES | GRAPPA | Private | MSE: 0.066; PSNR: 32.511; SSIM: 0.957 (Cartesian 1D AF×5)<br><br>MSE: 0.010; PSNR: 40.815; SSIM: 0.981 (Radial AF×5)<br><br>MSE: 0.001; PSNR: 49.186; SSIM: 0.993 (Spiral AF×5) | Not provided | Not provided | https://github.com/bbbbbbzhou/DuDoRNet |
| **Plug-and-Play Methods for Magnetic Resonance** | 2020 | Ahmad et al. | PnP-CNN | PnP | SC/MC, CNNs | Supervised | rSNR, SSIM | N.A. | YES | FISTA | FastMRI Dataset* | rSNR: 21.14; SSIM: 0.754 (Cartesian AF×4) | Not Provided | Not Provided | Not Provided |

| | | | | | | | | | | | | | | | | |
|---|---|---|---|---|---|---|---|---|---|---|---|---|---|---|---|---|
| **Imaging: Using Denoisers for Image Recovery** | | | | | | | | | | | | | | | | |
| **Deep Plug-and-Play Prior for Parallel MRI Reconstruction** | 2019 | Yazdanpanah et al. | N.A. | PnP | MC, U-Net, CNNs | Supervised | PSNR, SSIM | N.A. | YES | ADMM | Private | PSNR: 53.3; SSIM: 0.99 (AF×4, Brain dataset)<br><br>PSNR: 42.28; SSIM: 0.96 (AF×4, Knee dataset 1) | Not Provided | Not Provided | Not Provided | |
| **Plug and play augmented HQS: Convergence analysis and its application in MRI reconstruction** | 2023 | Rasti-Meymandi et al. | PnP-AugHQS | PnP | SC, U-Net, CNNs | Supervised | PSNR, SSIM, HFEN | Inference Time | YES | HQS | FastMRI Dataset* | PSNR: 30.29; SSIM: 0.835; HFEN: 1.613 (Radial 10% / AF×10, FastMRI Brain)<br><br>PSNR: 34.85; SSIM: 0.919; HFEN: 0.732 (Radial 20% / AF×5, FastMRI Brain)<br><br>PSNR: 37.18; SSIM: 0.944; HFEN: 0.416 (Radial 30% / AF×3.33, FastMRI Brain)<br><br>PSNR: 38.82; SSIM: 0.957; HFEN: 0.252 (Radial 40% / AF×2.5, FastMRI Brain)<br><br>PSNR: 35.21; SSIM: 0.874; HFEN: 0.862 (Radial 10% / AF×10, FastMRI Knee)<br><br>PSNR: 38.11; SSIM: 0.917; HFEN: 0.483 (Radial 20% / AF×5, FastMRI Knee)<br><br>PSNR: 39.71; SSIM: 0.939; HFEN: 0.319 (Radial 30% / AF×3.33, FastMRI Knee)<br><br>PSNR: 41.12; SSIM: 0.955; HFEN: 0.205 (Radial 40% / AF×2.5, FastMRI Knee) | Not Provided | 8.15s (NVIDIA GTX1060) | Not Provided | |
| **A Plug-and-Play Deep Denoiser Prior Model for Accelerated MRI Reconstruction** | 2022 | Karaoglu et al. | IRCNN-MRI | PnP | SC, CNNs | Supervised | NMSE, PSNR, SSIM | Inference Time | YES | HQS | Private | NMSE: 0.0480; PSNR: 37.2286; SSIM: 0.9492 (Radial 20% / AF×5, Brain Image)<br><br>NMSE: 0.0896; PSNR: 30.9532; SSIM: 0.8601 (Radial 20% / AF×5, Head Image)<br><br>NMSE: 0.1051; PSNR: 32.8249; SSIM: 0.9338 (Radial 20% / AF×5, Bust Image) | Not Provided | 11.577s~14.093s (NVIDIA GTX960) | Not Provided | |
| **Uncertainty Quantification in Deep MRI Reconstruction** | 2021 | Vineet et al. | N.A. | Generative Model (VAE) | SC, VAE | Supervised | MSE, RSS (residual sum of squares) , DOF (degrees of freedom) , SNR, SURE(stein's unbiased risk estimator) , SSIM | | YES | NA | mridata* | SURE-MSE R2: 0.97; MSE: 0.017; RSS: 0.061; DOF: 0.12; SNR: 20.7; SURE: 21.3 (Variable Density AF×2);<br><br>SURE-MSE R2: 0.90; MSE: 0.021; RSS: 0.065; DOF: 0.16; SNR: 19.1; SURE: 19.8 (Variable Density AF×4);<br><br>SURE-MSE R2: 0.92; MSE: 0.18; RSS: 0.11; DOF: 0.23; SNR: 16.8; SURE: 15.9 (Variable Density AF×8);<br><br>SURE-MSE R2: 0.84; MSE: 0.39; RSS: 0.13; DOF: 0.29; SNR: 15.5; SURE: 14.2 (Variable Density AF×16) | Not provided | Not provided | Not provided | RSS: Residual sum of squares DOF: Degrees of freedom SURE: Stein's unbiased risk estimator |
| **DAGAN: Deep De-Aliasing Generative Adversarial Networks for Fast Compressed Sensing MRI Reconstruction** | 2018 | Yang et al. | DAGAN | Generative Model (GAN) | SC, U-Net, CNNs | Supervised | NMSE, PSNR | Inference Time | NO | NA | BRaTS 2013 | NMSE: 0.17; PSNR: 33.79 (Gaussian 1D 10% / AF×10)<br><br>NMSE: 0.09; PSNR: 39.44 (Gaussian 1D 20% / AF×5)<br><br>NMSE: 0.08; PSNR: 40.20 (Gaussian 1D 30% / AF×3.33)<br><br>NMSE: 0.05; PSNR: 44.83 (Gaussian 1D 40% / AF×2.5)<br><br>NMSE: 0.04; PSNR: 47.83 (Gaussian 1D 50% / AF×2) | Not Provided | 0.2s (unknown GPU) | https://github.com/tensorlayer/DAGAN | |
| **Fast MRI Reconstruction: How Powerful Transformers Are?** | 2022 | Huang et al. | ST-GAN | Generative Model (GAN) | SC, SwinIR, Transformer | Supervised | SSIM, FID | #PARAMs, MACs, Inference Time | NO | NA | Calgary-Campinas Public Dataset* | FID: 18.86 (Gaussian 1D 10% / AF×10)<br><br>FID: 8.50 (Gaussian 1D 30% / AF×3.33) | #PARAMs(M): 11.40; MACs(G): 800.73 | 0.041s (NVIDIA RTX3090) | https://github.com/ayanglab/SwinGANMR | |
| **Magnetic resonance imaging reconstruction using a deep energy-based model** | 2023 | Guan et al. | EBMRec | Generative Model (DEM) | SC/MC, CNNs | Supervised | RMSE, PSNR, SSIM | N.A. | YES | NA | SIAT Dataset (Private) FastMRI Dataset* | RMSE: 0.0755; PSNR: 35.84; SSIM: 0.925; (Radial AF×3.3, SIAT Brain, SC)<br><br>RMSE: 0.0839; PSNR: 34.96; SSIM: 0.911; (Radial AF×4, SIAT Brain, SC)<br><br>RMSE: 0.0976; PSNR: 33.80; SSIM: 0.892; (Radial AF×5, SIAT Brain, SC)<br><br>RMSE: 0.1330; PSNR: 30.97; SSIM: 0.852; (Gaussian 2D (Random) AF×6.7, SIAT Brain, SC)<br><br>RMSE: 0.1569; PSNR: 30.78; SSIM: 0.766; (Cartesian AF×4, FastMRI Knee, SC)<br><br>RMSE: 0.2102; PSNR: 27.97; SSIM: 0.680; (Cartesian AF×8, FastMRI Knee, SC) | Not Provided | Not Provided | https://github.com/yqx7150/EBMRec | |
| **K-space and image domain collaborative energy-based model for parallel MRI reconstruction** | 2023 | Tu et al. | KI-EBM | Generative Model (DEM) | SC/MC, CNNs | Supervised | PSNR, SSIM | N.A. | YES | NA | SIAT Dataset (Private) | PSNR: 40.21; SSIM: 0.965; (Cartesian AF×2, Brain, MC)<br><br>PSNR: 38.09; SSIM: 0.945; (Cartesian AF×3, Brain, MC)<br><br>PSNR: 42.69; SSIM: 0.990; (Gaussian 2D AF=6, Brain, MC) | Not Provided | Not Provided | https://github.com/yqx7150/KI-EBM | |
| **Towards Performant and Reliable Undersampled MR Reconstruction via Diffusion Model Sampling** | 2022 | Peng et al. | DiffuseRecon | Generative Model (Diffusion Model) | SC, DDPM, U-Net, CNNs+Transformers | Supervised | PSNR, SSIM | Inference Time | YES | NA | FastMRI Dataset* SKMTEA Dataset* | PSNR: 30.56; SSIM: 0.648; (Cartesian AF×6, FastMRI Knee)<br><br>PSNR: 29.94; SSIM: 0.614; (Cartesian AF×8, FastMRI Knee)<br><br>PSNR: 32.58; SSIM: 0.743; (Cartesian AF×6, SKMTEA)<br><br>PSNR: 31.56; SSIM: 0.706; (Cartesian AF×8, SKMTEA) | Not Provided | Around 20s (unknown GPU) | https://github.com/cpeng93/DiffuseRecon | DDPM: Denoising diffusion probabilistic models |
| **Score-based diffusion models for accelerated MRI** | 2022 | Chung et al. | N.A. | Generative Model (Diffusion Model) | SC/MC, SMLD, U-Net, CNNs+Transformers | Supervised | PSNR, SSIM | Inference Time | YES | NA | FastMRI Dataset* | PSNR: 31.10; SSIM: 0.795; (Uniform 1D AF×4, FastMRI Knee, SC) PSNR: 28.37; SSIM: 0.771; (Uniform 1D AF×8, FastMRI Knee, SC) PSNR: 33.32; SSIM: 0.825; (Gaussian 1D AF×4, FastMRI Knee, SC) PSNR: 30.94; SSIM: 0.761; (Gaussian 1D AF×8, FastMRI Knee, SC) PSNR: 29.95; SSIM: 0.701; (Gaussian 2D AF×4, FastMRI Knee, SC) PSNR: 29.58; SSIM: 0.678; (Gaussian 2D AF×8, FastMRI Knee, SC) PSNR: 31.83; SSIM: 0.769; (VD Poisson Disk AF×4, FastMRI Knee, SC) PSNR: 30.46; SSIM: 0.709; (VD Poisson Disk AF×8, FastMRI Knee, SC) | Not Provided | Around 10~20mins (unknown GPU) | https://github.com/HJ-harry/score-MRI | SMLD: Score matching with Langevin dynamics |
| **Accelerating multi-echo MRI in k-space with complex-valued diffusion probabilistic model** | 2022 | Cao et al. | CDPM | Generative Model (Diffusion Model) | SC, DDPM, U-Net, CNNs+Transformers | Supervised | PSNR, SSIM | N.A. | YES | NA | Private | No Numerical Results Provided. (Provided in Plots.) | Not Provided | Not Provided | Not Provided | |
| **Adaptive diffusion priors for accelerated MRI reconstruction** | 2023 | Güngör et al. | AdaDiff | Generative Model (Diffusion Model) | SC/MC, DDPM, U-Net, CNNs+Transformers | Supervised | PSNR, SSIM | N.A. | YES | NA | IXI Dataset* FastMRI Dataset* | PSNR: 42.5; SSIM: 0.991; (Gaussian 2D AF×4, IXI T1, SC)<br><br>PSNR: 36.6; SSIM: 0.977; (Gaussian 2D AF×8, IXI T1, SC)<br><br>PSNR: 42.3; SSIM: 0.989; (Gaussian 2D AF×4, IXI T2, SC)<br><br>PSNR: 36.3; SSIM: 0.795; (Gaussian 2D AF×8, IXI T2, SC)<br><br>PSNR: 43.0; SSIM: 0.974; (Gaussian 2D AF×4, IXI PD, SC)<br><br>PSNR: 37.0; SSIM: 0.977; (Gaussian 2D AF×8, IXI PD, SC) | Not Provided | Not Provided | https://github.com/icon-lab/AdaDiff | |
| **High-Frequency Space Diffusion Model for Accelerated MRI** | 2024 | Cao et al. | HFS-SDE | Generative Model (Diffusion Model) | MC, Score-based SDE, U-Net, CNNs+Transformer | Supervised | NMSE, PSNR, SSIM | Inference Time | YES | NA | FastMRI Dataset* | NMSE: 0.0065; PSNR: 33.28; SSIM: 0.8409; (Cartesian Uniform AF×10, FastMRI Knee, MC)<br><br>NMSE: 0.0097; PSNR: 31.56; SSIM: 0.8060; (Cartesian Uniform AF×12, FastMRI Knee, MC) | Not Provided | 98.59s (NVIDIA A100) | https://github.com/Aboriginer/HFS-SDE | Score-based SDE: Score-based stochastic differential equation, a generalization from DDPM and SMLD. |
| **Conditional Invertible Neural Networks for Medical Imaging** | 2021 | Denker et al. | N.A. | Generative Model (NF) | SC, INNs | Supervised | PSNR, SSIM | #PARAMs | NO | NA | FastMRI Dataset* | PSNR: 29.15; SSIM: 0.777; (Cartesian Random AF×4, FastMRI Knee PD, SC)<br><br>PSNR: 23.18; SSIM: 0.536; (Cartesian Random AF×4, FastMRI Knee PDFS, SC) | #PARAMs(M): 2.5~2.6 | Not Provided | https://github.com/jleuschn/cinn_for_imaging | |
| **A Conditional Normalizing Flow for Accelerated Multi-Coil MR Imaging** | 2023 | Wen et al. | N.A. | Generative Model (NF) | MC, INNs | Supervised | PSNR, SSIM, FID, cFID | Inference Time | YES | NA | FastMRI Dataset* | PSNR: 35.23; SSIM: 0.8888; (Cartesian AF×4, AFFastMRI Knee, MC)<br><br>PSNR: 38.85; SSIM: 0.9495; (Cartesian AF×4, FastMRI Brain, MC) | Not Provided | 0.108s~0.177s (NVIDIA V100) | https://github.com/jwen307/mri_cnf | |

*** List of Datasets**

1. ADNI [1]
2. BraTS 2013, 2015, 2018 [2], [3]
3. CAF Project: Brain and Chest [4]
4. Calgary-Campinas Public Dataset [5]
5. FastMRI Dataset [6]
6. IXI Dataset [7]
7. Mridata [8]
8. SKMTEA Dataset [9]

## Supplementary Table

**TABLE S2.** COMMON EVALUATION METRICS FOR MRI RECONSTRUCTION

| Metric Type | Metric | Definition | Formula | Range | Reference Image Availability[1] |
|---|---|---|---|---|---|
| **Human-Perceived** | MOS | **Mean Opinion Score (MOS)** is a metric that evaluates human visual perception by collecting subjective assessments from skilled observers. | $\mathrm{MOS}(X) = \frac{1}{A}\sum_{a=1}^{A} \mathrm{Score}_a$ | $[\mathrm{Score}_{\mathrm{Min}}, \mathrm{Score}_{\mathrm{MAX}}]$ | **No-Reference Metric** |
| **Pixel-based** | SNR | **Signal-to-Noise Ratio (SNR)** is a metric for the ratio between the power of signal within a specific Region of Interest (ROI) and the power of corrupting noise that affects the fidelity of its representation. | $\mathrm{SNR}(X) = \frac{\mu_s}{\sigma_\mathrm{n}}$ | [0, +Inf] | **Full-Reference Metric** |
| | CNR | **Contrast-to-Noise Ratio (CNR)** is a metric that measures the ability to distinguish between two different signal intensities within specific ROI relative to the noise level in the image. | $\mathrm{CNR}(X) = \frac{\mu_s - \mu_b}{\sigma_\mathrm{n}}$ | [0, +Inf] | **Full-Reference Metric** |
| | NRMSE | **Normalized Root Mean Square Error (NRMSE)** is a metric that measures the differences between predicted and actual values, where normalization is performed by dividing the RMSE by the range of actual data values, allowing for comparability between datasets with different scales. | $\mathrm{NRMSE}(X,Y) = \frac{\sqrt{\frac{1}{MN}\sum_{i=1}^{M}\sum_{j=1}^{N}\left(X(i,j) - Y(i,j)\right)^2}}{\mathrm{MAX(Y)} - \mathrm{MIN(Y)}}$ | [0, +Inf] | **Full-Reference Metric** |
| | PSNR | **Peak Signal-to-Noise Ratio (PSNR)** is a metric for the ratio between the maximum possible power of a signal and the power of corrupting noise that affects the fidelity of its representation. | $\mathrm{PSNR}(X,Y) = 20\log_{10}\left(\frac{\mathrm{MAX}}{\sqrt{\frac{1}{MN}\sum_{i=1}^{M}\sum_{j=1}^{N}\left(X(i,j) - Y(i,j)\right)^2}}\right)$ | [0, +Inf] | **Full-Reference Metric** |
| | SSIM | **Structural Similarity Index (SSIM)** is a metric to measure the perceived quality by comparing the difference in structural information, luminance, and contrast between a reference and a reconstructed image. | $\mathrm{SSIM}(X,Y) = \frac{(2\mu_X\mu_Y + c_1)\,(2\sigma_{XY} + c_2)}{(\mu_X^2 + \mu_Y^2 + c_1)(\mu_X^2 + \mu_Y^2 + c_2)}$ | [-1, 1] | **Full-Reference Metric** |
| | ESSIM | **Edge Strength Similarity Index Metric (ESSIM)** is a metric to quantify the similarity in edge strength between a reference and a reconstructed image. | $\mathrm{ESSIM}(X,Y) = \frac{1}{MN}\sum_{i=1}^{M}\sum_{j=1}^{N}\frac{2E_X(i,j)E_Y(i,j) + C}{E_X^2(i,j) + E_Y^2(i,j) + C}$ | [-1, 1] | **Full-Reference Metric** |
| | EPI | **Edge Preservation Index (EPI)** is a metric for evaluating the extent of edge preservation by a denoising filter. | $\mathrm{EPI}(X,Y) = \frac{\sum_{i=1}^{M}\sum_{j=1}^{N}(\nabla^2 X(i,j) - \mu_{\nabla^2 X})(\nabla^2 Y(i,j) - \mu_{\nabla^2 Y})}{\sqrt{\sum_{i=1}^{M}\sum_{j=1}^{N}(\nabla^2 X(i,j) - \mu_{\nabla^2 X})^2\ \sum_{i=1}^{M}\sum_{j=1}^{N}(\nabla^2 Y(i,j) - \mu_{\nabla^2 Y})^2}}$ | [-1, 1] | **Full-Reference Metric** |
| | BRISQUE | **Blind/Referenceless Image Spatial Quality Evaluator (BRISQUE)** is a metric that uses natural scene statistics to predict perceived image quality without a reference image. The quality score is predicted by a learnt regression module, e.g., support vector machine (SVM) regressor. | $\mathrm{BRISQUE}(X) = \mathrm{SVM}([f_1, \ldots, f_{36}])$ <br> $[f_1, \ldots, f_{36}] = \mathrm{Feat}\left(\frac{X(i,j) - \mu(i,j)}{\sigma(i,j) + C}\right)$ | [0, 100] | **No-Reference Metric** |

[1] Reference Image Availability: whether or not a metric requires a ground truth reference image for evaluation of the result

| | | | | | |
|---|---|---|---|---|---|
| **Deep Feature-based** | LPIPS | **Learned Perceptual Image Patch Similarity (LPIPS)** is a data-driven metric that uses pre-trained neural networks to measure the perceptual similarity between two images. | $\mathrm{LPIPS}(\mathrm{X},\mathrm{Y})=\sum_{l}\frac{1}{H_l W_l}\sum_{h,w}\|w_l\odot(y^l_{Xhw}-y^l_{Yhw})\|_2^2$ | [0, 1] | **Full-Reference Metric** |
| | SSFD | **Self-Supervised Feature Distance (SSFD)** is a metric designed for assessing the quality of magnetic resonance imaging reconstructions. It uses a self-supervised learning approach to derive feature representations from MR images, which are then utilized to calculate the SSFD. | $\mathrm{SSFD}(\mathrm{X},\mathrm{Y})=\frac{1}{HW}\sum_{h,w}\|y_{Xhw}-y_{Yhw}\|_2^2$ | [0, +Inf] | **Full-Reference Metric** |
| | PaQ-2-PiQ | **PaQ-2-PiQ** is a series of metrics that leveraging deep learning models that predict the perceptual quality of pictures without reference images, including P2P-BM, P2P-RM and P2P-FM. | $\mathrm{P2PFM}(X)=\mathrm{F}\big(X(i,j)\big)$ | [0, 100] | **No-Reference Metric** |
| | FID | **Fréchet Inception Distance (FID)** is a data-driven metric that calculates the distance between latent features extracted from a pre-trained Inception-Net. FID measures the Fréchet distance between the distributions of reference and reconstructed images. | $\mathrm{FID}(\mathbb{I}_X,\mathbb{I}_Y)=\|\mu_X-\mu_Y\|^2+\mathrm{Tr}(\sigma_X+\sigma_Y-2(\sigma_X\sigma_Y)^{\frac{1}{2}})$ | [0, +Inf] | **Distribution-based Metric** |

**Notes for formulas**

$X$: reconstructed MR images; $Y$: reference MR images. $M$: height of MR images with index $i$ ; $N$: width of the MR images with index $j$.
**MOS**. $A$ is the number of rating radiologist; $\mathrm{Score}_{\mathrm{Min}}$ and $\mathrm{Score}_{\mathrm{MAX}}$ are preset rating boundary.
**SNR**. $\mu_{\mathrm{s}}$: mean of the pixel value within ROI of the signal ; $\sigma_{\mathrm{n}}$: standard deviation of the noise.
**CNR**. $\mu_{\mathrm{s}}$: mean of the pixel value within ROI of the signal ; $\mu_b$: mean of the pixel value within ROI of the background; $\sigma_{\mathrm{n}}$: standard deviation of the noise.
**NRMSE**. $\mathrm{MAX}(\cdot)$: maximum pixel value; $\mathrm{MIN}(\cdot)$: minimum pixel value.
**PSNR**. $\mathrm{MAX}$: theocratical maximum value.
**SSIM**. $\mu_X$, $\mu_Y$: mean value of the reconstructed and reference images; $\sigma_{XY}$: the covariance between the reconstructed and reference images; $c_1$, $c_2$: constants.
**ESSIM**. $E_X(i,j)$, $E_Y(i,j)$: local edge strengths of the reconstructed and reference images; $C$: a constant.
**EPI**. $\nabla^2 X(i,j)$, $\nabla^2 Y(i,j)$: the second derivatives of the reconstructed and reference images; $\mu_{\nabla^2 X}$, $\mu_{\nabla^2 Y}$: global mean of these second derivatives.
**BRISQUE**. $\mathrm{Feat}(\cdot)$ is a handcrafted feature extractor, producing 36 features $[f_1,\ldots,f_{36}]$ from an image, which includes parameters fit by generalized Gaussian distribution fitting and asymmetric generalized Gaussian distribution at original image scale and $\times$ 2 downsampled scale; SVM: a pre-trained support vector machine regressor to predict the final score. $\mu(i,j)$ and $\sigma(i,j)$: local mean and variance; $C$: a constant.
**LPIPS**. $y^l_{Xhw}$ , $y^l_{Yhw}$: feature maps of the reconstructed and reference images at layer $l$ using a pretrained neural network; $H_l$, $W_l$: height and width of feature maps at layer $l$; $w_l$: a pre-trained projection.
**SSFD**. $y_{Xhw}$, $y_{Yhw}$: feature maps of the reconstructed and reference images using a pretrained neural network; $H$, $W$: height and width of feature maps.
**PaQ-2-PiQ**. P2P-FM: one of the proposed PaQ-2-PiQ metrics; $\mathrm{F}(\cdot)$: a pretrained neural network to predict the final score.
**FID**. $\mathbb{I}_X$, $\mathbb{I}_Y$: the distribution of the reconstructed and reference images. $\mu_X$, $\sigma_X$: mean and variance of the multivariate normal distribution fit by features extracted from the distribution of reconstructed images, using a pre-trained Inception-v3 model; $\mu_Y$, $\sigma_Y$: mean and variance of the multivariate normal distribution fit by features extracted from the distribution of reference images; $\mathrm{Tr}(\cdot)$: the trace of a matrix.

## Supplementary Section

## S-I. Foundation of Generative Models for Fast MRI

### A. Generative Models

Conventional CS and deep learning-based models like PnP and Algorithm Unrolling for MRI reconstruction typically require computationally intensive iterative processes or cascaded network structures. These models are heavily influenced by the regularization term, rather than the actual data distribution. Conversely, enhancement-based models require a substantial amount of paired data to map k-space measurements or zero-filled images directly to fully sampled images, where such paired data is often scarce [10]. Generative models, in contrast, aim to learn the data distribution of fully sampled images and use image data priors to solve ill-posed inverse imaging problems.

*1) Variational Autoencoder*

Variational Autoencoders (VAEs) [11] are a group of probabilistic generative models that are designed to learn complex data distributions $p_{\boldsymbol{\theta}}(\mathbf{x})$ via variational inference. The core concept of VAE is to encode data $\mathbf{x}$ from data space $\mathcal{X}$ into a latent space $\mathcal{Z}$ and then reconstruct it back to the original data space $\mathcal{X}$ , which corresponds to two key components: a probabilistic encoder with parameters $\boldsymbol{\phi}$ and a probabilistic decoder with parameters $\boldsymbol{\theta}$.

In the inference stage where only the decoder is applied, the decoder samples the data point according to $p_{\boldsymbol{\theta}}(\mathbf{x} \mid \mathbf{z})$, using the latent vector $\mathbf{z}$ from a known prior distribution $p(\mathbf{z})$ , i.e., Gaussian distribution, and all these generated data points follow the distribution in (S1.1).

$$p_{\boldsymbol{\theta}}(\mathbf{x}) = \int_{\mathcal{Z}} p_{\boldsymbol{\theta}}(\mathbf{x} \mid \mathbf{z})\, \mathrm{d}\mathbf{z}. \tag{S1.1}$$

However, one of the critical challenges in training VAEs is the intractability of directly maximizing the likelihood of the data $p_{\boldsymbol{\theta}}(\mathbf{x})$ in (S1), as it involves high-dimensional integrals. To address this, VAEs employ the concept of the Evidence Lower Bound (ELBO). The ELBO provides a tractable lower bound on the log-likelihood of a data point, as in (S1.2).

$$\begin{aligned}\log p_{\boldsymbol{\theta}}(\mathbf{x}) \geq\ & \mathcal{L}_{\mathrm{ELBO}}(\boldsymbol{\phi}, \boldsymbol{\theta}; \mathbf{x}) \\ & \triangleq -\mathrm{D}_{\mathrm{KL}}(q_{\boldsymbol{\phi}}(\mathbf{z} \mid \mathbf{x}) \,||\, p(\mathbf{z})) \\ & + \mathbb{E}_{\mathbf{z}\sim q_{\boldsymbol{\phi}}(\mathbf{z} \mid \mathbf{x})}[\log p_{\boldsymbol{\theta}}(\mathbf{x} \mid \mathbf{z})],\end{aligned} \tag{S1.2}$$

where $\mathrm{D}_{\mathrm{KL}}(\cdot \,||\, \cdot)$ denotes the Kullback-Leibler (KL) divergence. $q_{\boldsymbol{\phi}}(\mathbf{z} \mid \mathbf{x})$ denotes an approximate posterior distribution for the latent variable, which is a multivariate normal distribution. This is because the encoder outputs its mean $\boldsymbol{\mu}_{\boldsymbol{\phi}}(\mathbf{x})$ and standard deviation $\boldsymbol{\sigma}_{\boldsymbol{\phi}}(\mathbf{x})$ instead of the latent vector itself, and $\mathbf{z}$ can be constructed via reparametrizing strategy in (S1.3).

$$\mathbf{z} = \boldsymbol{\mu}_{\boldsymbol{\phi}}(\mathbf{x}) + \boldsymbol{\sigma}_{\boldsymbol{\phi}}(\mathbf{x}) \odot \boldsymbol{\varepsilon}, \quad \text{s.t. } \boldsymbol{\varepsilon} \sim \mathcal{N}(0, \mathbf{I}), \tag{S1.3}$$

where $\odot$ denotes the element-wise product. The VAE can be trained by minimizing the negative ELBO with respect to $\boldsymbol{\phi}$ and $\boldsymbol{\theta}$, as specified in (S1.4).

$$\min_{\boldsymbol{\phi},\boldsymbol{\theta}} \mathcal{L}_{\mathrm{VAE}}(\boldsymbol{\phi}, \boldsymbol{\theta}) = \mathbb{E}_{\mathbf{x}\sim p(\mathbf{x})}[-\mathcal{L}_{\mathrm{ELBO}}(\boldsymbol{\phi}, \boldsymbol{\theta}; \mathbf{x})], \tag{S1.4}$$

In the context of MRI, VAEs are usually coupled with the Monte-Carlo sampling [12] for simultaneously MRI reconstruction [13] and uncertainty quantification [14] . Specifically, in [14], the proposed VAE takes the ZF $\mathbf{A}^{\mathrm{H}}\mathbf{y}$ as the input, and output the reconstructed results $\hat{\mathbf{x}}$ . Based on (S1.2) and (S1.4), the training loss includes a KL divergence and an L2 distance, as specified in (S1.5).

$$\begin{aligned}&\min_{\boldsymbol{\phi},\boldsymbol{\theta}} \mathcal{L}_{\mathrm{VAE}}(\boldsymbol{\phi}, \boldsymbol{\theta}) \\ &= \mathbb{E}_{\mathbf{x}\sim p(\mathbf{x})}\left[\mathrm{D}_{\mathrm{KL}}(\mathcal{N}(\boldsymbol{\mu}_{\boldsymbol{\phi}}(\mathbf{x}), \boldsymbol{\sigma}_{\boldsymbol{\phi}}(\mathbf{x})) \,||\, \mathcal{N}(0, \mathbf{I})) + \|\hat{\mathbf{x}} - \mathbf{x}\|_2^2\right],\end{aligned} \tag{S1.5}$$

where $p(\mathbf{z})$ is assumed to be a standard normal distribution. With the trained decoder, Monte Carlo sampling [12] is applied, where the pixel-wise mean $\frac{1}{K}\sum_{i=1}^{K}\hat{\mathbf{x}}_i$ and variance maps $\frac{1}{K}\sum_{i=1}^{K}(\hat{\mathbf{x}}_i - \hat{\mathbf{x}})^2$ are used for final reconstructed results and uncertainty maps respectively ($K$ is the number of iterations).

*2) Generative Adversarial Network*

Generative adversarial networks (GANs) [15] are a family of generative models that implicitly learn the data distribution $p(\mathbf{x})$ in an adversarial manner. A GAN usually involves two key components, the generator $G_{\boldsymbol{\theta}_G}(\cdot)$ and the discriminator, which are simultaneously trained and alternately updated through a competitive process in (S1.6).

$$\begin{aligned}\min_{\boldsymbol{\theta}_G} \max_{\boldsymbol{\theta}_D} & \mathcal{L}_{\mathrm{GAN}}(\boldsymbol{\theta}_G, \boldsymbol{\theta}_D) \\ & \triangleq \mathbb{E}_{\mathbf{x}\sim p(\mathbf{x})}[\log D_{\boldsymbol{\theta}_D}(\mathbf{x})] \\ & + \mathbb{E}_{\mathbf{z}\sim p(\mathbf{z})}\left[\log\left(1 - D_{\boldsymbol{\theta}_D}\left(G_{\boldsymbol{\theta}_G}(\mathbf{z})\right)\right)\right].\end{aligned} \tag{S1.6}$$

The generator aims to use a random vector $\mathbf{z}$ sampled from a known prior distribution $p(\mathbf{z})$, to create images $G_{\boldsymbol{\theta}_G}(\mathbf{z})$ that are indistinguishable from real images, while the discriminator is trained to differentiate between the "fake" creations from the generator and true images from the data distribution $p(\mathbf{x})$.

Generative adversarial networks have been widely applied for inverse problems [10], allowing for finer detail recovery and facilitating reconstruction even in cases of heavy degradation. Specifically in the context of MRI reconstruction [16], the generator $G_{\boldsymbol{\theta}_G}(\cdot)$ serves as the primary backbone network responsible for image reconstruction, which takes in zero-filled images $\mathbf{A}^{\mathrm{H}}\mathbf{y}$ as the input and is expected to produce reconstructed images $G_{\boldsymbol{\theta}_G}(\mathbf{A}^{\mathrm{H}}\mathbf{y})$ as the output. The generator can be any model from the image enhancement category as introduced above. The role of the discriminator $D_{\boldsymbol{\theta}_D}(\cdot)$, on the other hand, is to distinguish between the reconstructed image $G_{\boldsymbol{\theta}_G}(\mathbf{A}^{\mathrm{H}}\mathbf{y})$ and the ground truth $\mathbf{x}$, which is the fully sampled image. The loss function can be defined in (S1.7).

$$\begin{aligned}\min_{\boldsymbol{\theta}_G} \max_{\boldsymbol{\theta}_D} \mathcal{L}_{\mathrm{GAN}}(\boldsymbol{\theta}_G, \boldsymbol{\theta}_D) & \triangleq \mathbb{E}_{\mathbf{x}\sim p(\mathbf{x})}[\log D_{\boldsymbol{\theta}_D}(\mathbf{x})] \\ & + \mathbb{E}_{\mathbf{y}\sim p(\mathbf{y})}\left[\log\left(1 - D_{\boldsymbol{\theta}_D}\left(G_{\boldsymbol{\theta}_G}(\mathbf{A}^{\mathrm{H}}\mathbf{y})\right)\right)\right].\end{aligned} \tag{S1.7}$$

*3) Deep Energy-based Model*

Deep energy-based models (EBMs) [17] are a family of generative models that implicitly learn the data distribution through a learnable energy function. Such energy function is designed to learn the statistical properties of input data, mapping them to an energy representation. Specifically, the energy representation is mapped by assigning lower energy values to true data points and higher energy values to generated data points [18]. The energy function can be trained by minimizing the following loss function (S1.8).

$$\min_{\boldsymbol{\theta}} \mathcal{L}_{\mathrm{EBM}}(\boldsymbol{\theta}) = \min_{\boldsymbol{\theta}} \mathbb{E}_{\mathbf{x}^+\sim p(\mathbf{x})}[E(\mathbf{x}^+)] - \mathbb{E}_{\mathbf{x}^-\sim p_{\boldsymbol{\theta}}(\mathbf{x})}[E(\mathbf{x}^-)] \tag{S1.8}$$

where $E(\cdot)$ stands for the energy function $E(\cdot\,; \theta)$. $\mathbf{x}^+$ is the data point sampled from true data distribution $p(\mathbf{x})$, and $\mathbf{x}^-$ is the generated data from the model distribution $p_{\boldsymbol{\theta}}(\mathbf{x})$. For data point sampling, one commonly used strategy is Langevin dynamics for MRI reconstruction [19] (S1.9), as a Markov Chain Monte Carlo [12] gradient based sampling method.

$$\mathbf{x}_k = \mathbf{x}_{k-1} - \frac{\eta}{2}\nabla_{\mathbf{x}}E(\mathbf{x}_{k-1}) + \boldsymbol{\varepsilon}_k, \quad \text{s.t. } \boldsymbol{\varepsilon}_k \sim \mathcal{N}(0, \mathbf{I}), \tag{S1.9}$$

where $\nabla_{\mathbf{x}}(\cdot)$ denotes the gradient with respect to $\mathbf{x}$ and $\eta$ is the step size. During iteration, by adjusting the data points based on the energy function and incorporating random walk through the stochastic component $\boldsymbol{\varepsilon}$, EBMs can generate data points following the same distribution as the training data.

*4) Diffusion Model*

Diffusion models, which include Denoising Diffusion Probabilistic Models (DDPMs) [20] and Score Matching with Langevin Dynamics (SMLD) [21], have evolved into a unified framework known as Score-based Stochastic Differential Equation (SDE) [22]. These models facilitate the transformation of a simple prior distribution, often Gaussian, into a complex data distribution through a series of stochastic steps, referred to as diffusion.

The mathematical foundation of diffusion models is the score-based SDE theory [22]. This theory establishes a continuous diffusion process $\{\mathbf{x}(t)\}_{t=0}^{T}$ with $\mathbf{x}(t) \in \mathbb{R}^n$ as the solution to a specific SDE (S1.10).

$$\mathrm{d}\mathbf{x} = \boldsymbol{f}(\mathbf{x}, t)\mathrm{d}t + g(t)\mathrm{d}\boldsymbol{\omega}. \tag{S1.10}$$

In this SDE, $t$ represents the progression's time index, and $n$ denotes the data point's dimension. The SDE involves two crucial components: the drift coefficient $\boldsymbol{f}(\cdot, t)$ and the diffusion coefficient $g(t)$. $\boldsymbol{\omega}$ is a standard $n$-dimensional Brownian motion. These components govern the forward process, which progressively transforms an initial data sample $\mathbf{x}(0) \sim p_{\text{data}}$ into a data point $\mathbf{x}(T) \sim p_T$, belonging to the prior distribution, often a Gaussian distribution.

A reverse-time SDE, running backward from the prior distribution $p_T$ to the data distribution $p_{\text{data}}$, is also defined. This reverse-time SDE is described by (S1.11).

$$\mathrm{d}\mathbf{x} = [\boldsymbol{f}(\mathbf{x}, t) - g^2(t)\nabla_x \log p_t(\mathbf{x})]\mathrm{d}t + g(t)\mathrm{d}\overline{\boldsymbol{\omega}}, \tag{S1.11}$$

where $\mathrm{d}t$ denotes an infinitesimal negative time step, and $\overline{\boldsymbol{\omega}}$ represents the backward-running Brownian motion.

To solve the reverse-time SDE numerically, the actual score function can be approximated using a neural network, denoted as $\mathbf{s}_{\boldsymbol{\theta}}(\mathbf{x}, t) \cong \nabla_{\mathbf{x}} \log p_t(\mathbf{x})$. The objective function $\mathcal{L}(\boldsymbol{\theta})$ can be defined as follows (S1.12).

$$\min_{\boldsymbol{\theta}} \mathcal{L}(\boldsymbol{\theta}) = \min_{\boldsymbol{\theta}} \mathbb{E}_{t \sim U(0,1)} \Big[\lambda(t)\mathbb{E}_{\mathbf{x}(0)}\mathbb{E}_{\mathbf{x}(t)|\mathbf{x}(0)} \Big[\big\|\mathbf{s}_{\boldsymbol{\theta}}(\mathbf{x}(t), t) - \nabla_{\mathbf{x}} \log p_{0t}\big(\mathbf{x}(t)\big|\mathbf{x}(0)\big)\big\|_2^2\Big]\Big], \tag{S1.12}$$

where $\lambda(t)$ is a weighted function controlling the emphasis at different time steps $t$. The unknown true score can be technically replaced with $\nabla_{\mathbf{x}} \log p_{0t}\big(\mathbf{x}(t)\big|\mathbf{x}(0)\big)$ via denoising score matching [23] with $p_{0t}\big(\mathbf{x}(t)\big|\mathbf{x}(0)\big)$being the Gaussian perturbation kernel perturbing the probability density $p_0(x)$ to $p_t(x)$. With the trained neural network $\mathbf{s}_{\boldsymbol{\theta}}(\mathbf{x}, t)$ integrated into (S12), the SDE can be solved numerically.

*5) Normalizing Flow-based Model*

Normalizing Flow (NF)-based models [24] is a kind of generative models that can explicitly learn the data distribution $p(\mathbf{x})$, through an invertible process of transforming a known prior distribution $p_Z(\mathbf{z})$, e.g., Gaussian, into a more complex one that is closely resembling the target data distribution [18]. A key feature of NFs is the invertible transformation $\mathbf{f}(\cdot\,; \boldsymbol{\theta})$, where the transformation between a random vector $\mathbf{z} \sim p_Z(\mathbf{z})$ and a sample in the target domain can be defined as $\mathbf{x} = \mathbf{f}(\mathbf{z}; \boldsymbol{\theta})$ or $\mathbf{z} = \mathbf{f}^{-1}(\mathbf{x}; \boldsymbol{\theta})$. Normalizing Flow-based models aim to approximate the data distribution $p(\mathbf{x})$ with the resulting distribution $p_{\boldsymbol{\theta}}(\mathbf{x})$, which is defined in (S1.13).

$$p_{\boldsymbol{\theta}}(\mathbf{x}) = p_Z\big(\mathbf{f}^{-1}(\mathbf{x}; \boldsymbol{\theta})\big)\left|\det\left(\frac{\partial \mathbf{f}(\mathbf{x}; \boldsymbol{\theta})}{\partial \mathbf{x}}\right)\right|^{-1}, \tag{S1.13}$$

where $\det(\cdot)$ denotes the determinant of the matrix. With the exact formulation of the probability density $p_{\boldsymbol{\theta}}(\mathbf{x})$, NF-based models can effectively learn the parameters $\boldsymbol{\theta}$ by maximum likelihood estimation. In practice, this is usually achieved by minimizing the negative log-likelihood, as specified in (S1.14).

$$\begin{aligned}\min_{\boldsymbol{\theta}} \mathcal{L}_{\text{NF}}(\boldsymbol{\theta}) &= \min_{\boldsymbol{\theta}} \mathbb{E}_{\mathbf{x} \sim p(\mathbf{x})}[-\log p_{\boldsymbol{\theta}}(\mathbf{x})] \\ &= \min_{\boldsymbol{\theta}} \mathbb{E}_{\mathbf{x} \sim p(\mathbf{x})}\left[-\log p_Z\big(\mathbf{f}^{-1}(\mathbf{x}; \boldsymbol{\theta})\big) + \log\left|\det\left(\frac{\partial \mathbf{f}(\mathbf{x}; \boldsymbol{\theta})}{\partial \mathbf{x}}\right)\right|^{-1}\right],\end{aligned} \tag{S1.14}$$

where is proven to be equivalent to minimizing the KL divergence between data distribution $p(\mathbf{x})$ and approximated distribution $p_{\boldsymbol{\theta}}(\mathbf{x})$.

**Supplementary Section**

## S-II. Model Training Strategy

Within each model structure, data-driven fast MRI methods can be categorized into supervised and unsupervised learning models, depending on the training strategy—specifically, whether labeled data (in this case, fully-sampled MR images) is used.

It's important to note that GAN-based models, which are trained using adversarial loss, are treated as a distinct training strategy in this work. Since adversarial loss can be applied in both supervised and unsupervised learning, we discuss GAN-based models in both contexts.

### *A. Supervised Learning*

Most of the supervised fast MRI methods are trained by paired ground truth images $\mathbf{x}$ and undersampled images $\mathbf{A}^{\mathrm{H}}\mathbf{y}$ through a content loss function as in (S2.1):

$$\min_{\boldsymbol{\theta}} \mathcal{L}_{\mathrm{cont}}(\boldsymbol{\theta}) = L\left(g(\mathbf{x}), g\big(f_{\boldsymbol{\theta}}(\mathbf{A}^{\mathrm{H}}\mathbf{y})\big)\right), \quad \text{(S2.1)}$$

in which $g(\cdot)$ is a mapping to the different domain where the distance function $L(\cdot,\cdot)$ is performed.

The mapping $g(\cdot)$ are usually selected among (S2.2).

$$g(\mathbf{x}) = \begin{cases} \mathbf{x}, & \text{for pixel loss,} \\ \boldsymbol{\mathcal{F}}\mathbf{x}, & \text{for frequency loss,} \\ \mathrm{f_{nn}}(\mathbf{x}), & \text{for perceptual loss,} \end{cases} \quad \text{(S2.2)}$$

in which $\boldsymbol{\mathcal{F}}$ denotes the Fourier transform matrix and $\mathrm{f_{nn}}(\cdot)$ denotes a pre-trained deep neural network, e.g., VGG-16 [25].

The distance function $L(\cdot,\cdot)$ is typically selected among (S2.3).

$$L(\mathbf{x}_1, \mathbf{x}_2) = \begin{cases} \|\mathbf{x}_1 - \mathbf{x}_2\|_1, & \text{L1 loss,} \\ \|\mathbf{x}_1 - \mathbf{x}_2\|_2^2, & \text{L2 loss,} \\ \sqrt{\|\mathbf{x}_1 - \mathbf{x}_2\|_2^2 + \epsilon^2}, & \text{Charbonnier loss,} \end{cases} \quad \text{(S2.3)}$$

in which $\epsilon$ is a constant.

For supervised GAN-based fast MRI models, adversarial loss is also applied for training as in (S2.4).

$$\min_{\boldsymbol{\theta}_G} \max_{\boldsymbol{\theta}_D} \mathcal{L}_{\mathrm{GAN}}(\boldsymbol{\theta}_G, \boldsymbol{\theta}_D) \triangleq \mathbb{E}_{\mathbf{x}\sim\mathbf{p}(\mathbf{x})}\left[\log D_{\boldsymbol{\theta}_D}(\mathbf{x})\right] + \mathbb{E}_{\mathbf{y}\sim\mathbf{p}(\mathbf{y})}\left[\log\left(1 - D_{\boldsymbol{\theta}_D}\left(G_{\boldsymbol{\theta}_G}(\mathbf{A}^{\mathrm{H}}\mathbf{y})\right)\right)\right]. \quad \text{(S2.4)}$$

The training loss is usually designed as a linear combination of several different loss functions as in (S2.5)

$$\mathcal{L} = \sum_i \alpha_i \mathcal{L}_i. \quad \text{(S2.5)}$$

For instance, CNN-based models [26], [27] and Transformer-based models [28] are trained with supervised learning strategy using fully-sampling MR images as the ground truth.

### *B. Unsupervised Learning*

Supervised learning methods often require large, high-quality datasets as ground truth, which poses a significant challenge. To address this, many fast MRI methods based on unsupervised learning have been proposed. Unsupervised learning models offer more diverse loss functions. One straightforward approach is to minimize the difference between the $k$ -space data of the reconstructed images and the undersampled measurements $\mathbf{y}$ at the undersampled k-space locations [29] . This can be expressed as in (S2.6).

$$\min_{\boldsymbol{\theta}} \mathcal{L}_{\mathrm{cont}}(\boldsymbol{\theta}) = L\left(\mathbf{y}, \mathbf{A}\big(f_{\boldsymbol{\theta}}(\mathbf{A}^{\mathrm{H}}\mathbf{y})\big)\right), \quad \text{(S2.6)}$$

where $\mathbf{y}$ is the undersampled k-space data, and the undersampled images are given by $\mathbf{A}^{\mathrm{H}}\mathbf{y}$.

Guided by these loss functions, deep networks can learn a substantial amount of image statistics through their architecture, even without ground truth data [30], while preserving the realism of the reconstructed images.

For example, Cole et al. [31] proposed a CNN-based GAN training framework for dynamic contrast enhancement (DCE), 3D cardiac cine, and 4D flow data, where fully-sampled data is difficult to obtain. Additionally, Korkmaz et al. [32] introduced a novel unsupervised MRI reconstruction method using zero-shot adversarial Transformers.

## Supplementary References

[1] R. C. Petersen *et al.*, "Alzheimer's Disease Neuroimaging Initiative (ADNI)," *Neurology*, vol. 74, no. 3, pp. 201–209, Jan. 2010, doi: 10.1212/WNL.0b013e3181cb3e25.

[2] B. H. Menze *et al.*, "The Multimodal Brain Tumor Image Segmentation Benchmark (BRATS)," *IEEE Trans. Med. Imaging*, vol. 34, no. 10, pp. 1993–2024, Oct. 2015, doi: 10.1109/TMI.2014.2377694.

[3] S. Bakas *et al.*, "Identifying the Best Machine Learning Algorithms for Brain Tumor Segmentation, Progression Assessment, and Overall Survival Prediction in the BRATS Challenge," Nov. 2018.

[4] "CAF Project: Brain and Chest." [Online]. Available: https://masi.vuse.vanderbilt.edu/workshop2013/index.php/Segmentation_Challenge_Details

[5] R. Souza *et al.*, "An open, multi-vendor, multi-field-strength brain MR dataset and analysis of publicly available skull stripping methods agreement," *Neuroimage*, vol. 170, pp. 482–494, Apr. 2018, doi: 10.1016/j.neuroimage.2017.08.021.

[6] J. Zbontar *et al.*, "fastMRI: An Open Dataset and Benchmarks for Accelerated MRI," Nov. 2018.

[7] "IXI Dataset." [Online]. Available: https://brain-development.org/ixi-dataset/

[8] "mridata.org." [Online]. Available: mridata.org

[9] A. D. Desai *et al.*, "SKM-TEA: A Dataset for Accelerated MRI Reconstruction with Dense Image Labels for Quantitative Clinical Evaluation," Mar. 2022.

[10] Z. Zhao *et al.*, "Generative Models for Inverse Imaging Problems: From mathematical foundations to physics-driven applications," *IEEE Signal Process. Mag.*, vol. 40, no. 1, pp. 148–163, Jan. 2023, doi: 10.1109/MSP.2022.3215282.

[11] D. P. Kingma and M. Welling, "Auto-Encoding Variational Bayes," Dec. 2013.

[12] C. P. Robert and G. Casella, *Monte Carlo Statistical Methods*. in Springer Texts in Statistics. New York, NY: Springer New York, 2004. doi: 10.1007/978-1-4757-4145-2.

[13] C. Zhang *et al.*, "Conditional Variational Autoencoder for Learned Image Reconstruction," *Computation*, vol. 9, no. 11, p. 114, Oct. 2021, doi: 10.3390/computation9110114.

[14] V. Edupuganti *et al.*, "Uncertainty Quantification in Deep MRI Reconstruction," *IEEE Trans. Med. Imaging*, vol. 40, no. 1, pp. 239–250, Jan. 2021, doi: 10.1109/TMI.2020.3025065.

[15] I. J. Goodfellow *et al.*, "Generative adversarial nets," in *Advances in Neural Information Processing Systems*, Jan. 2014, pp. 2672–2680. doi: 10.3156/jsoft.29.5_177_2.

[16] M. Mardani *et al.*, "Deep Generative Adversarial Neural Networks for Compressive Sensing MRI," *IEEE Trans. Med. Imaging*, vol. 38, no. 1, pp. 167–179, Jan. 2019, doi: 10.1109/TMI.2018.2858752.

[17] Y. Lecun *et al.*, "A tutorial on energy-based learning," 2006.

[18] S. Bond-Taylor *et al.*, "Deep Generative Modelling: A Comparative Review of VAEs, GANs, Normalizing Flows, Energy-Based and Autoregressive Models," *IEEE Trans. Pattern Anal. Mach. Intell.*, vol. 44, no. 11, pp. 7327–7347, Nov. 2022, doi: 10.1109/TPAMI.2021.3116668.

[19] Y. Guan *et al.*, "Magnetic resonance imaging reconstruction using a deep energy-based model," *NMR Biomed.*, vol. 36, no. 3, Mar. 2023, doi: 10.1002/nbm.4848.

[20] J. Ho *et al.*, "Denoising Diffusion Probabilistic Models," in *Advances in Neural Information Processing Systems*, Curran Associates, Inc., 2020.

[21] Y. Song and S. Ermon, "Generative Modeling by Estimating Gradients of the Data Distribution," in *Advances in Neural Information Processing Systems*, Curran Associates, Inc., 2019.

[22] Y. Song *et al.*, "Score-Based Generative Modeling through Stochastic Differential Equations," in *International Conference on Learning Representations*, 2021.

[23] P. Vincent, "A Connection Between Score Matching and Denoising Autoencoders," *Neural Comput.*, vol. 23, no. 7, pp. 1661–1674, 2011, doi: 10.1162/NECO_a_00142.

[24] D. J. Rezende and S. Mohamed, "Variational Inference with Normalizing Flows," in *Proceedings of the 32nd International Conference on International Conference on Machine Learning - Volume 37*, in ICML'15. JMLR.org, 2015, pp. 1530–1538.

[25] K. Simonyan and A. Zisserman, "Very Deep Convolutional Networks for Large-Scale Image Recognition," in *International Conference on Learning Representations*, 2015.

[26] D. Lee *et al.*, "Deep Residual Learning for Accelerated MRI Using Magnitude and Phase Networks," *IEEE Trans. Biomed. Eng.*, vol. 65, no. 9, pp. 1985–1995, 2018, doi: 10.1109/TBME.2018.2821699.

[27] G. Yang *et al.*, "Multiview Sequential Learning and Dilated Residual Learning for a Fully Automatic Delineation of the Left Atrium and Pulmonary Veins from Late Gadolinium-Enhanced Cardiac MRI Images," in *Proceedings of the Annual International Conference of the IEEE Engineering in Medicine and Biology Society, EMBS*, 2018, pp. 1123–1127. doi: 10.1109/EMBC.2018.8512550.

[28] J. Huang *et al.*, "Swin transformer for fast MRI," *Neurocomputing*, vol. 493, pp. 281–304, Jul. 2022, doi: 10.1016/j.neucom.2022.04.051.

[29] Y. Chen *et al.*, "AI-Based Reconstruction for Fast MRI—A Systematic Review and Meta-Analysis," *Proc. IEEE*, vol. 110, no. 2, pp. 224–245, 2022.

[30] V. Lempitsky *et al.*, "Deep Image Prior," in *2018 IEEE/CVF Conference on Computer Vision and Pattern Recognition*, 2018, pp. 9446–9454. doi: 10.1109/CVPR.2018.00984.

[31] E. K. Cole *et al.*, "Unsupervised MRI Reconstruction with Generative Adversarial Networks," *arXiv e-prints*, p. arXiv:2008.13065, 2020.

[32] Y. Korkmaz *et al.*, "Unsupervised MRI Reconstruction via Zero-Shot Learned Adversarial Transformers," *IEEE Trans. Med. Imaging*, p. 1, May 2022, doi: 10.1109/TMI.2022.3147426.